\documentclass[12pt]{article}
\usepackage{amsmath,epsfig}
\usepackage{amsfonts}
\usepackage{amssymb}
\usepackage{amsthm}
\usepackage{mathrsfs}
\usepackage{epstopdf}
\usepackage{appendix}
\usepackage{setspace}
\usepackage{datetime}
\usepackage{booktabs}
\usepackage{subcaption}
\usepackage{caption}
\usepackage{titling}
\usepackage{array}
\usepackage{rotating}
\usepackage{mathtools}
\usepackage{float}

\usepackage{bm}
\usepackage{titlesec}
\titleformat*{\section}{\large\bfseries}
\titleformat*{\subsection}{\bfseries}
\titleformat*{\subsubsection}{\bfseries}
\usepackage[longnamesfirst]{natbib}
\usepackage[english]{babel}
\usepackage{hyperref}
\usepackage{enumerate}
\hypersetup{
	colorlinks,
	linkcolor={blue!50!black},
	citecolor={blue!50!black},
	urlcolor={blue!80!black}
}
\newcommand{\defeq}{\vcentcolon=}
\newdateformat{specialdate}{\monthname[\THEMONTH] \THEDAY, \THEYEAR}
\usepackage[left=1in,right=1in,top=1in,bottom=1in,footskip=0.5in]{geometry}

\usepackage{tikz}
\usetikzlibrary{positioning,shapes,shadows,arrows,calc,decorations.pathreplacing,calligraphy,angles,quotes}
\tikzstyle{line} = [draw, -latex']
\usetikzlibrary{positioning,shapes,shadows,arrows,calc,decorations.pathreplacing,calligraphy}
\usetikzlibrary{tikzmark,fit,shapes.geometric,decorations.shapes}
\tikzset{decorate sep/.style 2 args=
	{decorate,decoration={shape backgrounds,shape=circle,shape size=#1,shape sep=#2}}}

\usepackage{comment} %
\usepackage{CJKutf8}

\newtheorem{innercustomgeneric}{\customgenericname}
\providecommand{\customgenericname}{}
\newcommand{\newcustomtheorem}[2]{%
	\newenvironment{#1}[1]
	{%
		\renewcommand\customgenericname{#2}%
		\renewcommand\theinnercustomgeneric{##1}%
		\innercustomgeneric
	}
	{\endinnercustomgeneric}
}

\newcustomtheorem{customass}{Assumption}
\newcustomtheorem{customlemma}{Lemma}

\AtBeginDocument{
	
	}

\newcommand{\killpunct}[1]{}

\makeatletter
\renewenvironment{abstract}{%
	\if@twocolumn
	\section*{\abstractname}%
	\else %
	\begin{center}%
		{\bfseries \normalsize\abstractname\vspace{\z@}}%
	\end{center}%
	\quotation
	\fi}
{\if@twocolumn\else\endquotation\fi}
\makeatother

\definecolor{maroon}{rgb}{0.5, 0.0, 0.0}

\newcolumntype{P}[1]{>{\centering\arraybackslash}p{#1}}

\AtBeginDocument{
}

\global\long\def\E{\mathbb{E}}

\newcommand{\Paren}[1]{\left({#1}\right)}
\newcommand{\Brack}[1]{\left[{#1}\right]}
\newcommand{\Curly}[1]{\left\{{#1}\right\}}
\newcommand{\Abs}[1]{\left|{#1}\right|}
\newcommand{\covhat}{\widehat{\mathrm{cov}}}
\newcommand{\varhat}{\widehat{\mathrm{var}}}
\newcommand{\qtld}{\widetilde{q}}
\newcommand{\lamtld}{\widetilde{\lambda}}
\newcommand{\flowstock}{\$\textrm{flow}^{\textrm{stock}}}
\newcommand{\flowfund}{\$\textrm{flow}^{\textrm{fund}}}
\newcommand{\flowfactor}{\textrm{flow}^{\textrm{factor}}}
\newcommand{\rfund}{r^{\textrm{fund}}}
\newcommand{\weightfund}{\textrm{weight}^{\textrm{fund}}}
\newcommand{\citepossessive}[1]{\citeauthor{#1}'s (\citeyear{#1})}

\newcommand{\rhat}{\widehat{r}}
\newcommand{\omegamax}{\omega^{\textrm{max}}}
\newcommand{\lamstock}{\lambda^{\text{stock}}}
\newcommand{\qstock}{q^{\text{stock}}}
\newcommand{\qstockh}{q^{\text{stock}, h}}

\newcommand{\rf}{r_\textrm{rf}}
\newcommand{\ftld}{\widetilde{f}}
\newcommand{\Price}{\mathbb{P}\mathrm{rice}}
\newcommand{\betahat}{\widehat{\beta}}

\usepackage[utf8]{inputenc}  %
\newcommand{\block}[1]{%
  \raisebox{\dimexpr(\fontcharht\font`X-1em)/2}{\rule{1em}{#1\dimexpr1em/8}}%
}
\DeclareUnicodeCharacter{2581}{\block{1}}
\DeclareUnicodeCharacter{2582}{\block{2}}
\DeclareUnicodeCharacter{2583}{\block{3}}
\DeclareUnicodeCharacter{2584}{\block{4}}
\DeclareUnicodeCharacter{2585}{\block{5}}
\DeclareUnicodeCharacter{2586}{\block{6}}
\DeclareUnicodeCharacter{2587}{\block{7}}
\DeclareUnicodeCharacter{2588}{\block{8}}

\begin{document}
\sloppy

\setlength{\droptitle}{-2.5em}   %
\title{Quantity, Risk, and Return%
	\thanks{We thank 
		Caio Almeida, Federico Bandi, Hank Bessembinder, Andrew Chen, Hui Chen, Zhi Da, Darrell Duffie, Zhiyu Fu, Robin Greenwood, Zhiguo He, Ben H\'ebert, Shiyang Huang, Bryan Kelly, Ralph Koijen, 
        Nikolai Roussanov, 
        Serhiy Kozak, Martin Lettau, Jiacui Li, Dong Lou, Sydney Ludvigson, Jun Pan, Paolo Pasquariello, Nagpurnanand Prabhala, Seth Pruitt, Alessandro Rebucci, Paul Schultz, Dongho Song, Yang Song, Zhaogang Song, Semih \"Usl\"u, Wei Wu, Jun Yu; conference discussants, Andrea  Eisfeldt, Raymond Kan; 
		and participants at the NBER Asset Pricing Program Meeting and NFA for valuable comments and suggestions. 
	The previous version of the paper, titled ``A Factor Framework for Cross-Sectional Price Impacts,'' was presented at the Fed Board, Wolfe, MFA, Southern Methodist University, FMCG, DC Junior, SoFiE, CICF, U.\ of Macau, CityU HK, CUHK, SAFE Asset Pricing Workshop, UT Dallas Finance Conference, World Symposium on Investment Research, FMA Applied Finance, Michigan Mitsui Symposium, with discussants Aref Bolandnazar, Aditya Chaudhry, Thummim Cho, Fotis Grigoris, Badrinath Kottimukkalur, Xin Liu, Marcel M\"uller, Andrey Pankratov, Oleg Rytchkov, Andrea Vedolin. We also thank them for valuable feedback and suggestions.}}
\author{
\mbox{Yu An}\thanks{Carey Business School, Johns Hopkins University; \href{mailto:yua@jhu.edu}{yua@jhu.edu}.} \hspace{1cm}
\mbox{Yinan Su}\thanks{Carey Business School, Johns Hopkins University; \href{mailto:ys@jhu.edu}{ys@jhu.edu}.}  \hspace{1cm}
\mbox{Chen Wang}\thanks{Mendoza College of Business, University of Notre Dame; \href{mailto:chen.wang@nd.edu}{chen.wang@nd.edu}.} \\ \\
{\normalsize February 17, 2025}}
\date{}
\maketitle

\vspace*{-1cm}
\begin{abstract}

	\onehalfspacing

	\noindent We propose a new model of expected stock returns that incorporates quantity information from market trading activities into the factor pricing framework. We posit that the expected return of a stock is determined by not only its factor risk exposures ($\beta$) but also the factor's quantity fluctuations ($q$) induced by noise trading flows, and hence term the model \emph{beta times quantity} (BTQ). The rationale is that a factor's premium should be higher when sophisticated investors have absorbed flows of stocks with high exposure to that factor. The BTQ model provides a compelling risk-based explanation for stock returns, which is otherwise obscured without considering the quantity information. The cross-sectional risk-return association, which is nearly flat unconditionally, strongly depends on the quantity variable. The structured BTQ model reliably predicts monthly stock returns out of sample, and addresses the factor zoo problem by selecting a small number of factors.

    \pagenumbering{gobble}

    \vfill
    \vspace{8pt}

    \vspace{8pt}
\end{abstract}

\doublespacing

\clearpage

\setcounter{page}{1}
\pagenumbering{arabic}

\section{Introduction}

Explaining the expected returns of different stocks is a central question in asset pricing. The theoretical answer is clear---risk---investors are averse to risk and require compensation for bearing risk. Therefore, riskier investments should earn higher expected returns in equilibrium. However, the empirical answer has proven more complicated: evidence of the risk-return tradeoff, such as their positive association in the cross section, is elusive in data; and risk-based models hardly predict individual stock returns, in contrast to unstructured predictions using firm characteristics and machine learning models.\footnote{\citet{lopez2023common}, for example, question whether factor exposure can explain the cross section of expected stock returns. See Footnote \ref{fn:rr} for other papers reporting an elusive risk-return relationship and Footnote \ref{fn:pred} for those focused on predicting stock returns.} A revamped model is critically needed for the risk-based approach to expected returns.

This paper makes headway in this important area by incorporating a new aspect of risk's economic role in determining asset prices---the \emph{quantity} variation in investors' risk holdings induced by trading flows. Many existing efforts focus on the statistical aspects of risk, such as identifying the common factors and estimating factor premiums, and on the properties of the securities per se, such as risk exposures and firm characteristics.\footnote{These related topics constitute a large and growing body of literature. We contribute to three sub-areas with references listed in Footnotes \ref{fn:rr}, \ref{fn:pred}, and \ref{fn:zoo}, respectively.} In contrast, we show that the canonical risk framework equipped with the quantity variables, which are constructed from market trading activities and are about sophisticated investors' risk-holding conditions, yields a compelling explanation for the cross section of expected returns.

We integrate quantity into factor pricing by considering market trading activity's effect on sophisticated investors' risk holdings and, in turn, their required compensation for bearing risk. First, we acknowledge that the market is not populated with representative agents but is modeled with two groups of investors: noise investors (such as retail investors) and sophisticated investors (such as hedge funds and market makers). Noise investors generate large and correlated flows in individual stocks. Sophisticated investors take the other side of these trades, which causes fluctuations in the quantities of their holdings of the underlying systematic risks. For example, if noise investors sell a large quantity of value stocks with high HML (high-minus-low) loadings, sophisticated investors' holdings of the HML risk will increase. The sophisticated investors are the marginal investors whose demand determines asset prices. We posit that they require greater compensation for a systematic risk factor when they hold more of it. 
This leads to a key innovation in factor model specification: a factor's premium varies with the factor's quantity fluctuations induced by trading flows. Meanwhile, sophisticated investors enforce no-arbitrage pricing across stocks, so the canonical factor pricing condition still holds. These two forces combined give rise to our main empirical model, in which the expected return of a stock is determined by the interactions of its factor risk exposures ($\beta$) and the factors' quantity fluctuations induced by trading flows (shortened to ``quantity'' or $q$ throughout the paper), which we term the \emph{beta times quantity} (BTQ) model.

This framework, though abstracted from many details of the market microstructure, captures a significant economic force central to risk aversion that has long been missing in empirical studies of risk and return. Our approach draws from the literature that studies the price impacts of noise trading flows.%
\footnote{See \cite{gabaix2022search} for a review. We discuss related papers in detail further below.}
The novelty lies in integrating quantities into the factor pricing framework and adapting the ``price impacts'' mechanism to explain expected future stock returns.
This advance enhances the empirical power of workhorse methods in cross-sectional asset pricing, addresses previous limitations, and leads to important empirical discoveries in three aspects.

First, quantity information elicits risk-return tradeoff relationships that would otherwise be obscured. Previous studies report a flat security market line (SML, which plots expected return $\E r$ against market $\beta$), inconsistent with the theoretical premise of high-risk-high-return.\footnote{\cite{black1972}, \cite{black1972capital}, and \cite{frazzini2014betting} report a flat SML. Along this direction but with more involved investigations, \cite{lopez2023common} question whether common factor exposure ($\beta$) really explains the cross-sectional variation in expected returns.\label{fn:rr}} However, a significant positive $\beta$-$\E r$ association emerges {\it conditional on} high levels of market factor $q$. That is, the risk-aversion implied high-risk-high-return association holds when sophisticated investors have absorbed more market factor quantity. In this view, the previously reported flat SML is an unconditional average when the quantity information is ignored.\footnote{Relatedly, \cite{hong2016speculative}, \cite{jylha2018margin}, and \cite{hendershott2020asset} find varying slopes of the SML conditional on investor disagreement, margin requirements, and whether returns occur during the day or night.} This positive association between the factor-level $q$ and the factor's risk-return tradeoff (i.e., factor premium) holds across SMLs of other factors and in Fama-MacBeth regressions conditional on quantity information.

Second, quantity information enables a risk-based model that predicts individual stock returns. A central goal of asset pricing is to explain conditional expected returns, with the statistical prediction of individual stock returns serving as a touchstone for proposed explanations. This task is empirically difficult, and researchers have only recently achieved significant progress by using unstructured machine learning models designed for forecasting and a large number of firm characteristics, which inevitably sacrifice interpretability. The state-of-the-art machine learning methods can reliably predict stock returns at the monthly horizon, although the explained variation is small given the low signal-to-noise nature of market prices.\footnote{Studies on stock (and equity portfolio) return forecasting include \citet{fama2008dissecting}, \citet{welch2008comprehensive}, \citet{koijen2011predictability}, \citet{rapach2013forecasting}, and \citet{lewellen2015cross}. More recent advances with machine learning methods include \citet{gu2020empirical}, \citet{feng2018deep}, \citet{freyberger2020dissecting}, \citet{choi2023alpha}, and \citet{kelly2024virtue}.\label{fn:pred}} In contrast, we build an economically grounded predictor that interacts stock-level factor exposures ($\beta$) with factor-level quantity fluctuations ($q$). The resulting beta times quantity (BTQ) model reliably predicts the panel of monthly individual stock returns with an out-of-sample (OOS) $R^2$ of around 1\% in various robustness settings, a level comparable to high-dimensional machine learning models. The predictability is robust to different sample periods, firm size groups, and model specifications. Without quantity, the ``$\beta$-only'' model exhibits no predictive power, aligning with the reported null result that using risk alone hardly explains expected stock returns \citep{lopez2023common}.

Third, quantity offers a new perspective for addressing the factor zoo problem and providing new results on factor selection. The proliferation of proposed factors challenges the asset pricing literature in identifying factors that are important for expected returns and investors' pricing decisions. The traditional tests focus on the {\it existence} of factor premium: essentially, whether there is a positive spread in expected returns between stocks with high and low factor exposures in the cross section.\footnote{The proliferation of proposed factors to explain the cross section of expected stock returns (a.k.a.\ the factor zoo problem) is noted by \cite{cochrane2011presidential}, \cite{harvey2016and}, \cite{mclean2016does}, and \cite{hou2017comparison}. Existing studies address the problem by selecting or ``shrinking'' the factors (broadly speaking, estimating a low-dimensional factor space), including \citet{feng2020taming}, \citet{lettau2020factors}, \cite{kozak2020shrinking}, \cite{giglio2021thousands}, and \citet{giglio2021asset}. Essentially, they discipline a factor by whether its factor premium is positive (i.e., positive cross-sectional risk-return association). In this sense, these are developments of the more traditional \citet{fama1973risk} method.\label{fn:zoo}} Our new test asks an upgraded question about {\it changes} in factor premium driven by quantity: whether the expected return spread \emph{widens} when the sophisticated investors' factor quantity ($q$) is high (and vice versa).\footnote{The new test is analogous to the difference-in-differences (DID) analysis commonly used in applied microeconomics: $\beta$ captures the cross-sectional variation while $q$ provides the time-series variation in expected returns. In this analogy, the ``$\beta$-only'' model has only one dimension of ``difference'' and assumes constant factor premiums.}
For one, using quantity as an instrument for factor premium should provide more variation and, hence, greater identification power. More importantly, this upgrade is more informative of the economic mechanism through which risk aversion takes place and, therefore, should lead us closer to identifying the fundamental risks to investors. We find the market factor is the most prominent across various specifications, while other selected factors include betting-against-beta, volatility, idiosyncratic risk, and value. %
These results are obtained by conducting variable selection (Lasso) from a BTQ configuration that includes
a large number of candidate factors
(including 153 factors from \citealp{jensen2023there}, henceforth JKP). Alternatively, pre-processing the candidate factors with principal component analysis (PCA) to ``shrink the cross section'' \citep*{kozak2020shrinking} leads to a similar but even more parsimonious result in which only the first two principal components are selected, and the return predictive power is equally strong.

In summary, these three key results highlight the importance of incorporating quantity into the factor pricing framework to empirically establish a risk-based explanation of expected returns. To sharpen this argument, we compare the BTQ model with two alternative baseline models that contain only risk or only quantity, respectively.

The first alternative, the ``$\beta$-only'' model, represents the traditional factor pricing framework where risk (specifically, $\beta$) is the sole determinant of differences in expected stock returns. Our main results benchmark BTQ against the ``$\beta$-only'' baseline, and show compelling empirical improvements in various familiar workhorse asset pricing settings: the security market line (SML), Fama-MacBeth regressions, and stock return prediction. Moreover, as discussed earlier, incorporating quantity provides an additional perspective for selecting factors from the ``zoo'' based on their economic relevance. 
Future studies can easily test newly proposed factors, as a factor's BTQ term can be easily constructed from the factor's return series.
These properties highlight the advantages and broad applicability of incorporating quantity into factor pricing for future research.

Second, relative to the   ``quantity-only'' alternative, the emphasis on risk %
is embedded in our construction of the $q$ variables. They track the fluctuations of sophisticated investors' factor risk holdings induced by retail trading flows. This is achieved by aggregating stock-level flows to the factor level according to each stock's factor exposure ($\beta$), in a way similar to ``portfolio beta'' in risk management.\footnote{Stock-level noise trading flows from retail investors are constructed using mutual fund holdings and flow data, following standard procedures in the literature \citep{coval2007asset, froot2008institutional, lou2012flow}. See Section \ref{subsec:qconstruction} for the complete construction procedure of $q$.} For example, if noise investors sell a large quantity of value stocks with high HML loadings, then from the perspective of sophisticated investors, the $q$ of HML increases accordingly. This construction underscores the economic mechanism in which investors are averse to systematic risk, with their degree of aversion adjusting based on the amount of systematic risk they bear.\footnote{Appendix \ref{sec:microfound} provides the theoretical foundation that formalizes this statement.}

This setup is contrasted with the ``quantity-only'' model, where stock-level flows and quantity variations directly affect stocks' expected returns, bypassing the factor structure (see Figure \ref{fig:illustration} for a comparison of the architectures). This alternative model does not adhere to the cross-sectional no-(statistical)-arbitrage condition and implies that investors are averse to the physical quantity of stocks rather than the systematic risk they represent. Empirically, we find little to no predictive power for stock returns in various implementations of the ``quantity-only'' model. This comparison highlights the critical role of risk in the BTQ model. It is consistent with the view that statistical arbitrage activities by some sophisticated investors are effective in determining the cross section of expected returns, even in the presence of significant impacts of noise trading flows on prices \citep{kozak2018interpreting}. It is also related to the distinction between micro and macro elasticities: stocks with similar risk loadings are close substitutes, whereas the demand for systematic risks is more inelastic to price \citep{gabaix2022search, li2022price}.

We provide further evidence to support the economic interpretation that quantity explains expected stock returns through factor risk. We find that different factors' $q$ variables provide distinct pricing information along their respective risk dimensions, and that BTQ variants crossing one factor's $q$ with other factors' $\beta$ terms fail to predict future stock returns. This result highlights that the quantity-risk premium association is independently robust across factors.
We also find that no other conditioning variables---such as factor momentum signals and a comprehensive set of macroeconomic variables (dividend yield, default spread, income growth, etc.)---can substitute for $q$ in reproducing BTQ's predictive power.
This result rejects the idea that the factor premium variation we report is driven by other underlying economic forces and that the quantity variable is merely a facade.

In summary, the core message is that both quantity and risk matter for expected stock returns. At a high level, this naturally stems from the interaction between sophisticated investors and noise traders \citep{shleifer1990noise}. %
Considering the interaction allows us to bridge factor pricing (which emphasizes rational agents' aversion to risk) and the price impact of noise flows (which emphasizes noise traders' non-fundamental flows cause price dislocation). The contribution of this empirical paper is providing a playbook for integrating quantity information into the canonical factor framework and showing its significant improvement to factor pricing models' empirical relevance.

\paragraph{Literature.} This paper is related to two frameworks in the literature but has differences in its objective and approach. First, we do not treat flow or quantity fluctuations as a source of risk, and the constructed quantity time-series variables are not new risk factors, as in a recent paper by \citet{dou2022common}.\footnote{Other papers that treat flow or quantity information as sources of risk include \cite{delong1990noise}, \cite{shleifer1997limits}, \cite{lo2000trading}, \cite{hasbrouck2001common}, \cite{adrian2014financial}, and \cite{he2017intermediary}.} 
Instead, we still use previously proposed factors, and the newly proposed factor-level quantity variables work together with risks in the form of ``beta times quantity.''

Second, this paper belongs to the growing  literature on demand-based asset pricing, which shows that investor demand plays a critical role in determining asset prices and that incorporating flow and quantity data can improve empirical asset pricing research \citep[etc.]{koijen2019demand, gabaix2022search, koijen2024investors,haddad2024competitive}. In particular, a strand of the literature estimates factor-level price multipliers, including \cite{teo2004style}, \cite{peng2021factor}, \cite{ben2022ratings}, \cite{li2022drives}, \cite{li2022price}, and \cite{huang2024noise}. We focus on the empirical study of expected future stock returns rather than impacts on contemporaneous prices.%
\footnote{Appendix \ref{sec:microfound} discusses the connection between flow/quantity's impact on contemporaneous prices and expected future returns (risk premiums) with a formal theoretical microfoundation.} 
In this regard, our goal and approach align more closely with the factor pricing literature: we explicitly model the factor structure of returns; maintain the associated factor pricing condition; 
and take return prediction accuracy as the central criterion of empirical success.%
\footnote{\citepossessive{koijen2019demand} demand system models a stock's demand elasticities with respect to a) the stock's price (or the market capitalization) and b) the stock's factor risk exposures (proxied by the stock's characteristics). Neither is exactly our channel: a) operates at the stock level, rather than the factor level, and b) is about the cross-sectional demand variation related to a stock's factor loadings or characteristics, rather than time-series demand variation driven by aggregated factor risk quantity. They use the factor framework as a microfoundation for the characteristic-based demand system. Related to our research objective, one of their applications shows that mean reversion in latent demand introduces a new source of predictability for cross-sectional variation in stock returns.} 

Additionally, our use of the no-arbitrage factor pricing (APT) condition to link the cross-sectional quantity-return relationship also differs from existing approaches, such as nested logit demand systems in \cite{koijen2020exchange}, \cite{bretscher2024institutional}, and \cite{jiang2024portfolio}, controlling for close substitutes as in \cite{chaudhary2023corporate}, and mean-variance optimization as in \citet{vayanos2021preferred}, \cite{davis2024asset}, and \cite{jansen2024granular}.\footnote{Relatedly, \cite{berk2016assessing}, \cite{barber2016factors}, and \cite{ben2022mutual} use a revealed preference approach to determine which factors investors care about, and \cite{bretscher2023investor} show that betas measured relative to institutional investor portfolios explain stock returns.
}

In the remainder of the paper, Section~\ref{sec:theory} provides the theoretical motivation, empirical model, and methods; Section~\ref{sec:construct} constructs the quantity and other empirical measures; Section~\ref{sec:emp_res} presents the main empirical results; Section~\ref{sec:mustcombine} shows that quantity must be combined with risk to forecast returns; Section~\ref{sec:channels} investigates alternative economic channels; Section~\ref{sec:conclusion} concludes.

\section{Theoretical motivation, empirical model, and methods} \label{sec:theory}

\subsection{Theoretical motivation}
\label{subsec:theoretical_motivation}

The theoretical rationale for integrating quantity information into factor pricing to explain \textit{expected} stock return is that market trading activities affect sophisticated investors' risk holdings and, in turn, their required compensation for bearing risks. We focus on a prominent channel where the noise trading flows---a significant type of trading activities---matter for the central element of asset pricing, namely the factor premium, although there can be many other market microstructure mechanisms in which trading activities generate contemporaneous price impacts. We outline this theoretical channel below, and %
Appendix \ref{sec:microfound} provides the formal microfoundation.

Suppose the market is populated with two groups of investors: noise investors and sophisticated investors. Noise investors, such as retail traders, generate uninformed flows in and out of individual stocks over time. The noise flows are large and correlated across stocks, which can induce significant fluctuations when aggregated to the factor level.\footnote{Previous studies report  (which we also confirm empirically) that the retail flows are not only significant in magnitude but also correlated across stocks due to the commonality in retail investors' trading behaviors. The correlation aligns with investment styles, such that, say in one period, retails tend to sell growth stocks and in the next, they buy small stocks (\citealp{li2022drives}; \citealp*{huang2024noise}). This fact supports that retail flows can induce significant fluctuations in the quantity of risk when aggregated to the factor level.}

Sophisticated investors, such as hedge funds and market makers, take the other side of the retail trades by absorbing the noise flows and supplying liquidity. Therefore, noise flows induce fluctuations in the sophisticated investors' holding quantities of the underlying systematic risks. For example, if retail investors sell lots of value stocks with high HML exposures, then sophisticated investors will accumulate more HML risk holdings. The aggregation from stock-level flows to factor-level quantities accounts for each stock's factor exposure ($\beta$) in the fashion of ``portfolio beta'' commonly used in risk management practice (see Section \ref{subsec:qconstruction} for aggregation details). The sophisticated investors are the marginal investors whose risk-holding conditions drive asset prices.
They have limited capacity to bear risk and absorb flows, and require greater compensation for a systematic risk factor when they hold more of it.\footnote{Limited risk-bearing capacity can stem from liquidity or balance-sheet constraints (e.g., \citealp*{adrian2014financial, gabaix2015international, he2017intermediary, kondor2019liquidity, haddad2021intermediaries}). In particular, \citet{eisfeldt2024interdealer} and \citet{kargar2021heterogeneous} emphasize that heterogeneity within the intermediary sector can further lead to risk misallocation, offering a novel explanation for why liquidity is priced.}
This gives rise to the key model specification that a factor's premium varies with the factor's quantity fluctuations induced by trading flows, and we hypothesize that the relationship is positive.%
\footnote{Appendix \ref{sec:microfound} provides the formal theoretical model to microfound the quantity-factor premium association (Eq.\ \ref{eq:fp_specification} specified further below). This specification is related to the demand-based literature, which emphasizes the ``price multiplier'' is high, or, in other words, the demand is inelastic to price \citep{gabaix2022search}. The empirical distinction is that our goal is explaining the expected future returns, rather than the contemporaneous price impact (although the two are theoretically connected as high expected returns imply low current prices).}
Meanwhile, sophisticated investors enforce no-arbitrage pricing across stocks, so the canonical factor pricing condition still holds.\footnote{Enforcing the cross-sectional APT condition is consistent with \citet*{kozak2018interpreting}, who argue that cross-sectional no-arbitrage conditions are still valid in the presence of noise traders as long as there exist some sophisticated investors. This is in contrast to those models that directly link each individual stock's flow to its price. See Section \ref{subsec:qonly} for the comparison against this benchmark.} These two forces combined imply the main empirical model specified below, in which both the stock's factor risk exposures ($\beta$) and factor quantity ($q$) determine its expected return.

\subsection{Empirical model}
\label{subsec:model}

The empirical model starts with the canonical factor pricing framework, in which the cross section of stock returns follows a factor structure
\begin{align}
	r_{i,t+1} = \sum_{k=1}^K \beta_{i,k,t}  f_{k,t+1}  + \epsilon_{i,t+1}, &  & \forall i, t,
\end{align}
where $r_{i,t+1}$ is the excess return of stock $i$ in month $t+1$, $k$ indexes factors, $f$ is factor return (zero-cost or excess return), and $\beta$ is the stock's factor exposure, which is subsequently estimated using realized daily returns. According to the APT \citep{ross1976arbitrage}, the cross section of expected return follows the factor pricing condition,
\begin{align}\label{eq:betaprice}
	\E_t[r_{i,t+1}] = \sum_{k=1}^K \beta_{i,k,t} \mu_{k, t}, &  & \forall i, t,
\end{align}
where $\mathbb{E}_t[r_{i,t+1}]$ is the conditional expected stock return, our research object, and $\mu_{k, t}$ is the factor premium conditional on time-$t$ information.

The departure from the canonical framework lies in the modeling of the factor premium. According to the theoretical motivation above, we specify that the factor premium is not a constant but varies with the factor's quantity fluctuations induced by trading flows.
\begin{align}
	\mu_{k, t}=\mu_k(q_{k,t}) = \mu_k + \lambda_k q_{k,t}, &  & \forall k, t,\label{eq:fp_specification}
\end{align}
where the first is a general non-parametric form in which $\mu_k(~\cdot~)$ is an unspecified function of $q_{k,t}$, while the second is the parametric linear specification, which is implemented in most empirical settings.\footnote{%
The various parametric and non-parametric empirical methods are detailed further below in Section \ref{subsec:methods}.}
Parameter $\mu_k$ corresponds to the constant factor premium, which is the key interest of estimation in traditional factor pricing tests. The linear coefficient $\lambda_k$ is the new central parameter of interest, which measures the sensitivity of the factor premium to the factor's quantity fluctuations.%
\footnote{Appendix \ref{sec:inelastic} provides a microfoundation for the linear specification and the economic interpretation of its parameters. The parameter $\lambda_k$ reflects the inelasticity of sophisticated investors' demand for factor risk. This inelasticity is further attributed to two primitives: high risk aversion and the limited capital of sophisticated investors relative to the aggregate stock market.}

Plugging the factor premium specification into the factor pricing condition (Eq.\ \ref{eq:fp_specification} into Eq.\ \ref{eq:betaprice}), we arrive at the main empirical model, the beta times quantity (BTQ) model of expected stock returns:
\begin{align}
	\mathbb{E}_t[r_{i,t+1}]  =  \Paren{\sum_{k=1}^K \mu_k \beta_{i,k,t}}  + \sum_{k=1}^K \lambda_k  \beta_{i,k,t} q_{k,t}, &  & \forall i, t.\label{eq:BTQ}
\end{align}
The first summation term is the traditional factor pricing model, which we refer to as the ``$\beta$-only'' model, serving as the baseline in empirical comparisons. The second is the new beta times quantity (BTQ) term. In empirical implementation, we often find the $\beta$-only term is so close to zero (and so noisy for explaining expected returns) that including it in the BTQ model can even hurt the empirical fit. Therefore, the BTQ model typically omits the $\beta$-only term in parentheses and only includes the beta times quantity term.

The key hypothesis implied by the theoretical motivation is that, for a ``true'' fundamental risk factor $k$, $\lambda_k > 0$. The hypothesis means that the cross-sectional return dispersion between high and low $\beta$ stocks widens when the factor's quantity is high. This is similar to the difference-in-differences (DID) analysis: $\beta$ captures the cross-sectional variation in expected returns while $q$ provides the time-series variation. In other words, the observed factor risk aversion is stronger when $q$ is high.
This offers a new perspective compared to the traditional hypothesis $\mu_k > 0$, which asks whether higher exposure to that factor is associated with higher \emph{average} returns, i.e., only the first ``difference'' in the DID analysis. The new test has more identification power provided by the time-series variation in $q$. More importantly, this test has more economic relevance since the $q$ variation tracks sophisticated investors' risk-holding conditions. Hence, we are no longer inferring investors' risk pricing process from asset and asset price information alone. Therefore, the new framework can lead us closer to identifying the fundamental risks that investors care about.

The model allows for multiple factors and allows each to have a different $\lambda_k$ coefficient. This is useful for testing each factor's marginal importance in a joint setting, controlling for other factors' contribution to expected returns.\footnote{
	The model (Eq.\ \ref{eq:fp_specification}) specifies that factor $k$'s premium $\mu_{k,t}$ is affected only by its own quantity $q_{k,t}$, not by the quantities of other factors $q_{j,t}$. %
	Allowing for cross-factor impacts would complicate the model, increasing the number of parameters from $K$ to $K^2$, which becomes impractical for large $K$. Our most salient empirical results are attained with single-factor settings, where cross-factor impacts are irrelevant. %
    }

An important property of the sign of $\lambda_k$ is noted. Regardless of the sign of the factor (e.g., small-minus-big or big-minus-small), the sign of $\lambda_k$ should, theoretically speaking, always be positive. This is because when factor return $f$ flips its sign, both $\beta$ and $q$ flip their signs, and $\beta$ times $q$ remains unchanged. A positive $\lambda_k$ estimate, nonetheless, is not empirically guaranteed. Thus, it provides another layer of testing for the risk-based theory, regardless of the specification of the factor's sign. A negative $\lambda_k$ estimate would be an unambiguous rejection of the risk-based theory, and the empiricist could not blame the ``wrong'' sign of the factor as an excuse. Notice that $\mu_k$ in the traditional $\beta$-only model does not have this property: big-minus-small would have a negative $\mu_k$.

We focus on testing the hypothesis ``$\lambda_k > 0$'' in the cross-sectional setting of the BTQ model (Eq.\ \ref{eq:BTQ}), not in the time series context of predicting factor returns $f_{k,t+1}$ with $q_{k,t}$. Although the BTQ model is theoretically motivated by the time-series specification of factor premium (Eq.\ \ref{eq:fp_specification}), empirically, a positive time-series predictive coefficient between $q_{k,t}$ and $f_{k,t+1}$ is far from implying the cross-sectional hypothesis of $\lambda_k > 0$. The gap between the two is the cross-sectional variation of the risk exposures ($\beta$), which is not present in the time series setting. A similar gap is familiar in the traditional factor pricing framework: a long-short portfolio with a high average return does not guarantee that it is a priced factor in cross-sectional tests, such as the Fama-MacBeth regressions.

\subsection{Empirical methods}\label{subsec:methods}

We use a series of empirical methods to estimate and test the BTQ model. The methods are presented as upgrades of familiar procedures in asset pricing, such as the security market line, Fama-MacBeth factor premium estimates, and return prediction exercises, for ease of comparison and to demonstrate the value of incorporating quantity information into the factor model. We present an overview of the methods here, while the details are provided when presenting the empirical results in Section \ref{sec:emp_res}.

From the methodological perspective, the progression of the methods can be seen as gradually adding parameterization to the model of expected stock returns. To start with, the familiar security market line (SML) can be seen as a simple non-parametric model, $\E_t[r_{i,t+1}] = \textit{Er}(\beta_{i,k,t})$, where $\textit{Er}(~\cdot~)$ is an unspecified function. (The SML is typically estimated with the market beta, i.e.,  $k$ = MKT, but we implement it with other factors as well.) The conditional SML (Section \ref{subsec:sml}) upgrades it to a bi-variate non-parametric model that includes $q$, $\E_t[r_{i,t+1}] = \textit{Er}(\beta_{i,k,t}, q_{k, t})$. We estimate this non-parametric model with a simple kernel method by binning observations of $\beta$ and $q$. This method is easy to interpret via the familiar SML plot, and clearly shows that $q$ is a highly relevant variable in the expected return function (\textit{Er}) with significant effects on the risk-return ($\beta$-$\E r$) relation.

The second method, the quantity upgraded Fama-MacBeth factor premium estimates, is semi-parametric (Section \ref{subsec:FM}). It imposes a linear relationship between risk ($\beta$) and expected return according to APT, but is still non-parametric about $q$'s effect: $Er(\beta_{i,k,t}, q_{k, t}) = \beta_{i,k,t}\mu_k(q_{k, t})$, where the factor premium function $\mu_k(~\cdot~)$ is left unspecified. It is still estimated non-parametrically by binning $q$ and then averaging the returns of the Fama-MacBeth factor mimicking portfolio (FMP, which are coefficients from the cross-sectional regression $r_{i,t+1}$ on $\beta_{i,k,t}$) within each bin.

Third, once the $\mu_k(~\cdot~)$ function is also specified as linear, we arrive at the parametric BTQ model $Er(\beta_{i,k,t}, q_{k, t}) = \lambda_k \beta_{i,k,t} q_{k,t}$. The parametric setting easily accommodates multiple factors, and is estimated with a linear predictive regression on the panel of monthly stock returns $r_{i, t+1} = \sum_{k = 1}^K \lambda_k \beta_{i,k,t}q_{k, t} + error_{i, t+1}$ (Section \ref{subsec:retpred}). Notice that each factor's beta times quantity (BTQ) term together serves as a predictor, and the BTQ terms of different factors serve as multivariate predictors. Predicting stock returns has experienced significant progress with firm characteristics and machine learning models. We follow the literature's stock-month panel setup and use the same measure of empirical success: the monthly stock return predictive $R^2$ evaluated out-of-sample (OOS). This is our (and also the literature's) key evaluation metric for ``explaining expected stock returns.''

Lastly, in response to the factor zoo problem, when the number of candidate factors ($K$) is large, the number of BTQ predictors grows accordingly to more than 100. In such a setting, we use machine learning methods designed for high-dimensional prediction, such as Lasso, to select a small number of priced factors (Section \ref{subsec:factorzoo}). By inducing sparsity in the $\lambda_k$ coefficients, Lasso allows us to select a small number of BTQ terms and reveal which factors are priced in a joint setting, controlling for other factors. Additionally, we follow \cite*{kozak2020shrinking} and pre-process the candidate factors with principal component analysis (PCA). Then, we supply the principal component factors to the same BTQ construction and Lasso prediction exercise (Section \ref{subsec:pca}). The potential benefit of this method is to ``shrink the cross section'' of factors and elicit latent factors that capture most of the time-series return variation among the many candidates. According to existing literature, such latent factors are often more reliable for explaining expected returns.

In summary, we put forward the message that integrating quantity information into various empirical methods can lead to significant empirical discoveries. We implement the methods outlined above to support this message, but they are far from exhaustive given the vast asset pricing literature.  We believe these quantity variables can similarly interact with many other existing methods, opening a broad avenue for further empirical discoveries.

\section{Constructing quantity ($q$) and other variables}\label{sec:construct}

The data to run a BTQ predictive regression include the (unbalanced) panel of monthly excess stock returns $r_{i, t+1}$, along with a panel of $\beta_{i, k, t}$ and a time series of $q_{k, t}$ for each factor $k$, which serve as right-hand side predictors. Among these, $\beta_{i, k, t}$ is constructed from the time series of factor return $f_{k, t}$ as in the first stage of the Fama-MacBeth procedure. The construction of $q_{k, t}$ is new. It requires the stock-level retail flow in the same unbalanced panel structure as the returns, which is then aggregated to the factor level according to each stock's factor exposure measures. In summary, the source data are only the panel of returns and the panel of flows at the stock level, from which one can calculate both $\beta$ and $q$ for any factor, given the time series of factor returns $f_{k, t}$.

\subsection{Return, risk, and flow variables constructed with standard procedures}\label{subsec:construct1}

The factor and stock return, risk exposure, and stock-level dollar flow variables are all constructed using data sources and procedures standard in the literature.

We use delisting-adjusted stock returns from CRSP. The six Fama-French-Carhart \citep[i.e.,][]{fama1993common, fama2015five,carhart1997persistence} factors are from Kenneth French's website, and the 153 \citet*[JKP]{jensen2023there} factors are from the authors' website. All returns are obtained in both daily and monthly frequencies in excess of the risk-free rate.

Each stock's exposure to factor $k$ in month $t$ is
\begin{align}
	\betahat_{i,k,t} \defeq \frac{\covhat_t(r_{i,t}, f_{k,t})}{\varhat_t(f_{k,t})}, &  & \forall i, t, k,\label{eq:betadef}
\end{align}
where $\covhat_t$ and $\varhat_t$ are realized covariance and variance estimated with daily returns in a 12-month rolling window up to month $t$.\footnote{Notice $\betahat_{i,k,t}$ corresponds to the regression coefficient of a single-factor model. This differs from the original Fama-MacBeth procedure, where the first stage is a multi-factor regression. A single-factor beta is simply the realized covariance normalized by scalar variance and offers two advantages. First, multi-factor regressions can be unreliable even with a moderately high number of factors. Second, a single-factor beta, and consequently each factor's BTQ term, can be constructed independently of other factors in the model, allowing for a more convenient empirical procedure. See \citet*{feng2020taming} for a related discussion, who also use covariances rather than multi-variate betas.}

We construct the stock-level dollar flow $\flowstock_{i,t}$ panel using the mutual fund flow-induced trading (FIT) metric, proposed by \cite{coval2007asset}, \cite{froot2008institutional}, and \cite{lou2012flow}. We use the standard mutual fund data source but carefully clean data errors by cross-validating several sources. In particular, we obtain monthly mutual fund returns and characteristics from the CRSP Survivorship-Bias-Free Mutual Fund database and quarterly holdings data from the Thomson/Refinitiv Mutual Fund Holdings Data (S12). Our sample period spans from January 2000 through December 2022.\footnote{We start the sample period in 2000, following the convention in the literature. The mutual fund industry experienced significant growth and sustained inflows throughout the 1990s \citep*{lou2012flow, ben2022ratings}. Since 2000, the size of the mutual fund sector has remained stable relative to the total equity market, resulting in stationary monthly flow shocks.}  The mutual fund sample comprises both active and passive mutual funds. To ensure accuracy in our flow measure, we cross-validate mutual funds' monthly returns and total net assets (TNA) obtained from the CRSP database with corresponding data from Morningstar and Thomson/Refinitiv. In the process, we manually correct several data input inaccuracies. Details regarding this process are in Appendix \ref{sec:flowcleaning}.

The standard $\flowstock_{i,t}$ construction procedure has three steps. First, dollar mutual fund flows are
\begin{equation}\label{eq:mutualfundflow}
	\flowfund_{m,t} \defeq \mbox{TNA}_{m,t} - \mbox{TNA}_{m,t-1} (1+\rfund_{m,t}),
\end{equation}
where $\mbox{TNA}_{m,t}$ is the total net assets of mutual fund $m$ at the end of month $t$, and $\rfund_{m,t}$ is mutual fund $m$'s net-of-fee return in month $t$.

Second, we allocate mutual fund flows to dollar stock-level flows, based on the established assumption in the literature that mutual funds buy or sell stocks in proportion to their prior holdings,
\begin{align}\label{eq:flowhat}
	\flowstock_{i,t} \defeq -\sum_{\mbox{\scriptsize fund } m}\flowfund_{m,t}\weightfund_{i,m,\text{quarter}(t)-2}~.
\end{align}
The negative sign is used to shift the perspective from retail investors to sophisticated investors when accounting for the flow. Specifically, a positive $\flowstock_{i,t}$ dollar value indicates that retail investors are selling stock $i$ in month $t$, while sophisticated investors are buying. Moreover, we use the two-quarter-lagged mutual fund holding weight, denoted as $\weightfund_{i,m,\text{quarter}(t)-2}$. For instance, $\text{quarter}($July$)-2 = $ Q1.\footnote{The use of a two-quarter lag deviates from the conventional one-quarter lag \citep{lou2012flow} to be more conservative and ensures that the constructed $\flowstock_{i,t}$ is observable with information up to month $t$. In particular, mutual fund holding is reported with a maximum statutory delay of 45 days (\citealp{christoffersen2015institutions}), which means the end of Q2 holdings may not be observable in July. By using a two-quarter lag, July relies on the end of Q1 holdings, which are guaranteed to be available. Our results remain robust when we apply the one-quarter lag commonly used in the literature. These results are available upon request.}

In total, we have around 1,644,000 stock-month observations in a full sample of 276 months from January 2000 to December 2022, or on average around 6,000 stock-month observations per month.

\subsection{Constructing quantity variables}\label{subsec:qconstruction}

The construction of $q_{k,t}$ is guided by the theoretical motivation outlined in Section \ref{subsec:theoretical_motivation} and the microfoundation detailed in Appendix \ref{sec:inelastic}. It involves two steps. First, we aggregate stock-level flows to the factor level, using the same risk measures, $\covhat_t(r_{i,t}, f_{k,t})$, from Eq.\ \ref{eq:betadef}:
\begin{align}
	\flowfactor_{k,t}\defeq \sum_{i} \flowstock_{i,t} \covhat_t(r_{i,t}, f_{k,t})= \sum_{i} \flowstock_{i,t} \betahat_{i,k,t}\varhat_t(f_{k,t}), &  & \forall k, t.\label{eq:porttheory}
\end{align}
The aggregation accounts for each stock's factor exposure, in a similar spirit to calculating the portfolio beta commonly used in risk management. The second expression in Eq. \ref{eq:porttheory} is for explaining the intuition: every month, sophisticated investors add a marginal portfolio to their existing holdings in response to retail flows, and $\flowstock_{i,t}$ is the dollar weights of this portfolio. The portfolio's risk characteristics are determined by its composition (portfolio weights $\flowstock_{i,t}$), as well as each constituent stock's factor exposures ($\betahat_{i,k,t}$). For example, if retail investors sell a large quantity of value stocks with high HML loadings, the sophisticated investors' HML quantity would experience a positive flow shock.\footnote{Notice we aggregate flow to the factor level (HML in this example) based on each stock's HML exposure ($\beta$), not on the stock's characteristics (the book-to-market ratio) or its weight in the HML portfolio. This choice is based on the theoretical motivation that sophisticated investors are {\it averse to factor risk}, not the factor portfolio itself. The goal is to measure the quantity variation in each factor's risk, not the factor portfolio itself. \cite{li2022drives} aggregates using portfolio weights, which can be reconciled with our framework if characteristics are viewed as proxies for factor exposures.}
Moreover, multiplying by $\varhat_t(f_{k,t})$ modulates the portfolio's risk by the time-series fluctuation in factor return variance.\footnote{More specifically, the variance term arises in the theoretical model that assumes CARA utility for sophisticated investors (see Appendix \ref{sec:inelastic}).} 
In this sense, we are indeed tracking the quantity of factor \emph{risk}, not the physical quantity of securities or portfolios.\footnote{Appendix \ref{subsec:altermutual} discusses an alternative method that directly constructs factor-level flows from mutual fund flows.}

Second, the flow shocks $\flowfactor_{k,t}$ are normalized by the lagged total US stock market capitalization and accumulated in a six-month lookback window,
\begin{align}\label{eq:qtildeconstruction}
	\qtld_{k,t} \defeq \frac{~1~}{h}\sum_{h'=0}^{h-1} \frac{	\flowfactor_{k,t-h'}   }{\textrm{total stock market cap}_{t- h' - 1}}, &  & \forall k, t, &  & \text{with} \ h = 6.
\end{align}
This normalization accounts for the upward trend in dollar flows, which reflects the overall growth of the equity market, as well as the increasing capacity of sophisticated investors to absorb these flows.\footnote{Appendix \ref{sec:inelastic} provides a theoretical justification for normalizing by the total stock market capitalization under the assumption that the fraction $\pi$ of sophisticated investors relative to the total stock market remains constant over time. The smaller this fraction ($\pi$), the more sensitive the risk premium (the higher $\lambda$).}
Accumulating $\flowfactor_{k,t}$ over time accounts for the persistent effects of older flows on future returns. What matters for the expected return in month $t+1$ is the factor quantity held at the end of month $t$, which is impacted by flow shocks in all previous periods, $\flowfactor_{k,t}, \flowfactor_{k,t-1}, \flowfactor_{k,t-2}\dots$ The speed at which sophisticated investors can absorb these shocks and eliminate their effect on risk premiums is not our research focus. We accumulate past flows in a 6-month lookback window for simplicity and transparency to avoid a more involved study of the speed. The empirical results are robust to alternative specifications (see Section \ref{sec:robustness}).

In many empirical exercises, we standardize the raw $\qtld_{k,t}$ time series as $q_{k,t} \defeq \qtld_{k,t}/\sigma(\qtld_{k,t})$, where $\sigma(\qtld_{k,t})$ is the full-sample time-series standard deviation, for ease of interpreting the regression coefficients.

\subsection{Basic properties of the constructed quantity variables}

\begin{figure}[t!]
	\centering
	\caption{Quantity ($\qtld_{k,t}$) time series plot}
	\centering
	\includegraphics[scale=0.8]{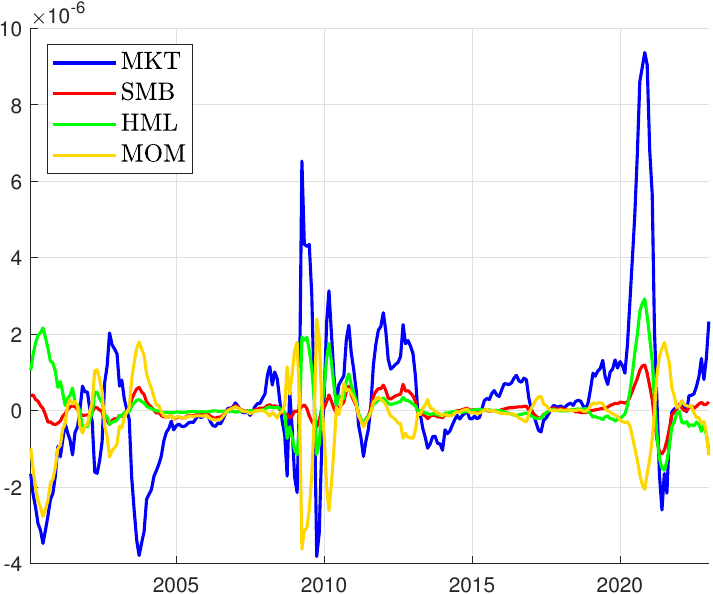}
	\begin{minipage}{\textwidth}
		\bigskip
		{\footnotesize Note: Time series of the constructed quantity ($\qtld_{k,t}$) variables for the Fama-French-Carhart factors. The monthly observations span from January 2000 to December 2022.\par}
	\end{minipage}
	\label{fig:ff3c_flow}
\end{figure}

Next, we present the summary statistics of the flow-induced quantity, $\qtld_{k,t}$, the key new variable introduced in this paper. Figure \ref{fig:ff3c_flow}  shows the time-series plots of $\qtld_{k,t}$ for the Fama-French-Carhart (FF3C) factors. We plot the pre-standardized series $\qtld$ to show magnitudes.\footnote{The magnitudes of $\qtld$ are in the unit of $10^{-6}$. The absolute level is irrelevant for empirical analysis, as the variables are standardized in regressions. To understand this magnitude, we know the monthly mutual fund flows are in the order of tens of billions of dollars, and the total market capitalization is in the order of tens of trillions of dollars (see Appendix Figure \ref{fig:flow_comp}). So the first term in Eq.\ \ref{eq:porttheory} is in the order of $10^{-3}$ (given market $\beta$ around 1). The last term, monthly $\varhat_t(f_{k,t})$ is in the order of $10^{-3}$, so $\qtld$ is in the order of $10^{-6}$.} Table \ref{tab:factorflow_stats} presents the full-sample statistics of FF3C's $\qtld$ and summaries of these statistics across the 153 JKP factors.

\begin{table}[t!]
	\caption{Summary statistics of quantity $\qtld_{k,t}$ (unit: $10^{-6}$)}
	\vspace*{-1em}
	\renewcommand{\arraystretch}{1.3}
	\begin{center}
		\scalebox{1}{ \begin{tabular*}{\textwidth}{l@{\extracolsep{\fill}}ccccccc}
    \toprule
    & \multicolumn{4}{c}{Fama-French-Carhart factors} & \multicolumn{3}{c}{Across 153 JKP factors} \\
    \cmidrule(lr){2-5} \cmidrule(lr){6-8}
    & MKT & SMB & HML & MOM & Q25 & Median & Q75 \\
    Mean	&0.29	&0.04&	0.13&	-0.15	&-0.05	&-0.01	&0.03\\
    Std	&1.88	&0.29&	0.65&	0.82&	0.23&	0.39	&0.76\\
    \bottomrule
\end{tabular*}}
	\end{center}
	\label{tab:factorflow_stats}
	\footnotesize
	{Note: The mean and standard deviation of the constructed quantity time series $\qtld_{k,t}$ for the Fama-French-Carhart factors and JKP factors.}
\end{table}%

Examining the basic time series properties of $\qtld_{k,t}$, we find that variation dominates its trend, making quantity fluctuation the primary feature compared to the secular trend in retail flows. The series also exhibits dynamic volatility clustering, similar to that seen in more familiar factor return time series.

Among the four factors plotted in Figure \ref{fig:ff3c_flow}, MKT's quantity (in blue) has the most time-series variation. The reason is that most stocks have positive market beta centered around one, so $\qtld_{\text{MKT}, t}$ roughly aggregates the \emph{overall} retail flows into (and out of) the entire mutual fund sector. In contrast, the three long-short factors have stock betas that are more evenly distributed around zero, so their $\qtld_{k, t}$ series reflect the \emph{net} retail flows into (and out of) stocks of particular investment styles. Therefore, these series are not mechanically correlated, even though they are all constructed from the same retail flow panel data.

Appendix \ref{appsec:qproperties} reports that the pairwise correlations of the four $q_{k,t}$ series are far from $\pm 1$, indicating that series are not collinear. It also reports a principal component analysis (PCA) on the $q_{k,t}$ series for the 153 JKP factors. These $q$ series have a multi-factor structure with independent variation along various principal dimensions as well as substantial idiosyncratic variation. 
Section \ref{subsec:mustpair} further shows each factor's $q$ provides distinct and independent pricing information along its respective risk dimension.
These results suggest BTQ's consistent predictive power across different factors is not mechanically driven by one (or a few) special ``secrete sauce'' $q$ series, highlighting the robustness of the underlying economic mechanism.

Turning to notable spikes in the plot, we note $\qtld_{\text{MKT}, t}$ experiences significant increases during the Global Financial Crisis and the COVID-19 pandemic in the spring of 2020. These spikes are attributed to significant outflows from mutual fund investors during these periods. As a result, the sophisticated investors' risk holding quantity increases, making them more ``averse'' to the market risk, which can be related to market crashes and subsequent rebounds. However, this is a highly simplified and anecdotal explanation of the main economic mechanism, as it does not consider cross-sectional variation in factor exposures, more nuanced fluctuations, or factors beyond MKT. Next, we turn to formal empirical analysis.

\section{Main empirical results}\label{sec:emp_res}

\subsection{Security market line (SML) depends on quantity}\label{subsec:sml}

The security market line is a simple and commonly used tool to visualize the relationship between systematic risk exposure and expected return ($\beta$-$\E r$) in the cross section of stocks, without relying on parametric modeling. We construct the empirical SML and its conditional versions based on factor $q$. We show that the $\beta$-$\E r$ relationship is nearly flat unconditionally, which is consistent with the existing empirical results that factor exposure alone cannot adequately explain the cross-sectional variation in stock returns. However, once conditional on quantity information, the SML reveals interesting risk-return patterns that strongly support a risk-based explanation.

The unconditional SML displays the $\beta$-$\E r$ relationship in the non-parametric regression model: $\E_t[r_{i,t+1}] = Er(\beta_{i,k,t})$. We estimate it with a simple kernel method by sorting stock-month observations into twenty quantile bins by $\betahat_{i,k,t}$, and then plotting the average of $r_{i,t+1}$ against the average $\betahat_{i,k,t}$ within each bin. Notice that return $r_{i,t+1}$ leads $\betahat_{i,k,t}$ by one month, so that it estimates conditional expected returns.

The upgraded SML {\it conditional on quantity} estimates the bi-variate non-parametric model: $\E_t[r_{i,t+1}] =  Er(\beta_{i,k,t}, q_{k,t})$. Our purpose is to show that the second entry, $q$, matters for the risk-return relationship. Again, we conduct a simple non-parametric estimation for transparency and intuitiveness. The estimation procedure is the same as the unconditional SML, but we further split each bin of stock-month observations into two sub-bins by the time-series median of $q_{k,t}$, and plot sub-bin average $r_{i,t+1}$ against average $\betahat_{i,k,t}$.%
\footnote{Formally, an unconditional bin is defined as $\{\Paren{i, t} \text{ s.t. } \betahat_{i,k,t} \in [a, b)\}$, where $a$ and $b$ are boundaries of the 20 quantiles of $\betahat_{i,k,t}$, for example, the first pair is $\big[\textrm{quantile}(\betahat_{\cdot, k, \cdot}, 0\%), \textrm{quantile}(\betahat_{\cdot, k, \cdot}, 5\%)\big)$. A ``high $q$'' bin is defined as $\{\Paren{i, t} \text{ s.t. } \betahat_{i,k,t} \in [a, b) \text{ and } q_{k,t} \geq \text{median}(q_{k,t})\}$, where $\text{median}(q_{k,t})$ is the time-series median of $q_{k,t}$. And, ``low $q$'' is the same as ``high $q$'' but with ``$\geq$'' replaced by ``$<$''.}

\begin{figure}[t!]
	\centering
	\caption{Security market line (SML) conditioning on quantity: $\E_t[r_{i,t+1}] =  Er(\beta_{i,k,t}, q_{k,t})$}
	\hspace*{-1em}%
	\begin{subfigure}[t]{0.5\textwidth}
		\captionsetup{justification=centering}
		\centering
		\includegraphics[scale=0.62]{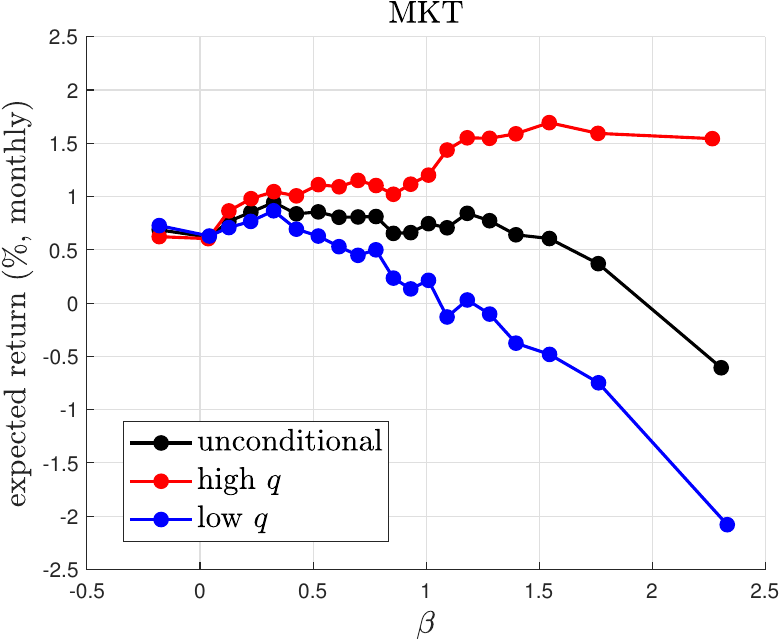}
	\end{subfigure}%
	~
	\begin{subfigure}[t]{0.5\textwidth}
		\captionsetup{justification=centering}
		\centering
		\includegraphics[scale=0.62]{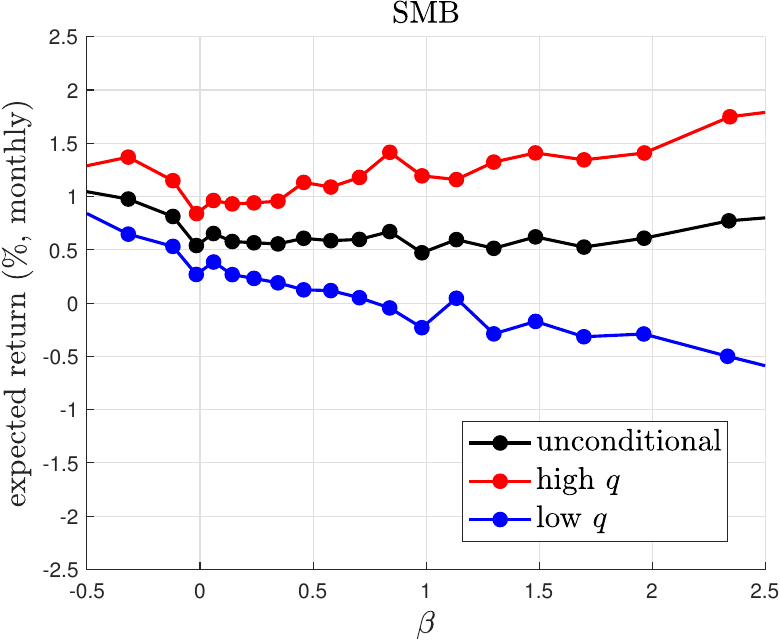}
	\end{subfigure}\\[6pt]
	\hspace*{-1em}%
	\begin{subfigure}[t]{0.5\textwidth}
		\captionsetup{justification=centering}
		\centering
		\includegraphics[scale=0.62]{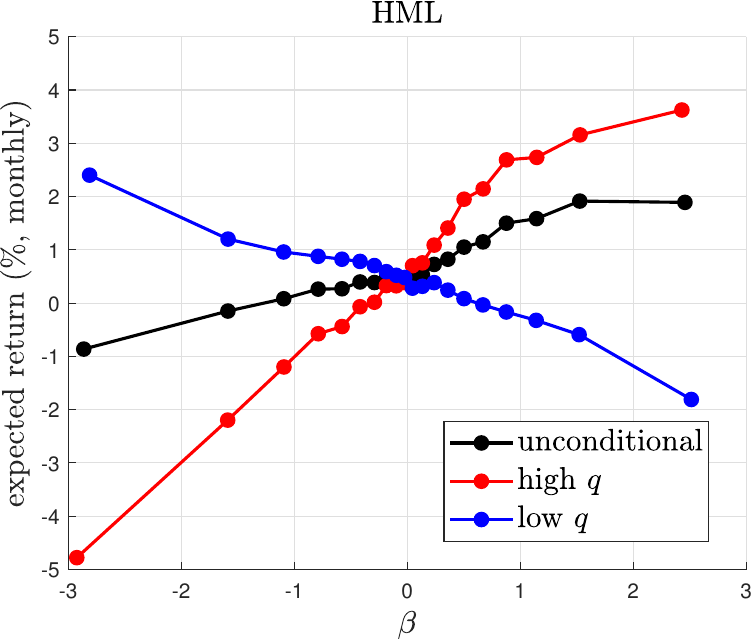}
	\end{subfigure}%
	~
	\begin{subfigure}[t]{0.5\textwidth}
		\captionsetup{justification=centering}
		\centering
		\includegraphics[scale=0.62]{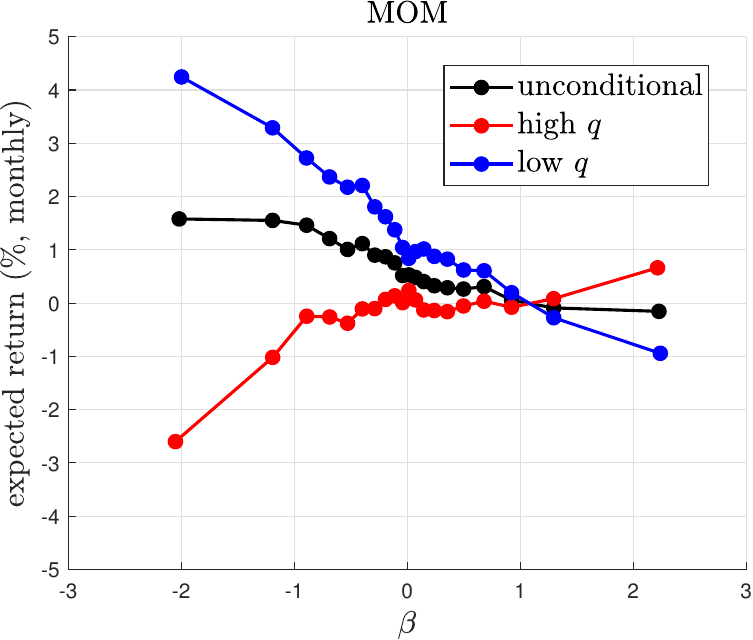}
	\end{subfigure}\\
	\bigskip
	\begin{minipage}{\textwidth} %
		{\footnotesize Note: Security market line (SML) plots expected stock returns against $\beta$. The unconditional SML (black): sorts the stock-month observations into twenty quantile bins of $\betahat_{i, k, t}$ and plots the average return $r_{i, t+1}$ against average $\betahat_{i, k, t}$ within each bin. The conditional SMLs (red for high $q$, blue for low $q$): the same process but split bins by the time-series median of $q_{k,t}$. Notice the scales of the x- and y-axes in the bottom two panels are zoomed out by a factor of two to accommodate the larger ranges of HML and MOM $\beta$'s.}
	\end{minipage}
	\label{fig:ff3c_flow6}
\end{figure}

Figure \ref{fig:ff3c_flow6} presents single-factor models using the Fama-French-Carhart factors (MKT, SMB, HML, MOM). The black curves represent the unconditional SMLs, while the red and blue curves correspond to conditional SMLs for high and low $q_{k,t}$, respectively.

We find that the unconditional SML is nearly flat for the market factor, with a slight downward slope in the higher beta range. This implies that the market beta \emph{alone} cannot explain the cross-sectional variation in expected returns, which is consistent with similar reports in the existing literature. Similar null results for unconditional SMLs are observed for SMB and MOM, while HML's SML is slightly upward-sloping.

In contrast, the conditional SMLs show interesting risk-return patterns that are not observable without conditioning on $q$. The high-$q$ SMLs (red) exhibit a strong positive slope, while the low-$q$ (blue) SMLs are downward sloping. The unconditional SML (black) lies in between these conditional SMLs as the mixed average. The gaps in the slopes suggest that sophisticated investors' risk-holding conditions matter for their demand for risk, which in turn significantly impacts the pricing of factor risks in the cross section. Notice the four plots are produced with different $q_{k,t}$ time series and $\betahat_{i,k,t}$ panels, yet the slope patterns are consistent across factors. This consistency suggests that quantity's effects on factor premiums are general and robust, reflecting a stable underlying economic mechanism.

The positive high-$q$ slope suggests that sophisticated investors demand higher additional compensation for high systematic risk in high-$q$ environments. Conversely, the negative low-$q$ slope indicates high-risk investments have low expected returns (or high concurrent prices).
This is likely because sophisticated investors are more willing to hold high-risk investments when they are required to sell lots of such stocks to retail traders in low-$q$ months, i.e., when they are in a relatively short position of the factor.%
\footnote{%
The sophisticated investors' risk management mechanisms as described in \cite{frazzini2014betting} can provide a potential explanation for the pricing behavior in the low-$q$ months.} 

The magnitude of $q$'s effects is economically large. For instance, a market beta-neutral stock has an unconditional expected return of around 0.75\% per month. In contrast, for a stock with a market beta of 1, the expected return is as high as 1.25\% in high-$q$ months or as low as 0.25\% in low-$q$ months, with the average still around 0.75\%. The high vs.\ low-$q$ gap is around 1\% per month or more than 10\% annualized. This gap is even greater for stocks with higher market betas. For HML, the gap is even more pronounced: a $\beta_{\text{HML}}=1$ stock is expected to earn around 30\% annually, while an HML-neutral stock's expected return remains unaffected by $q$, as evidenced by the crossing of the three curves at $\beta_{\text{HML}}=0$. This result reveals that HML is a salient fundamental factor for sophisticated investors, as both high $\beta$ exposure and high quantity holdings are compensated by significantly higher risk premiums. For the SMB factor, while the general patterns of SML slopes remain consistent, the effects of both $\beta$ and $q$ are smaller in magnitude compared to the other factors.
We provide additional support for these findings and present more precise point estimates using parametric estimations further below.\footnote{It is also interesting to note that the crossings of the high/low-$q$ and unconditional SMLs are almost exactly at $\beta = 0$ for MKT and HML, and somewhat near zero for SMB and MOM. Crossing at $\beta=0$ is consistent with the parametric BTQ model and the theoretical motivation: the expected return of a factor risk-neutral stock should not be affected by that factor's quantity fluctuations. %
}

All SMLs, regardless of their slopes, are approximately straight lines, regardless of their slopes, particularly around the central range of $\beta$, where most stocks are concentrated, and sampling noise is less pronounced. This linearity in $\beta$ is consistent with the cross-sectional law of one price (LOOP), even as the slope (risk premium) varies significantly with $q$. Next, we specify the linearity of expected returns in $\beta$, while still leaving the effect of $q$ non-parametric. %

\subsection{Fama-MacBeth factor premium increases with quantity}\label{subsec:FM}

We specify a linear relationship between factor exposure ($\beta$) and expected return, where the linear coefficient (factor premium) is allowed to vary with quantity: $Er(\beta_{i,k,t}, q_{k,t}) = \beta_{i,k,t} \mu_k(q_{k,t})$.

To estimate this model, the first stage of the Fama-MacBeth regressions provides factor risk exposures $\betahat_{k, i, t}$ from time-series regression (already detailed in Section \ref{subsec:construct1}). The second stage of the Fama-MacBeth regressions runs cross-sectional regression for each $t$:
\begin{align}
	r_{i, t+1} = \gamma_{0,t+1}+\gamma_{k, t+1} \betahat_{i, k, t} + error_{i, t+1}, &  & \forall i,
\end{align}
where $\gamma_{k, t+1}$ is the Fama-MacBeth factor mimicking portfolio (FMP) return. Canonically, the factor premium is estimated as the time-series average of $\gamma_{k, t+1}$. It measures the average cross-sectional association between factor loading and stock return. The average factor premiums are often found to be close to zero, challenging factor pricing (\citealp{lopez2023common}).

\begin{figure}[t!]
	\centering
	\caption{Fama-MacBeth factor premium conditioning on quantity, $\mu_k(q_{k,t})$}
	\label{fig:factorflow6_vs_por}
	\hspace*{-1em}%
	\begin{subfigure}[t]{0.5\textwidth}
		\captionsetup{justification=centering}
		\centering
		\includegraphics[scale=0.62]{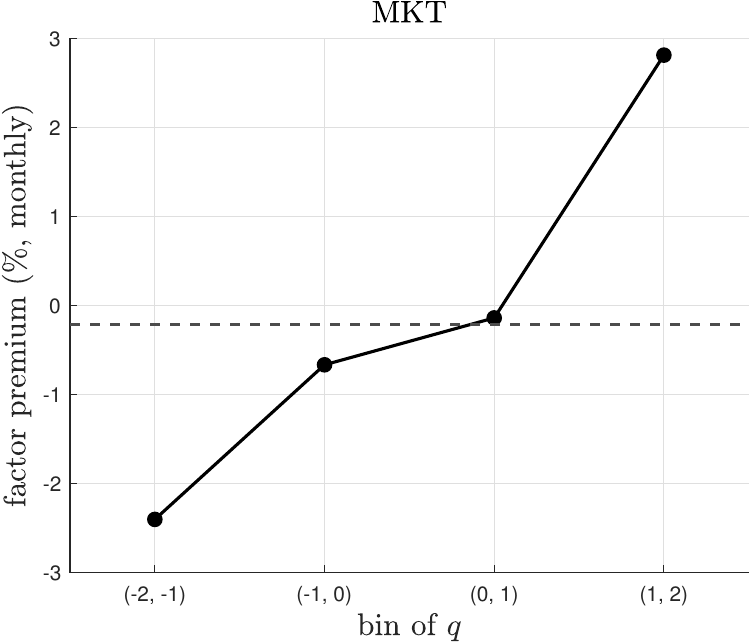}
	\end{subfigure}%
	~
	\begin{subfigure}[t]{0.5\textwidth}
		\captionsetup{justification=centering}
		\centering
		\includegraphics[scale=0.62]{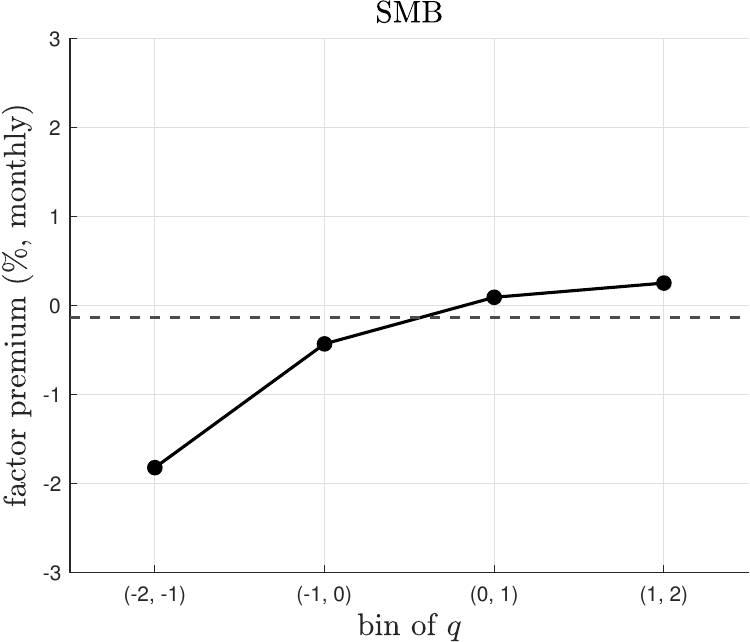}
	\end{subfigure}
	\\[8pt]
	\hspace*{-1em}%
	\begin{subfigure}[t]{0.5\textwidth}
		\captionsetup{justification=centering}
		\centering
		\includegraphics[scale=0.62]{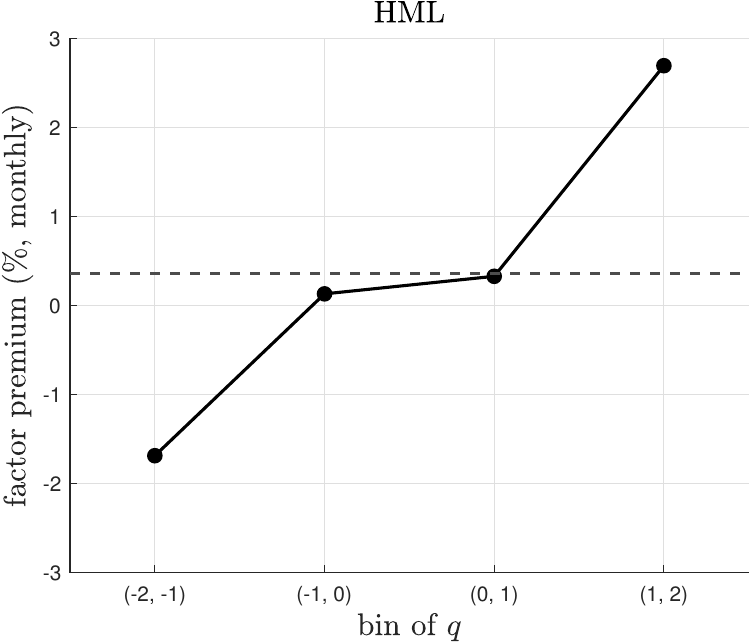}
	\end{subfigure}%
	~
	\begin{subfigure}[t]{0.5\textwidth}
		\captionsetup{justification=centering}
		\centering
		\includegraphics[scale=0.62]{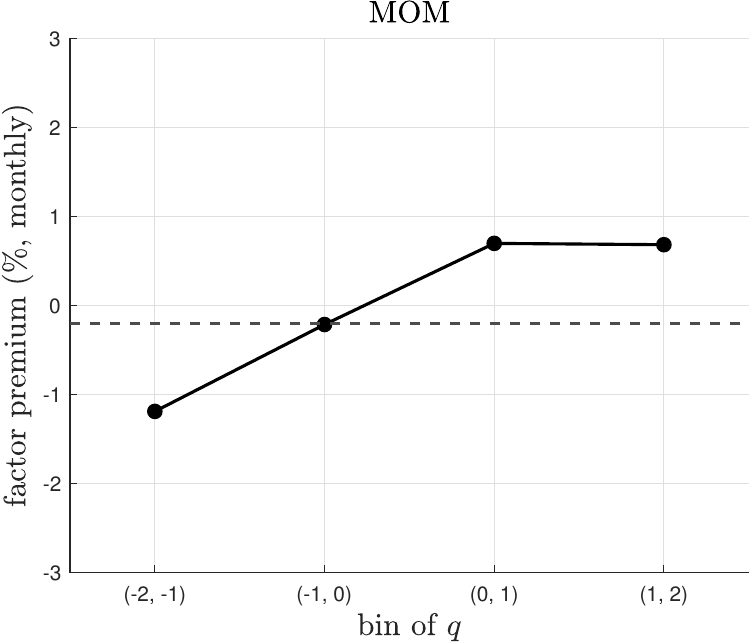}
	\end{subfigure}\\
	\bigskip
	\begin{minipage}{\textwidth}
		{ \footnotesize Note: Fama-MacBeth factor mimicking portfolio returns (FMP, $\gamma_{k,t+1}$) averaged unconditionally (dashed line) and averaged within unit bins of $q_{k,t}$ (solid line).\par}
	\end{minipage}
\end{figure}

The innovation of our approach is to estimate the mean of $\gamma_{k, t+1}$ conditional on $q_{k, t}$. To achieve this, we form four unit bins of $q_{k, t}$ (which is already standardized) and calculate the average of $\gamma_{k, t+1}$ within each bin. Figure \ref{fig:factorflow6_vs_por} presents the conditional (solid lines) and the unconditional (dashed lines) factor premiums for each of the four FF3C factors.

The plot shows strong and consistent evidence that the Fama-MacBeth factor premium is not zero but increasing in factor quantity $q_{k, t}$. Specifically, the cross-sectional risk-return relationship is strong and positive when quantity $q_{k, t}$ is high, while the factor premium becomes negative when $q_{k, t}$ is low, suggesting that the risk-return tradeoff is reversed in low-$q$ environments. On average, the unconditional premium is close to zero, but this masks the significant dynamics that only unfold when we condition on quantity information.

The increasing relationship in $\mu_k(q_{k,t})$ is consistent across the four factors, with the market factor exhibiting the most substantial variation. The market factor premium varies from less than $-2\%$ per month when market $q_{k,t}$ is in the lowest $(-2, -1)$ standard deviation range to nearly $+3\%$ per month when market $q$ is in the $(1, 2)$ range. Consistent with the SML results, the magnitude of factor premium fluctuation driven by $q_{k,t}$ can reach double-digit annualized percentages, highlighting the economic relevance of quantity in driving factor premiums.

\subsection{Beta times quantity (BTQ) forecasts individual stock returns}\label{subsec:retpred}

The empirical results so far from non-parametric plots show that the quantity information significantly impacts the cross-sectional risk-return relationship. We now turn to the parametric BTQ model, which allows us to include multiple factors, provide more formal point estimates, and conduct OOS model fit evaluation and factor selection tests. We show that the BTQ model provides a compelling explanation for the expected return of individual stocks.

Once the factor premium function $\mu_k(q_{k,t})$ is specified in linear form as $\mu_k(q_{k,t}) = \lambda_k q_{k,t}$, we arrive at the parametric BTQ model, which is estimated using the following return predictive regression with a panel of individual stocks:
\begin{align}
	r_{i,t+1} = \sum_{k=1}^K \lambda_k q_{k,t} \betahat_{i,k,t} + error_{i,t+1}, &  & \forall i, t.\label{eq:btqreg}
\end{align}
We compare it with the ``$\beta$-only" model, which is implied by a constant factor premium $\mu_k$:
\begin{align}
	r_{i,t+1} = \sum_{k=1}^K \mu_k \betahat_{i,k,t} + error_{i,t+1}, &  & \forall i, t.\label{eq:betaonlyreg}
\end{align}

We first present the results of the single-factor predictive regressions ($K = 1$), using each of the four Fama-French-Carhart factors (MKT, SMB, HML, MOM) and the 153 JKP factors (Table \ref{tab:xsreg_individual_factor}).

\begin{table}[t!]
	\caption{{Predicting stock returns with and without quantity, single factor}}
	\vspace*{-1em}
	\renewcommand{\arraystretch}{1.3}
	\begin{center}
		\begin{tabular*}{\textwidth}{c@{\extracolsep{\fill}}ccccccc}
    \toprule
    & \multicolumn{4}{c}{Fama-French-Carhart factors} & \multicolumn{3}{c}{Across 153 JKP factors} \\
    & MKT & SMB & HML & MOM & Q25 & Median & Q75 \\
    \midrule
    \multicolumn{8}{l}{Panel A: IS $R^2$ comparison, full sample 2000-2022 (\%)}
    \\[2pt]
    BTQ  & 1.01&	0.30&	1.00&	0.91	&0.39	&0.62&	0.95\\
    $\beta$-only  &0.05&	0.05&	0.12&	0.06&	0.02&	0.06&	0.10\\
    \midrule
    \multicolumn{8}{l}{Panel B: OOS $R^2$ comparison, evaluation window 2010-2022 (\%)}
    \\[2pt]
    BTQ    &0.75&	0.60&	0.84&	0.65&	0.20&	0.38&	0.67 \\
    $\beta$-only  & 0.05&	-0.10&	0.15	&0.02&	-0.03&	0.04&	0.11\\
    \midrule
    \multicolumn{8}{l}{Panel C: full-sample coefficient comparison: 2000-2022}\\[4pt]
    \multicolumn{8}{l}{~BTQ}
    \\[2pt]
    $\lambda_k $ (\%)& 1.80&	0.72&	1.48&	1.77&	0.62	&0.99	&1.48\\
    $t$-stat &(4.18)&	(2.76)&	(3.52)	&(3.38)&	(2.24)&	(2.96)&	(3.69)
    \\[6pt]
    \multicolumn{8}{l}{$\beta$-only}
    \\[2pt]
    $\mu_k$ (\%)& 0.38&	0.31&	0.56&	-0.50	&-0.33	&-0.14	&0.22\\
    $t$-stat &(1.07)	&(1.25)&	(1.71)&	(-1.23)&	(-1.52)&	(-0.71)	&(1.11)
    \\
    \bottomrule
\end{tabular*}
	\end{center}
	\label{tab:xsreg_individual_factor}
	\footnotesize
	{Note: BTQ and $\beta$-only return predictions (Eq.\ \ref{eq:btqreg} and \ref{eq:betaonlyreg}), single-factor models ($K=1$). The first four columns repeat the same prediction exercises with $k = $ MKT, SMB, HML, MOM, respectively. The last three columns report the summary statistics across the 153 JKP factors. The $t$-statistics (in parentheses) are calculated using standard errors clustered by month. Return prediction $R^2$ is calculated without demeaning  ($R^2 \defeq 1 - \sum_{i,t}\Paren{r_{i, t+1} - \widehat r_{i, t+1}}^2 / \sum_{i,t}r_{i, t+1}^2$, where $\widehat r_{i, t+1}$ is predicted return) throughout the paper following \cite*{gu2020empirical}.}
\end{table}%

The key finding is that the BTQ model significantly outperforms the $\beta$-only model in predicting stock returns, with substantial $R^2$ improvements across different factor choices and in both in-sample and out-of-sample evaluations.\footnote{For OOS evaluations, we estimate the model parameters ($\lambda_k$ and $\mu_k$) using the sample period from 2000 to 2009 and apply these estimates to calculate the OOS $R^2$ for the period from 2010 to 2022. Return prediction $R^2 \defeq 1 - \sum_{i,t}\Paren{r_{i, t+1} - \widehat r_{i, t+1}}^2 / \sum_{i,t}r_{i, t+1}^2$, where $\widehat r_{i, t+1}$ is the predicted return. These standards are maintained throughout the paper.} Even with only one factor, the BTQ model's OOS return predictive $R^2$'s are around $0.8\%$ for MKT and HML, which are among the highest fit within the 153 JKP factors. The median OOS $R^2$ across the 153 JKP factors is around $0.4\%$, and 139 out of the 153 factors yield a positive OOS $R^2$.\footnote{Appendix \ref{sec:comparelambda} provides further interpretation of the economic magnitude of these $R^2$ values. Roughly speaking, with various simplifications, a one-monthly standard deviation shock in quantity corresponds to 1\% of the mutual fund sector's market capitalization or about 0.2\% of the total U.S. stock market capitalization. Assuming a price multiplier of 5 \citep{gabaix2022search}, this translates to approximately $1\% = 5 \times 0.2\%$ of expected return fluctuation, which fits about $R^2 = 1\%$ of the monthly variation of realized stock returns.} This return predictability at the individual stock level is economically significant and comparable to unstructured state-of-the-art machine learning models that use a large number of firm characteristics to predict stock returns, which typically achieve an OOS $R^2$ of $1\%$ to $2\%$.
In contrast, the $\beta$-only models have a low $R^2$ close to zero, with the median OOS $R^2$ across the 153 JKP factors at $0.04\%$ and even the 75th percentile reaching only $0.11\%$. 

Turning to the coefficients estimates, the BTQ model's $\lambda_k$ are significantly positive for all four Fama-French-Carhart factors and for most of the 153 JKP factors. The economic magnitude of the $\lambda_k$ estimates is substantial. For example, $\lambda_{\text{MKT}} = 1.8\%$, meaning that for one standard deviation increase in market factor $q$, the expected return of a stock with a market beta of 1 increases by $1.8\%$ per month, or $1.8\% \times 2 = 3.6\%$ per month for a stock with a market beta of 2, and so on.\footnote{See Appendix \ref{sec:comparelambda} for additional details showing that the magnitude of the $\lambda_{\text{MKT}}$ estimates is comparable to those reported in the literature.} In contrast, the $\beta$-only model's $\mu_k$ coefficients are mostly statistically insignificant, with 90 out of the 153 JKP factors even exhibiting negative coefficient point estimates.

In summary, the single-factor results show that the BTQ model reliably predicts stock returns, with coefficients consistent with the risk-based explanation, while the $\beta$-only model fails in both model fit and coefficient estimates.

In addition, Appendix Table \ref{tab:factor_predict_results} presents an incidental empirical finding regarding factor returns: each factor's return $f_{k,t+1}$ is predictable by its quantity $q_{k,t}$, with the predictive coefficients predominantly positive and statistically significant. However, the OOS $R^2$'s are unstable and mostly negative, due to the limited statistical power of the simple time-series prediction of factor returns. As discussed in Section \ref{subsec:model}, while this time-series predictability is consistent with the BTQ model's cross-sectional return predictability, it is a much weaker argument for the pricing power of quantity and is peripheral to our primary research focus (see further discussion in Appendix \ref{sec:factor_predict}).

Moving onto multi-factor models, Table \ref{tab:xsreg_ff_factors} presents the results for these models while maintaining a relatively low dimensionality with $K \leq 6$. This is achieved by using various combinations of the Fama-French-Carhart (FF5C) factors. The BTQ model continues to significantly outperform the $\beta$-only model across all multi-factor specifications. Allowing for multiple factors further boosts BTQ's predictive accuracy, with the best OOS $R^2$ values exceeding $1\%$. In contrast, the $\beta$-only model still struggles to predict stock returns, with low $R^2$ values even within the sample.

\begin{table}[t!]
	\caption{{Predicting stock returns with and without quantity: multi-factor models}}
	\vspace*{-1em}
	\renewcommand{\arraystretch}{1}
	\begin{center}
		\newcolumntype{C}{>{\centering\arraybackslash}p{0.13\textwidth}}

\begin{tabular*}{\textwidth}{@{\extracolsep{\fill}}CCCCCC}
    \toprule
    &CAPM& FF3 & FF3C & FF5 & FF5C\\
    & $K=1$~~~~~~~~ & 3 & 4 & 5 & 6\\
    \midrule
    \multicolumn{6}{l}{Panel A: IS $R^2$ comparison, full sample 2000-2022 (\%)}
    \\[2pt]
    BTQ  & 1.01&	1.17&	1.19	&1.17&	1.21\\
    $\beta$-only &0.05	&0.17&	0.21&	0.18	&0.22\\
    \midrule
    \multicolumn{6}{l}{Panel B: OOS $R^2$ comparison, evaluation window 2010-2022 (\%)}
    \\[2pt]
    BTQ  &0.75&	1.03&	1.07&	0.44	&0.65\\
    $\beta$-only & 0.05&	0.15&	0.22&	-0.26&	-0.05\\
    \midrule
    \multicolumn{6}{l}{Panel C: coefficients, full sample 2000-2022}
    \\[4pt]
    \multicolumn{6}{l}{~~~BTQ, $\lambda_k$ (\%) and $t$-statistics in parentheses} \\[6pt]
    ~~~~~~~~MKT& 1.80&	1.27&	1.15&	1.28&	1.16\\
    &(4.18)&	(2.08)&	(1.96)&	(2.00)	&(1.98)
    \\[2pt]
    ~~~~~~~~SMB&&-0.23	&-0.16&	-0.20&	-0.10\\
    &&(-0.77)&	(-0.59)&	(-0.69)&	(-0.38)
    \\[2pt]
    ~~~~~~~~HML&& 0.82&	0.50&	0.80&	0.50\\
    &&(1.43)	&(0.70)&	(1.55)&	(0.73)
    \\[2pt]
    ~~~~~~~~MOM&&&	0.53	&&	0.74\\
    &&	&(0.71)	&&	(0.93)\\[2pt]
    ~~~~~~~~CMA&&		&&0.10&	0.08\\
    &&	&&	(0.35)&	(0.28)\\[2pt]
    ~~~~~~~~~~RMW&&	&&	-0.09	&-0.25\\
    &	&&&	(-0.28)&	(-0.68)\\[2pt]
    \multicolumn{6}{l}{~~~$\beta$-only}
    \\
    \multicolumn{6}{c}{--- see Appendix Table \ref{tab:xsreg_ff_factors_coef} ---}
    \\
    \bottomrule
\end{tabular*}

	\end{center}
	\label{tab:xsreg_ff_factors}
	\footnotesize
	{Note: BTQ and $\beta$-only return predictions (Eq.\ \ref{eq:btqreg} and \ref{eq:betaonlyreg}). Same as Table \ref{tab:xsreg_individual_factor} but with multi-factor models ($K \geq 1$). The coefficients of the $\beta$-only model are relegated to Appendix Table \ref{tab:xsreg_ff_factors_coef}.}
\end{table}%

Regarding factor importance, MKT stands out as the most prominent after controlling for the contributions of other factors. It has the highest and most statistically significant coefficients across all multi-factor models, despite an attenuation in  $\lambda_{\text{MKT}}$ as more factors are included. HML and MOM also have positive coefficients but lack statistical significance. The inclusion of these factors in the BTQ model increases both IS and OOS $R^2$, indicating that their BTQ terms provide additional predictive power and that they are priced factors. Conversely, the coefficients for SMB, CMA, and RMW are either near zero or negative, indicating they are not priced factors according to the BTQ model. This is also evidenced in the fact that the OOS $R^2$ drops when these factors are added to the model. The $\beta$-only model's $\mu_k$ coefficients are all insignificant or negative. (These numbers are relegated to Appendix Table \ref{tab:xsreg_ff_factors_coef}.)

Comparing BTQ's IS vs.\ OOS model fits, we observe slight reductions in $R^2$ when moving from IS to OOS for CAPM, FF3, and FF3C models, indicative of mild overfitting or parameter instability. This underscores the robustness of the BTQ model's predictive power, especially considering the inherent difficulty of forecasting monthly stock returns due to the low signal-to-noise ratio in stock prices. For FF5 and FF5C, the IS $R^2$ continues to increase slightly, while the OOS $R^2$ reverses to lower values of $0.5\%$ and $0.7\%$. These levels of prediction accuracy are still economically significant, but the gap between IS and OOS $R^2$ indicates an overfitting issue. It suggests the ordinary least squares (OLS) estimation method has limitations for moderately higher-dimensional BTQ models. The additional factors might be noisy or redundant and introduce sample estimation errors. Next, we adopt a regularization method to select factors from a much greater number of candidates.

\subsection{Taming the factor zoo with BTQ}\label{subsec:factorzoo}

The proliferation of proposed factors challenges the asset pricing literature, and the BTQ model offers a new method to select factors. This method has stronger identification power and economic relevance than traditional factor premium tests.

To implement this approach, we use the same return prediction framework (Eq.\ \ref{eq:btqreg}) but overload it with a large number of proposed factors ($K = 159$, including six from FF5C and 153 from JKP). It is well expected that many of these factors are noisy or redundant when controlling for other factors for pricing stock returns. To address this, we use the Lasso method to induce sparsity in the predictive model and filter out the factors that are not priced according to the BTQ model.

Lasso is a regularization method that adds a penalty term to the OLS objective function to shrink and threshold the coefficients towards zero. Specifically, the parameter estimates solve the following optimization problem:
\begin{equation}
	\min_{\lambda_1 \dots \lambda_K} \frac{1}{2|\text{IS}|}\sum_{i, t \in \text{IS}}\Paren{r_{i,t+1}-\sum_{k=1}^K \lambda_k \betahat_{i,k,t} q_{k,t}}^2+ \omega \sum_{k=1}^K \frac{1}{\sigma(\qtld_{k,t})}\Abs{\lambda_k},
	\label{eq:lasso}
\end{equation}
where $|\text{IS}|$ is the number of stock-month observations in the training sample, and $\omega$ is the regularization parameter that controls the strength of the penalty term.\footnote{The penalty on $\lambda_k$ is normalized by the standard deviation of $\qtld_{k,t}$ for technical reasons. It allows the economic interpretation of $\lambda_k$ with respect to the standardized $q_{k,t}$ as used throughout the paper. See technical details in Appendix \ref{app:technical_details_lasso}, where the Lasso essentially is conducted with the pre-standardized $\qtld_{k,t}$, and these two forms are mathematically equivalent.}

\begin{figure}[t!]
	\centering
	\caption{{Return prediction with factor selection from the factor zoo}}
	\hspace*{-1em}%
	\begin{subfigure}[t]{0.5\textwidth}
		\captionsetup{justification=centering}
		\caption{BTQ, $R^2$}
		\centering
		\includegraphics[scale=0.62]{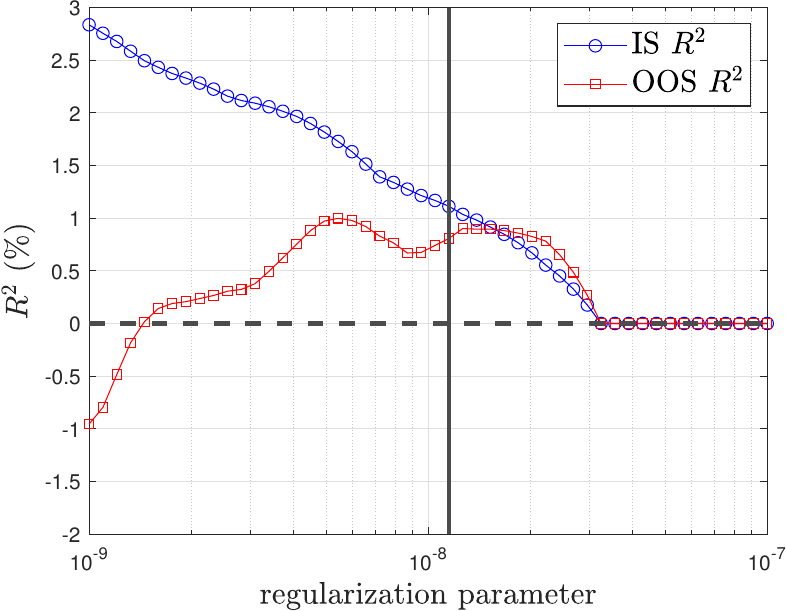}
	\end{subfigure}%
	~
	\begin{subfigure}[t]{0.5\textwidth}
		\captionsetup{justification=centering}
		\caption{$\beta$-only, $R^2$}
		\centering
		\includegraphics[scale=0.62]{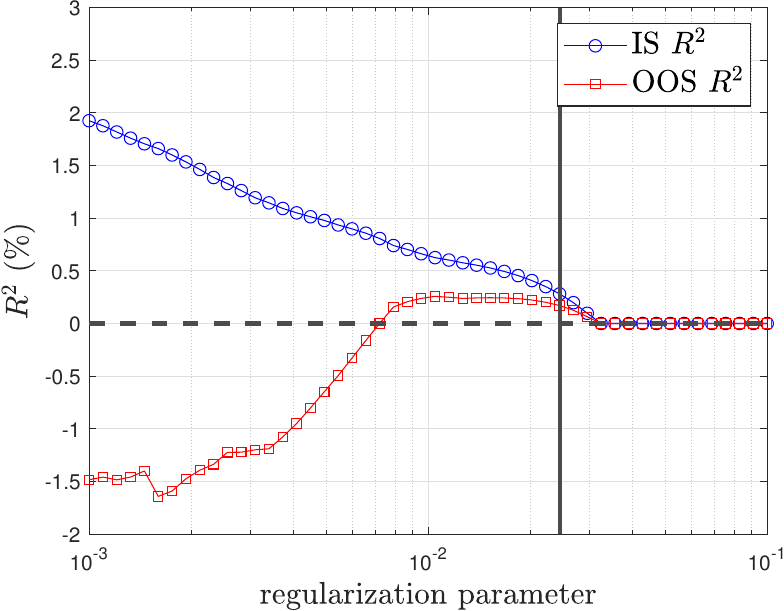}
	\end{subfigure}
	\\[10pt]
	\hspace*{-1em}%
	\begin{subfigure}[t]{0.5\textwidth}
		\captionsetup{justification=centering}
		\caption{BTQ, factor selection}
		\centering
		\includegraphics[scale=0.62]{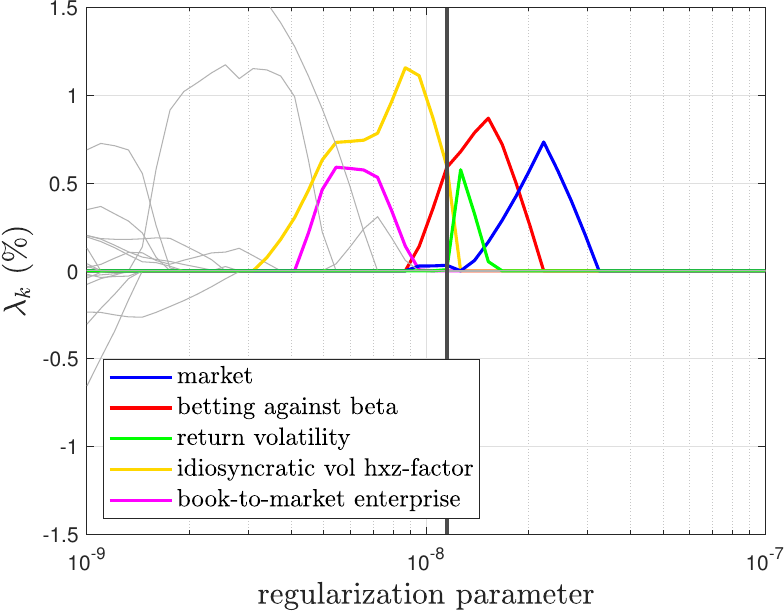}
	\end{subfigure}%
	~
	\begin{subfigure}[t]{0.5\textwidth}
		\captionsetup{justification=centering}
		\caption{$\beta$-only, factor selection}
		\centering
		\includegraphics[scale=0.62]{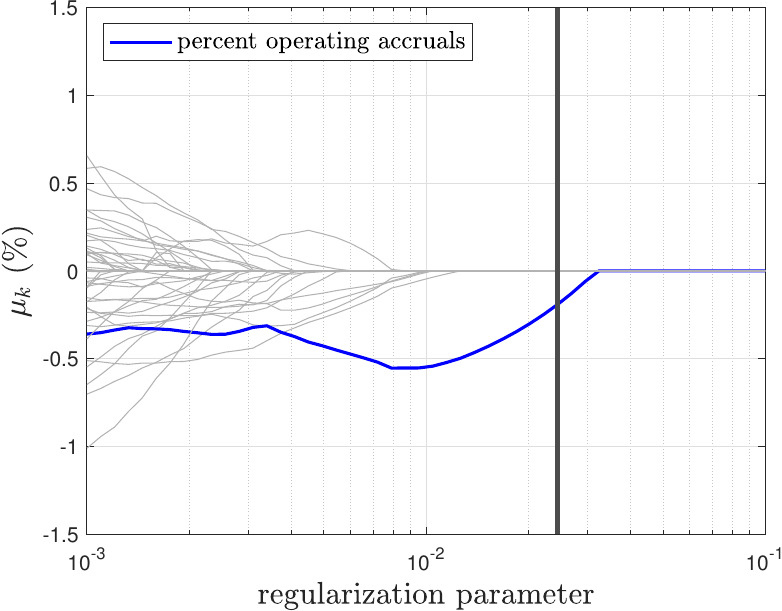}
	\end{subfigure}\\
	\bigskip
	\begin{minipage}{\textwidth} %
		{ \footnotesize Note: Model fit and parameter estimates as the regularization parameter ($\omega$, horizontal axis) varies. In Panels A and B, the IS $R^2$ is evaluated in the training window (2000-2009), and the OOS $R^2$ is the same model evaluated in the testing window (2010-2022). Panels C and D plot the parameter estimates from the training window, which are also brought out of the sample for evaluating the OOS $R^2$ in Panels A and B. The selected factors (colored curves) are, for BTQ: market (\texttt{mkt}), betting against beta (\texttt{betabab\_126d}), return volatility (\texttt{rvol\_21d}), idiosyncratic volatility from HXZ q-factor model (\texttt{ivol\_hxz4\_21d}), and book-to-market enterprise value (\texttt{bev\_mev}); and for $\beta$-only, percent operating accruals (\texttt{oaccuruals\_ni}). The unselected factors are in gray, reported in Appendix \ref{appsubsec:factorzoo} with factor importance measures. The vertical black line indicates the tuned $\omega$ based on cross-validation; see Appendix \ref{app:technical_details_lasso} for details.\par}
	\end{minipage}
	\label{fig:lasso_full}
\end{figure}

Figure \ref{fig:lasso_full} plots the model fit and factor selection results for the BTQ and $\beta$-only models as the regularization parameter ($\omega$) varies. (The $\beta$-only model's Lasso implementation is similar; see technical details in Appendix \ref{app:technical_details_lasso}.) As $\omega$ increases, the fitted BTQ model becomes more parsimonious, as shown by the decreasing IS $R^2$ (Panel A, blue curve) and the decreasing number of selected factors (those with non-zero $\lambda_k$ in Panel C). This behavior is expected from Lasso. More importantly, the OOS $R^2$ (Panel A, red curve) displays a hump shape, with a broad and relatively stable peak that reaches around $1.0\%$. This suggests that the BTQ model's predictive power is strong and robust to the choice of $\omega$. In contrast, the $\beta$-only model's OOS $R^2$ never exceeds $0.3\%$ and is only positive in a smaller range of $\omega$ values. This comparison once again highlights that quantity is essential for a risk-based explanation of expected stock returns.

The most important application of the BTQ + Lasso setup is a new way to investigate which factors are important for pricing stock returns. We find that only a few factors out of the factor zoo are sufficient for the models' high predictive power. The selected factors (those with non-zero $\lambda_k$ when OOS $R^2$ peaks) are colored in Panel C. We find that MKT is the first and most important factor, consistent with the observations in previous sections (\ref{subsec:sml} to \ref{subsec:retpred}). The MKT factor is central to multi-factor pricing theories such as \citepossessive{merton1973intertemporal} ICAPM model, and has historically been the most important factor in workhorse empirical models such as the CAPM and Fama-French models. Nevertheless, some research casts doubt on whether market beta is indeed related to expected returns (\citealp{black1972}; \citealp*{black1972capital}; \citealp{frazzini2014betting}). Our results show that the market factor \emph{equipped with quantity} variation remains highly effective in explaining expected stock returns. However, this conclusion cannot be reached with $\beta$-only models.

The other selected factors include three based on technical information, betting against beta, return volatility, and idiosyncratic volatility from \citepossessive{hou2015digesting} q-factor model, and one based on fundamental information, book-to-market enterprise value (which is a variant of the HML factor). These are among the usual suspects in the literature, while our results reinforce their importance when considering quantity. 
Moreover, it is worth noting that the $\lambda$ estimates of these selected factors from the BTQ model are all positive, which is consistent with the risk-based explanation discussed in Section \ref{subsec:model}.
On the other hand, SMB and other size-related factors are excluded by the Lasso selection process. The unselected factors are in gray and can be found in Appendix \ref{appsubsec:factorzoo} with factor importance measures.

The $\beta$-only model only selects one factor, percent operating accruals (Panel D), with a negative coefficient. This result is inconsistent with the risk-based explanation and likely reflects the model's misspecification, as indicated by its low model fit.

Additionally, choosing $\omega$ based on the OOS $R^2$ peak is sufficient for the purpose of interpreting the BTQ model's factor selection. However, for the purpose of forecasting stock returns, it has a look-ahead bias. To address the problem, we provide the tuned $\omega$ using only IS information via ten-fold cross-validation, as shown by the vertical black lines in Figure \ref{fig:lasso_full} (see technical details in Appendix \ref{app:technical_details_lasso}). The IS tuned $\omega$ is close to the OOS $R^2$ peak, suggesting the robustness of prediction and selection results.

\subsection{BTQ with latent factors}\label{subsec:pca}

Latent factors estimated using statistical methods to fit the realized time-series variation of returns have shown superior explanatory power for expected returns.\footnote{See, e.g.,  \cite*{kelly2019characteristics, kelly2020instrumented}, \cite*{kozak2020shrinking}, \cite*{lettau2020factors}, \cite{chen2023semiparametric}, and \cite{chen2024common}.} We demonstrate that the BTQ framework can be applied to latent factors as well, and it leads to a strong two-factor structure with high predictive power for stock returns that is unattainable with the $\beta$-only counterpart.

We extract the principal components (PC) of the factor zoo portfolio returns, which are the linear combinations of the factor returns that capture the most time-series variation.\footnote{Specifically, we use the first 50 principal components estimated from the monthly returns of the FF5C and 153 JKP factors from 1970 to 2009.} Then, we construct $\betahat$ and, in turn, the quantity $q$ for each of these PC factors from scratch, following the same procedure reported in Section \ref{sec:construct}. Based on these variables, we conduct the same BTQ predictive regression with Lasso as in the previous section. The new set of $\betahat$ and $q$ variables provides some external validation of our method's robustness and generalizability.

\begin{figure}[t!]
	\centering
	\caption{{Return prediction with PC and factor selection}}
	\hspace*{-1em}%
	\begin{subfigure}[t]{0.5\textwidth}
		\captionsetup{justification=centering}
		\caption{BTQ, $R^2$}
		\centering
		\includegraphics[scale=0.62]{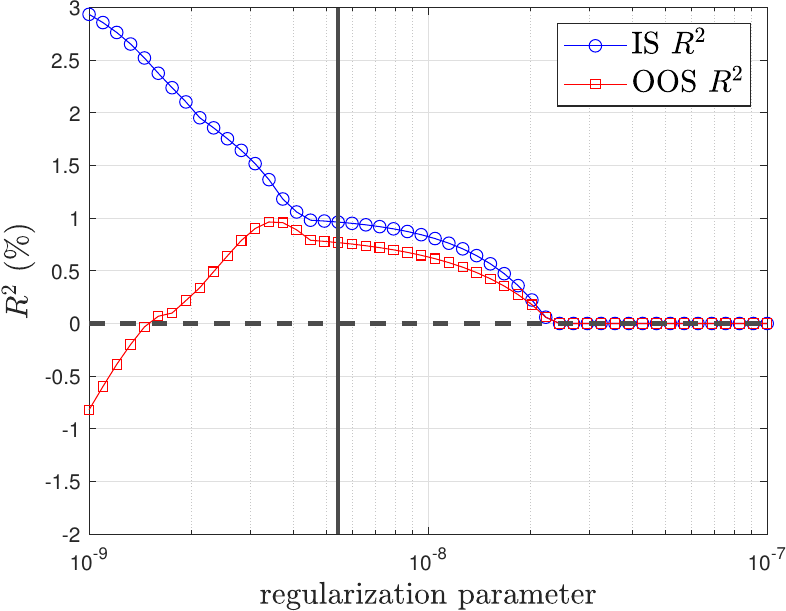}
	\end{subfigure}%
	~
	\begin{subfigure}[t]{0.5\textwidth}
		\captionsetup{justification=centering}
		\caption{$\beta$-only, $R^2$}
		\centering
		\includegraphics[scale=0.62]{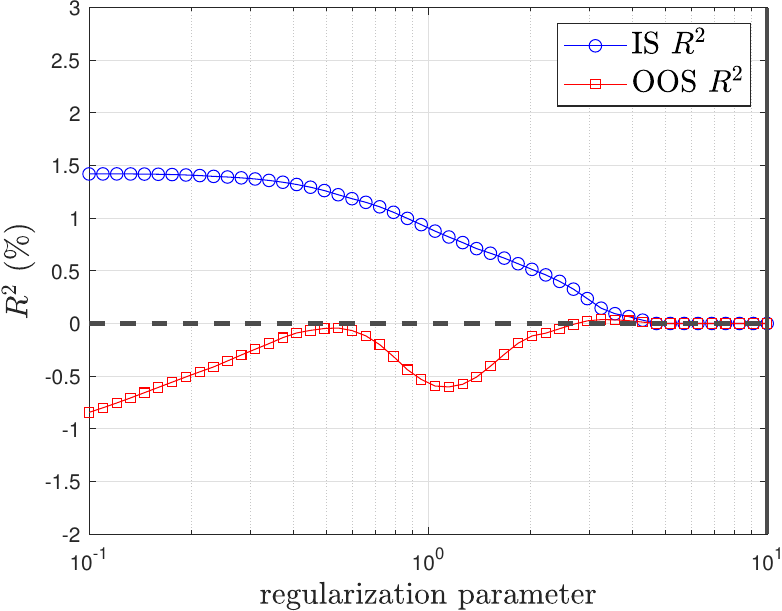}
	\end{subfigure}
	\\[10pt]
	\hspace*{-1em}%
	\begin{subfigure}[t]{0.5\textwidth}
		\captionsetup{justification=centering}
		\caption{BTQ, factor selection}
		\centering
		\includegraphics[scale=0.62]{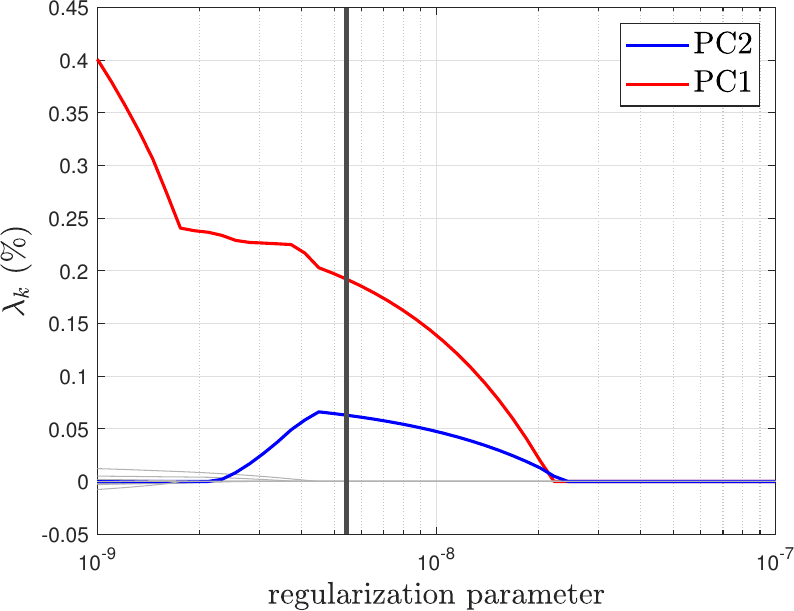}
	\end{subfigure}%
	~
	\begin{subfigure}[t]{0.5\textwidth}
		\captionsetup{justification=centering}
		\caption{$\beta$-only, factor selection}
		\centering
		\vspace*{-0.5em}%
		\includegraphics[scale=0.62]{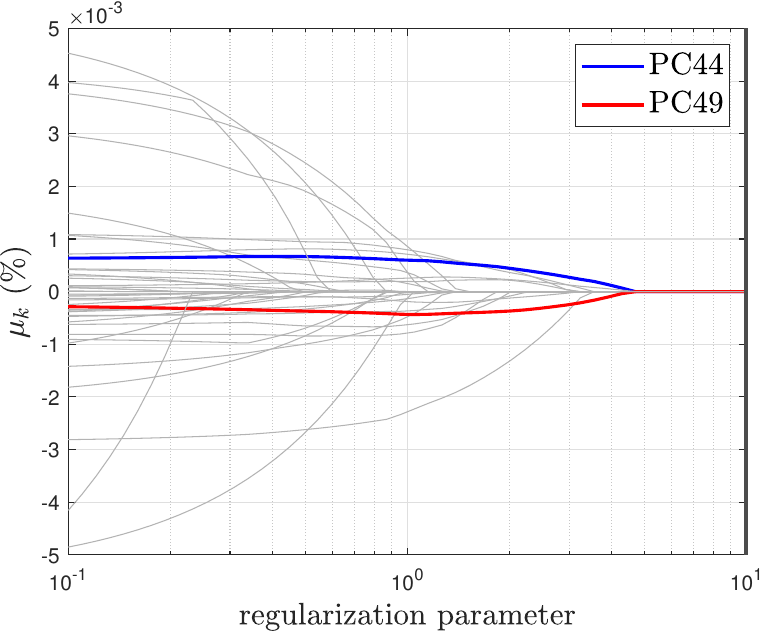}
	\end{subfigure}\\
	\bigskip
	\begin{minipage}{\textwidth} %
		{ \footnotesize Note: Model fit and parameter estimates as the regularization parameter ($\omega$, horizontal axis) varies. In Panels A and B, the IS $R^2$ is evaluated in the training window (2000-2009), and the OOS $R^2$ is the same model evaluated in the testing window (2010-2022). Panels C and D plot the parameter estimates from the training window, which are also brought out of the sample for evaluating the OOS $R^2$ in Panels A and B. We perform Lasso regression using the first 50 principal components derived from the monthly returns of the FF5C and JKP factors from 1970 to 2009. The unselected factors are in gray. The vertical black line indicates the tuned $\omega$ based on ten-fold cross-validation; see Appendix \ref{app:technical_details_lasso} for tuning details.\par}
	\end{minipage}
	\label{fig:lasso_pc}
\end{figure}

Figure \ref{fig:lasso_pc} shows that the BTQ model with PC factors has strong predictive power for stock returns, with the OOS $R^2$ peaking at around $1.0\%$, similar to the previous Figure \ref{fig:lasso_full} using original factors. The high OOS $R^2$ is, once again, robust to the choice of $\omega$, as evidenced by the broad peak of the OOS $R^2$ hump-shaped curve. In contrast, the $\beta$-only model with PC factors hardly delivers any predictive power, with OOS $R^2$ remaining below zero for almost all $\omega$ values.

More importantly, Panel C reveals a strong two-factor structure, with PC1 and PC2 emerging as the most important factors for predicting stock returns. The magnitude of their $\lambda$ estimates dominates those of subsequent PC factors (depicted as gray curves). This parsimonious structure attained with the BTQ model with latent factors can explain expected stock returns well with high OOS $R^2$. This is consistent with the literature that suggests latent factors are helpful in ``shrinking the cross section'' and reducing the dimensionality of the factor zoo \citep*{kozak2020shrinking}.

Notably, the signs of the $\lambda$ estimates for the selected factors, PC1 and PC2, are both positive. This is required by the risk-based theory, regardless of how the signs of the PCs are specified, and further reinforces the validity of the BTQ model. In contrast, the $\beta$-only model's selection and parameter estimates exhibit no discernible pattern, which likely stems from estimation noise, as the $\beta$-only model is misspecified.

\subsection{Robustness of the main predictive results above}\label{sec:robustness}

This subsection reports robustness checks that validate the BTQ model's predictive power reported above. We have already shown the BTQ model is robust to different factor specifications, including single-factor, multi-factor, selected factors, and latent factors extracted from the factor zoo. We further change the specifications in different dimensions, including various sub-sample evaluations and alternative constructions of the quantity variable.

\paragraph{Subsamples.} We first evaluate the forecasts of the BTQ models reported above in different size and time sub-samples. Table \ref{tab:OOS_R2_LASSOPC} Panel A breaks down the OOS panel into five size groups according to NYSE market capitalization quintiles and reports the OOS $R^2$ in each size group. Panel B similarly breaks down the OOS evaluation into three sub-periods: 2010-2014, 2015-2018, and 2019-2022. Panel C reports the original joint OOS (2010-2022) evaluation for reference. This table evaluates the BTQ models with factors selected from the factor zoo (initially reported in Section \ref{subsec:factorzoo}) and with selected PC factors (in Section \ref{subsec:pca}).\footnote{Table \ref{tab:OOS_R2_LASSOPC} evaluates the OOS forecasts ($\rhat_{i,t+1}$) produced with the in-sample cross-validated hyperparameter $\omega$. That is, Panel C reports the same OOS $R^2$ values at the vertical black line in Figures \ref{fig:lasso_full} and \ref{fig:lasso_pc} Panel A.}
Appendix \ref{app:add_rob} contains the same sub-sample robustness evaluations for the Fama-French-Carhart factors (in Section \ref{subsec:retpred}), %
and the results are mostly the same.

\begin{table}[t!]
	\caption{{BTQ OOS prediction accuracy ($R^2$ in \%) in size and time sub-samples} }
	\label{tab:OOS_R2_LASSOPC}
	\vspace*{-1em}
	\renewcommand{\arraystretch}{1.3}
	\begin{center}
		\scalebox{1}{ \begin{tabular*}{\textwidth}{l@{\extracolsep{\fill}}ccc}
    \toprule
    evaluation sample & \# of obs.  &selection & PC+selection\\
    \midrule
    \multicolumn{4}{l}{
        Panel A: size group evaluation}\\
    1 (small)	&323,617&	0.46&0.44\\
    2	&165,059	&1.12&1.07\\
    3	&141,153	&1.48&1.40\\
    4	&115,763&	2.02&1.91\\
    5 (big)	&103,927	&2.16&2.09\\
    \midrule
    \multicolumn{4}{l}{
        Panel B: sub-period evaluation}\\
    2010-2014	&321,425	&1.16&1.18\\
    2015-2018	&255,959	&0.15&0.14\\
    2019-2022	&272,135&	1.00&0.92\\
    \midrule
    \multicolumn{4}{l}{
        Panel C: original benchmark OOS evaluation}\\
    OOS (2010-2022) & 849,519  & 0.81&0.77\\
    \bottomrule
\end{tabular*}}
	\end{center}
	\footnotesize
	{Note:} OOS $R^2$ evaluated in different size and time sub-samples for the BTQ models with factors selected from the factor zoo (in Section \ref{subsec:factorzoo}) and with selected PC factors (in Section \ref{subsec:pca}). Panel A breaks down the OOS panel into five size groups according to NYSE market capitalization quintiles and reports the OOS $R^2$ in each size group. Panel B breaks down the OOS evaluation into three sub-periods: 2010-2014, 2015-2018, and 2019-2022. Panel C reports the original joint OOS (2010-2022) evaluation for reference.
\end{table}%

Table \ref{tab:OOS_R2_LASSOPC} shows the BTQ model's predictive results reported above are consistent in most size and time sub-samples. In particular, Panel A shows the accuracy is higher in large stocks, which are usually the most challenging section for stock return prediction. Characteristics-based anomalies and machine learning models typically find stronger predictive power in the small groups due to stronger limits to arbitrage in small stocks, including illiquidity and information asymmetry. This result indicates that BTQ's predictive power can be more reliably implemented in investment strategies in practice, given liquidity costs and trading constraints are typically weaker for larger stocks.\footnote{Cf.\ \citet{jensen2024machine,goyenko2024trading}.}

Regarding sub-periods, the BTQ model's predictive power is mostly stable over time. The first and the last sub-periods (2010-2014 and 2019-2022) have higher $R^2$ values than the middle sub-period (2015-2018) in both model specifications. We attribute this to the fact that quantity fluctuations in the middle sub-period are less volatile, as shown in Figure \ref{fig:ff3c_flow}.\footnote{Notice for each model specification, the predictive model is trained once with the 2000-2009 training sample (IS). Repeated size group-specific training (a.k.a.\ expert models) and rolling-window training have the potential to further improve the $R^2$ in OOS sub-samples above. We leave these extensions for future research due to their focus on forecasting engineering.}

\paragraph{Alternative constructions of the quantity variable.} Next, we evaluate the robustness of the BTQ model to alternative specifications in constructing the quantity time series $q_{k, t}$. In particular, there is no explicit theoretical guidance on whether factor-level flows have immediate or lagged effects on factor premium, or how fast past flows' effects decay. The benchmark specification of the quantity variables (in Section \ref{sec:construct}) accumulates past flows in a six-month lookback window, which aligns with the common expectation. We now change the specification by constructing the quantity variables using lookback windows ranging from 1 to 12 months. That is, in Eq.\ \ref{eq:qtildeconstruction}, $h=6$ is replaced by $h=1$ to $h=12$. The same empirical analyses from the previous sections are re-run with these alternative quantity variables.

\begin{table}[t!]
	\caption{BTQ OOS $R^2$ (\%) robustness to lookback window length in $q_{k,t}$ construction}
	\renewcommand{\arraystretch}{1.3}
	\begin{center}
		\vspace*{-18pt}
		\begin{minipage}{0.5\textwidth}
    \begin{flushright}
        \begin{tabular}{lcc}
            \toprule
            lookback ($h$) & selection & PC+selection \\
            \midrule
            1              & 0.36      & 0            \\
            2              & 0.41      & 0.42         \\
            3              & 0.65      & -0.15        \\
            4              & 0.49      & 0.38         \\
            5              & 0.85      & 0.82         \\
            6 (benchmark)  & 0.81      & 0.77         \\
            \bottomrule
        \end{tabular}
    \end{flushright}
\end{minipage}%
~~~~~
\begin{minipage}{0.5\textwidth}
    \begin{flushleft}
        \begin{tabular}{lcc}
            \toprule
            lookback ($h$) & selection & PC+selection \\
            \midrule
            7              & 0.78      & 0.88         \\
            8              & 0.62      & 0.70         \\
            9              & 0.62      & 0.10         \\
            10             & 0.33      & 0.11         \\
            11             & 0.48      & 0.19         \\
            12             & 0.48      & 0.25         \\
            \bottomrule
        \end{tabular}
    \end{flushleft}
\end{minipage}
\\[10pt]
	\end{center}
	\footnotesize
	\label{tab:lookback}
	{Note: OOS $R^2$ evaluated for BTQ models with $q_{k,t}$ constructed with alternative lookback window lengths ($h$) using factors selected from the factor zoo (in Section \ref{subsec:factorzoo}) and selected PC factors (in Section \ref{subsec:pca}).}
\end{table}

Table \ref{tab:lookback} shows the BTQ model's predictive accuracy is robust to alternative lookback window lengths in constructing the quantity variables. Certain perturbations (such as $h = 5$ or $7$) can even improve the $R^2$, meaning the benchmark results are not sensitive to the exact specification of the quantity variable. Having an $h$ that is too short or too long will hurt the predictive performance, but the OOS $R^2$ values are mostly significant and positive, especially for the method that directly selects factors from the factor zoo.

Additionally, Table \ref{tab:lookback} offers suggestive evidence regarding the speed and persistence with which factor flows influence sophisticated investors' pricing of factor risks. Flow shocks likely have an immediate impact on the factor premium next month, given that $h=1$ already has some predictive power. The $R^2$ is higher with an intermediate window ($h = $ 5, 6, or 7), suggesting the lagged flows in the recent few months also have impacts on factor premium, and that accumulating flows in a lookback window has, at least, statistical benefits in smoothing the predictors. On the other hand, longer windows near one year suppress prediction accuracy, suggesting that flows older than seven months have attenuated impacts on factor premiums. The attenuation is likely related to mechanisms through which sophisticated investors can gradually unwind their absorbed positions and adjust their risk holdings over time. A more detailed investigation of the dynamics between factor flows and factor premiums is left for future research, which likely requires models and data more focused on investor holdings.

\section{Quantity must be combined with risk to forecast returns}\label{sec:mustcombine}

This paper emphasizes a {\it risk-based} explanation of expected stock returns that incorporates quantity information. But is the risk modeling essential? Can quantity information alone explain expected stock returns? This section makes the case that risk and quantity must operate in tandem to forecast returns effectively. First, stock-level quantity information must be aggregated at the factor level to predict stock returns. Second, almost all factor-level $q$ variables in the ``factor zoo'' exhibit predictive power; however, each $q$ is effective only in explaining the cross-sectional return dispersion along its corresponding factor's own risk dimension---any mismatch between $\beta$ and $q$ significantly diminishes prediction accuracy. These results show that previously reported empirical success is unique to the factor risk structure, supporting our economic interpretation.

\subsection{``Quantity-only'' models do not explain expected returns}\label{subsec:qonly}

We examine an alternative economic model in which stock-level flow and quantity variations directly affect the expected stock returns without considering the factor structure and the arbitrage pricing condition. This exercise is important for understanding the joint economic role of quantity and risk in asset pricing. The main results presented earlier compare the benchmark BTQ model against the ``$\beta$-only'' baseline that accounts for risk without quantity; here, we demonstrate that an alternative ``quantity-only" baseline---relying solely on quantity while disregarding risk---also falls far short in explaining expected stock returns.

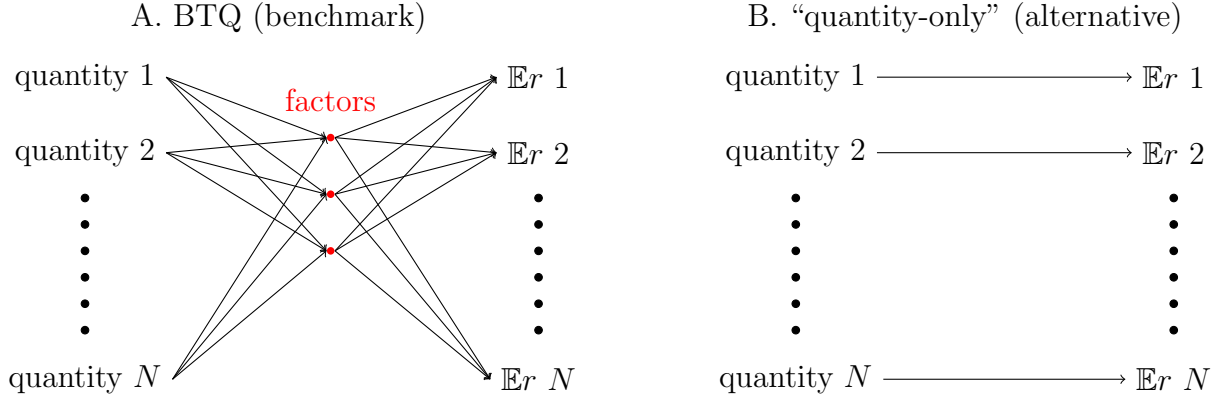
\begin{figure}[t!]
	\begin{center}
		\caption{Comparison of predictive architectures of the two models}
		\label{fig:illustration}
		\begin{minipage}{0.45\textwidth}
			\centering
			A. BTQ (benchmark)
			\\[4pt]
			\begin{tikzpicture}
				\node[rectangle,draw=none,minimum width=0.1cm,minimum height=0.8cm,align=center] at (-1,1) (q_1) {quantity $1$};
				\node[rectangle,draw=none,minimum width=0.1cm,minimum height=0.8cm,align=center] at (-1,0) (q_2) {quantity $2$};
				\node[rectangle,draw=none,minimum width=0.1cm,minimum height=0.8cm,align=center] at (-1,-3) (q_3) {quantity $N$};
				\draw[decorate sep={1mm}{3.5mm},fill] (-1,-0.6) -- (-1,-2.5);

				\node[rectangle,draw=none,minimum width=0.1cm,minimum height=0.8cm,align=center] at (2.25,.7) (q_x) {\color{red}factors};

				\node[circle,fill=red,draw=none,minimum size=1mm, inner sep=0pt] at (2.25,0.2) (q_4) {};
				\node[circle,fill=red,draw=none,minimum size=1mm, inner sep=0pt] at (2.25,-0.55) (q_5) {};
				\node[circle,fill=red,draw=none,minimum size=1mm, inner sep=0pt] at (2.25,-1.30) (q_6) {};

				\node[rectangle,draw=none,minimum width=0.1cm,minimum height=0.8cm,align=center] at (5,1) (q_7) {$\E r$ $1$};
				\node[rectangle,draw=none,minimum width=0.1cm,minimum height=0.8cm,align=center] at (5,0) (q_8) {$\E r$  $2$};
				\node[rectangle,draw=none,minimum width=0.1cm,minimum height=0.8cm,align=center] at (5,-3) (q_9) {$\E r$  $N$};
				\draw[decorate sep={1mm}{3.5mm},fill] (5,-0.6) -- (5,-2.5);

				\draw[->] (q_1.east)-- (q_4.west);
				\draw[->] (q_2.east)-- (q_4.west);
				\draw[->] (q_3.east)-- (q_4.west);
				\draw[->] (q_4.east)-- (q_7.west);
				\draw[->] (q_4.east)-- (q_8.west);
				\draw[->] (q_4.east)-- (q_9.west);

				\draw[->] (q_1.east)-- (q_5.west);
				\draw[->] (q_2.east)-- (q_5.west);
				\draw[->] (q_3.east)-- (q_5.west);
				\draw[->] (q_5.east)-- (q_7.west);
				\draw[->] (q_5.east)-- (q_8.west);
				\draw[->] (q_5.east)-- (q_9.west);

				\draw[->] (q_1.east)-- (q_6.west);
				\draw[->] (q_2.east)-- (q_6.west);
				\draw[->] (q_3.east)-- (q_6.west);
				\draw[->] (q_6.east)-- (q_7.west);
				\draw[->] (q_6.east)-- (q_8.west);
				\draw[->] (q_6.east)-- (q_9.west);

			\end{tikzpicture}
		\end{minipage}
		~~~~~~~~~~~
		\begin{minipage}{0.45\textwidth}
			\centering
			B. ``quantity-only'' (alternative)
			\\[4pt]%
			\begin{tikzpicture}
				\node[rectangle,draw=none,minimum width=0.1cm,minimum height=0.8cm,align=center] at (0,1) (q_1) {quantity $1$} ;
				\node[rectangle,draw=none,minimum width=0.1cm,minimum height=0.8cm,align=center] at (0,0) (q_2) {quantity $2$} ;
				\node[rectangle,draw=none,minimum width=0.1cm,minimum height=0.8cm,align=center] at (0,-3) (q_3) {quantity $N$} ;
				\draw[decorate sep={1mm}{3.5mm},fill] (0,-0.6) -- (0,-2.5) ;
				\node[rectangle,draw=none,minimum width=0.1cm,minimum height=0.8cm,align=center] at (5,1) (q_4) {$\E r$ $1$} ;
				\node[rectangle,draw=none,minimum width=0.1cm,minimum height=0.8cm,align=center] at (5,0) (q_5) {$\E r$ $2$} ;
				\node[rectangle,draw=none,minimum width=0.1cm,minimum height=0.8cm,align=center] at (5,-3) (q_6) {$\E r$ $N$} ;
				\draw[decorate sep={1mm}{3.5mm},fill] (5,-0.6) -- (5,-2.5) ;
				\draw[->] (q_1.east)-- (q_4.west)  ;
				\draw[->] (q_2.east)-- (q_5.west)  ;
				\draw[->] (q_3.east)-- (q_6.west)  ;
			\end{tikzpicture}
		\end{minipage}
	\end{center}
	\footnotesize Note: A. BTQ model: stock-level quantity variations affect expected stock returns {\it via} quantities of factor risks and factor premiums. B. ``quantity-only'' alternative model: stock-level quantity {\it directly} affects expected stock returns, short-circuiting the factor premium mechanism.
\end{figure}

In the benchmark model (BTQ), stock-level quantity variations are first aggregated to the factor-level quantities, which affect factor premiums, and then feed back to stock-level expected returns. In contrast, the ``quantity-only'' model specifies that stock-level flow and quantity variations {\it directly} affect expected stock returns, short-circuiting the factor premium adjustment mechanism (see the contrast in Figure \ref{fig:illustration}). Specifically, the alternative model is:
\begin{align}\label{eq:alternative}
	\E_t r_{i,t+1} = %
	\lamstock_i \qstock_{i,t}, &  & \forall i,t,
\end{align}
where $\qstock_{i,t}$ is a stock-specific flow or quantity measure, and $\lamstock_i$ is the sensitivity coefficient of stock returns to $\qstock_{i,t}$. ($\lamstock_i$ may or may not vary across stocks, to be specified below.)

This ``quantity-only'' model implies a fundamentally different economic mechanism, although Eq.\ \ref{eq:alternative} is similar in form to the main model's factor premium specification in Eq.\ \ref{eq:fp_specification}. In the main model, factor premiums vary dynamically, yet the cross-sectional no-(statistical) arbitrage pricing condition (with respect to the factors) holds each period. In contrast, the alternative model dispenses with the APT condition. For instance, two stocks with identical risk exposures but differing noise flow shocks are priced differently by the alternative model, creating an immediate arbitrage opportunity. 
The ``quantity-only'' model implies a lack of cross-sectional substitution, such that each stock is priced independently of its factor exposures. This might be the case if rigid frictions prevent cross-sectional arbitrage; or if stocks' idiosyncratic risks are not diversifiable and individually priced.

We experiment with various specifications of Eq.\ \ref{eq:alternative} and find that none come close to the BTQ model's explanatory power for expected stock returns. Specifically, stock-level $\qstockh_{i,t}$ is constructed similarly to that of the factor-level in Eq.\ \ref{eq:qtildeconstruction}:
\begin{equation}\label{eq:qstockh}
	\qstockh_{i,t} \defeq \frac 1 {~h~}\sum_{h'=0}^{h-1} \frac{\flowstock_{i,t-h'}}{\mbox{market\_cap}_{i,t-1-h'}}, \qquad \forall i,t, \text{ and } h = 1, \dots, 12.\footnote{We normalize the dollar stock-level mutual fund flow ($\flowstock_{i,t}$, see Eq.\ \ref{eq:flowhat}) by the stock's one-month-lagged market capitalization, so that the sensitivity coefficients are more interpretable. We accumulate past flows over various lookback windows, since we are agnostic about whether flow shocks have immediate or lagged effects on expected returns.}
\end{equation}
We explore different specifications of the sensitivity coefficient $\lamstock$, with varying degrees of parameter freedom. Table \ref{tab:totalflow} Panel A specifies $\lamstock$ as a constant for all stocks: $r_{i,t+1} = \lamstock \qstockh_{i,t} + error_{i,t+1}$. Panel B allows a size-dependent sensitivity coefficient such that $\lamstock$ is indexed by the NYSE size quintile of the stock: $r_{i,t+1} = \lamstock_{\text{size\_quintile}(i, t)} \qstockh_{i,t} + error_{i,t+1}$. Panel C allows stock-specific $\lamstock_i$, which is the most flexible specification: $r_{i,t+1} = \lamstock_i \qstockh_{i,t} + error_{i,t+1}$.\footnote{The specification in Panel B (size-dependent $\lamstock$) effectively runs five separate univariate predictive regressions, one for each size bin (``mixture of experts'' in machine learning terms). The specification in Panel C (stock-specific $\lamstock_i$) effectively runs stock-by-stock time-series predictive regressions. To address the unbalanced panel, we restrict the analysis to stocks with more than 80\% of monthly observations available in both the in-sample and out-of-sample windows. Stocks with fewer observations would be even more challenging to forecast.}

\begin{table}[t!]
	\caption{{``Quantity-only'' alternative model does not forecast stock returns} }
	\vspace*{-1em}
	\renewcommand{\arraystretch}{1.3}
	\begin{center}
		\scalebox{1}{ \begin{tabular*}{\textwidth}{c@{\extracolsep{\fill}}cccccccc}
    \toprule
    & \multicolumn{4}{l}{A. constant $\lamstock$} & \multicolumn{2}{l}{B. $\lamstock$ by size quintile}& \multicolumn{2}{l}{C. $\lamstock_i$ by stock}\\
    \cmidrule(lr){2-5} \cmidrule(lr){6-7} \cmidrule(lr){8-9}
    $h$~ &IS $R^2$(\%) \hspace*{-12pt} & OOS $R^2$(\%) \hspace*{-12pt} & $\lamstock$ & \hspace*{6pt}$t$-stat  \hspace*{10pt}
    &IS $R^2$(\%) \hspace*{-16pt}  & OOS $R^2$(\%)
    &\hspace*{-6pt}IS $R^2$(\%) \hspace*{-12pt}  & OOS $R^2$(\%) \hspace*{-18pt}  \\
    \midrule
    1	&		0.000	&0.000&	-0.10	&-0.27&0.003&	-0.001&0.47	&-232\\
    2	&		0.000&	-0.001&	0.05	&0.12&0.002&	-0.003&0.44&	-215\\
    3		&	0.000	&0.000	&0.17&	0.29&0.003&	0.000&0.39	&-155\\
    6		&	0.005&	0.006	&0.75&	1.01&0.007&	0.006&0.41	&-107\\
    9	&		0.004	&0.004	&0.75&	0.99&0.006&	0.006&0.38	&-96\\
    12		&	0.003	&-0.007	&0.77	&1.01&0.006	&-0.004&0.38&	-81\\
    \bottomrule
\end{tabular*}}
	\end{center}
	\label{tab:totalflow}
	\footnotesize
	Note: Panel A: univariate predictive regression, $r_{i,t+1} = \lamstock \qstockh_{i,t} + error_{i,t+1}$. B: size-dependent predictive regression, $r_{i,t+1} = \lamstock_{\text{size\_quintile}(i, t)} \qstockh_{i,t} + error_{i,t+1}$, where $\lamstock_{\text{size\_quintile}(i, t)}$ is indexed by the NYSE size quintile of the stock. C: stock-specific predictive regression, $r_{i,t+1} = \lamstock_i \qstockh_{i,t} + error_{i,t+1}$. The $R^2$ values are in percentages, e.g., $0.005$ in row $h=6$ means $0.005\%$, a very small value. The table skips some rows to save space, see complete results ($h = 1\sim 12$) in Appendix Table \ref{tab:totalflowfull}.
\end{table}%

The ``quantity-only'' models are too weak and unreliable to predict stock returns in any alternative specification, as shown in Table \ref{tab:totalflow}. In Panel A, the constant $\lamstock$ specification's in-sample $R^2$ values are about 100 times smaller than the BTQ model's. The out-of-sample $R^2$ values are not only small in magnitude, but also negative for some lookback lengths ($h$). The $\lamstock$ estimates are mostly positive, consistent with the existing literature---outflows from noise traders have negative concurrent price impacts and positively predict future returns.\footnote{See Appendix \ref{sec:comparelambda} for additional details showing that the magnitude of these $\lamstock$ estimates is comparable to those reported in the literature.} However, the estimates are statistically insignificant, and too weak to offer a meaningful $R^2$ in predicting stock returns. Allowing size-dependent $\lamstock$ slightly improves these predictive power evaluations, but no qualitative changes (Panel B). The low $R^2$ values suggest that these two specifications are too restrictive, and that the model is underfitting the data. In Panel C, stock-specific $\lamstock_i$ allows a much greater degree of freedom in parameterization (thousands of stocks vs.\ one or five parameters). The in-sample $R^2$ mechanically increases but is still smaller than the BTQ model. More importantly, the out-of-sample $R^2$ values are extremely negative, suggesting the in-sample $R^2$ values are greatly exaggerated by overfitting.

The poor performance of the ``quantity-only'' alternative model underscores that
the empirical success of the BTQ model is not driven by quantity alone, highlighting the necessity of combining quantity and risk to explain expected stock returns. In particular, the comparison implies that the factor structure is still essential in modeling expected stock returns, and that cross-sectional quantity-driven mispricing (alpha) is too weak to detect.\footnote{The ``quantity-only'' model (Eq.\ \ref{eq:alternative}) can be viewed as a quantity-driven alpha, especially when viewed in conjunction with the BTQ predictors: $\E_t r_{i,t+1} = \lamstock_i \qstock_{i,t} + \sum_k \lambda_k \beta_{i,k,t} q_{k,t}$. The term $\lamstock_i \qstock_{i,t}$ captures dynamic {\it alpha}, the part of the quantity-driven expected stock return that is not explained by the risk channel (BTQ, or the second term).} This is consistent with the view that statistical arbitrage activities by some sophisticated investors are effective in enforcing the cross-sectional APT condition,
even in the presence of noise traders (\citealp*{kozak2018interpreting}). The differing performance of BTQ and ``quantity-only'' models mirrors the contrast of macro vs.\ micro elasticities in \citet{gabaix2022search}. At the stock level, securities are highly substitutable, whereas 
quantity's effect on prices is more salient at the factor level due to greater inelasticity of factor demand  (\citealp[e.g., ][]{peng2021factor, li2022price}).

From a statistical and machine learning perspective, we can view both the BTQ and the ``quantity-only'' alternative as prediction models of stock returns based on the stock-level quantity information, i.e., they use the same predictors to predict the same targets. The difference is that the BTQ model employs a dimension reduction, aggregating predictors to the factor level, which, in turn, are used to predict the whole cross section. This is an encoder-decoder architecture in machine learning terms, where the low-dimensional ``code'' is the factor-level $q$ (see Figure \ref{fig:illustration} Panel A for the encoder-decoder illustration). In this perspective, BTQ performs well in forecasting because encoding reduces the noise in the predictors and captures the economically meaningful quantity variation aggregated at the factor level. In contrast, the ``quantity-only'' model is limited by the noisy inputs at the stock level.\footnote{See \citet{gu2021autoencoder} and \citet{kelly2023principal} for applications of encoder-decoder structures in asset pricing. Notice that BTQ specifies both the encoding and decoding weights as $\beta$ according to economic theory, rather than solely relying on statistical estimation.}

\subsection{Different factors' $q$'s explain risk-return tradeoffs along different dimensions}\label{subsec:mustpair}

The analysis above highlights the need to aggregate the quantity information at the factor level to effectively predict stock returns, consistent with our risk-based explanation. In addition, the theory also imposes a corresponding structure between a factor's exposure and the factor's quantity---BTQ must be built with each factor's own $\beta$ times its own $q$. We show that dispensing with this corresponding structure does not unleash more statistical power but, to the contrary, reduces it significantly, further validating the risk-return tradeoff channel of quantity's role in expected stock returns.

\begin{figure}[t!]
	\centering
	\caption{BTQ vs.\ mismatched BTQ variants in predicting stock returns}
	\label{fig:mismatch}
	\centering
	\vspace*{-40pt}
	\hspace*{-20pt}
	\includegraphics[scale=0.9]{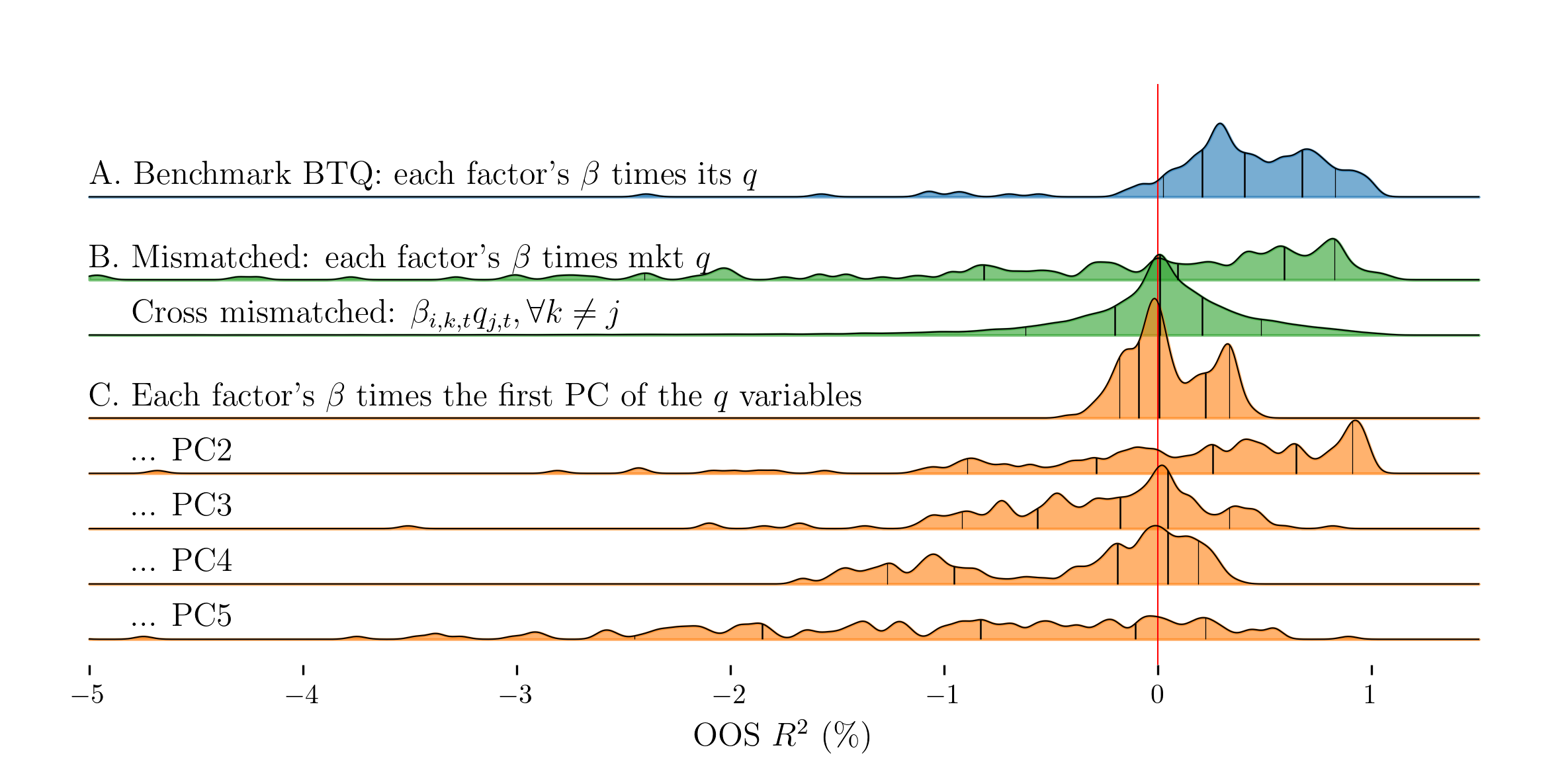}
	\begin{minipage}{\textwidth}
		{\footnotesize Note:
			Each distribution represents the outcomes (OOS $R^2$ values) of (A) 159 single-factor BTQ models, or (B, C) 159 mismatched BTQ variants, except Line ``Cross mismatched'' has $159 \times 159 - 159 =$ 25,122 OOS $R^2$ values. Kernel density estimates' (KDE) areas under the curve are standardized across all the 8 distributions. Vertical lines, thin: 10, 90 quantiles; thick: 25, 50, 75  quantiles.}
	\end{minipage}
	\label{fig:commonq}
\end{figure}

We know that stock returns exhibit a multi-factor risk structure. A remarkable aspect of BTQ is that its pricing power holds independently across a diverse array of factors in the ``factor zoo'' %
(see Table \ref{tab:xsreg_individual_factor} and repeated in Figure \ref{fig:mismatch} Panel A). We now show that each factor's $q$ provides distinct pricing information along its respective risk dimension. Mismatching one factor's $\beta$ to another factor's $q$ significantly reduces the prediction accuracy, meaning that a factor's $q$ pertains to explaining the cross-sectional return dispersion {\it only} along its own risk direction. Each dimension of risk provides independent evidence on the $q$-$\mu$ association, supporting the robustness of the economic interpretation. This result refutes the idea that the main results are driven by one (or a few) special ``secret sauce'' $q$ series.

In detail, Figure \ref{fig:mismatch} Panel A plots the benchmark distributions of the 159 OOS $R^2$ values for the single-factor BTQ models, with each factor's $\beta$ times its own $q$ (repeating the result in Table \ref{tab:xsreg_individual_factor}). More than 90\% of these factors yield positive OOS $R^2$ values, demonstrating the effectiveness of the BTQ model when correctly specified. 

The subsequent panels are BTQ variants where $\beta$ and $q$ are mismatched in various ways. The first variant in Panel B pairs each factor's $\beta_{i, k, t}$ with the market factor's quantity $q_{\text{mkt}, t}$. The second distribution is for the ``cross mismatched'' BTQ variants: each factor's $\beta$ is paired with the $q$ of every other factor ($\beta_{i,k,t}\times q_{j,t}, \forall j \neq k$), resulting in $159 \times 159 - 159 =$ 25,122 mismatched models. Panel C examines any underlying common signal among the 159 $q$ series, by pairing each factor's $\beta$ with a principal component of the $q$ series.\footnote{Notice, we conduct PCA of the 159 $q$ series here, rather than calculating the quantities for the PCs of the factor returns as in Section \ref{subsec:pca}. Appendix Figure \ref{fig:qcomove} Panel B reports the properties of the PC $q$ series.}

The mismatched variants perform significantly worse than the benchmark BTQ model. 
When correctly specified, more than 90\% of the factors in the ``zoo'' yield BTQ models with positive OOS $R^2$ values. %
However, the mismatched variants have approximately half of their density below zero, often with long left tails of highly negative $R^2$ values. 
In particular, the market factor's $q$ does not generalize to other factors. This means that the market factor's $q$ is not a special series that gives rise to other factors' BTQ results, despite its dominant time-series variation (Figure \ref{fig:ff3c_flow}) and leading prominence in factor selection (Figure \ref{fig:lasso_full}).
Other factors' $q$'s are essential to explain the risk-return tradeoff along their respective risk dimensions. The same conclusion holds for each principal component $q$ variant.\footnote{In terms of the right tails, the market $q$ and the second principal component $q$ retain some factors' predictive power from the benchmark BTQ model, likely due to the two's commonality with those factors.}
These results highlight the independence of each factor’s risk-return tradeoff and its unique association with its own $q$, providing robust evidence against the idea that our main results are driven by some particular signals in the $q$ series.

\section{Quantity or alternative channels driving factor premiums?}\label{sec:channels}

In the economic framework of the BTQ model, we argue that quantity is associated with factor premiums, representing the degree of cross-sectional risk-return trade-offs. 
Can alternative economic channels explain the observed empirical success of the BTQ model? 
What if some other underlying state variables drive the reported variation in factor premiums and, with quantity variables merely reflecting those state variables? If that is the case, the BTQ's reported empirical performance is merely a facade, and quantity would not be the direct driver of factor premiums.

To investigate this possibility, we examine alternative variables, including factor momentum signals and a large set of macroeconomic variables, both of which are well-documented in previous studies as predictors of factor returns. We find that their empirical performance is far behind the BTQ model across a wide range of empirical specifications.
These findings suggest factor quantity variation is directly associated with factor premiums and further validate the BTQ model's economic channel.

The first exercise considers factor momentum and retail flow's ``performance chasing'' behavior. Factor momentum implies past factor performances are positively associated with future factor returns. Meanwhile, retail mutual fund flows also positively respond to past performances.\footnote{Factor momentum is well-documented in the literature (e.g., \citealp{moskowitz2012time, gupta2018factor, ehsani2022factor, arnott2023factor}). Relatedly, \citet{huebner2023making} demonstrates the generation of price momentum in equilibrium driven by investor demands. Mutual fund performance chasing has also been extensively studied (e.g., \citealp{lou2012flow, ben2024discontinued}).} Can these two forces combined explain BTQ's reported empirical performance?

\begin{table}[t!]
	\caption{{BTQ vs. ``beta times momentum''} }
	\label{tab:alterq}
	\vspace*{-1em}
	\renewcommand{\arraystretch}{1.3}
	\begin{center}
		\scalebox{1}{ \newcolumntype{C}{>{\centering\arraybackslash}p{0.1\textwidth}}

\begin{tabular*}{\textwidth}{@{\extracolsep{\fill}}CCCCCCC}
    \toprule
    CAPM& FF3 & FF3C & FF5 & FF5C& selction & PC+selection\\
    \midrule
    \multicolumn{7}{l}{BTQ, benchmark results reported in Section \ref{sec:emp_res}}\\
    0.75&	1.03&	1.07&	0.44	&0.65	& 0.81&	0.77\\
    \multicolumn{7}{l}{``beta times momentum'', best of 1- to 12-month momentum signal formation periods}\\
    0.01 & -0.05 & -0.04 & 0.12 & 0.09 & 0 & 0 \\
    \bottomrule
\end{tabular*}
}
	\end{center}
	\footnotesize
	{Note:} All numbers are OOS predictive $R^2$ in \%. To save space, we only report the best $R^2$ values across the lookback window lengths. The complete results are available upon request, where most of the $R^2$ values are negative and with no discernible pattern across varying formation length $h = 1\sim 12$. $R^2 = 0$ is because Lasso selects no predictors, and the model forecasts zero return for all stock-months.
\end{table}%

The answer is no. First, this channel would imply a negative $q$-$\mu$ association, which is opposite to the BTQ model's prediction and our empirical findings. Performance-chasing behavior of retail investors would reduce sophisticated investors' holdings ($q$) when past factor returns are high. Under factor momentum, this reduction would be associated with high future factor returns---contradicting the observed positive association.
Regardless of the sign restriction, the data show that replacing $q_{k, t}$ in the BTQ model with a variety of factor-level momentum/reversal signals fails to yield reliable predictions of stock returns. Even the best ``beta times momentum'' predictor specifications ($\beta_{i,k,t} \times \text{mom}_{k,t}$, where $\text{mom}_{k,t}$ is the past return of factor $k$ in 1- to 12-month windows) have only slightly positive OOS $R^2$ values, with most cases yielding negative values (see Table \ref{tab:alterq}).
Our interpretation is that while factor momentum and performance-chasing are respectively valid phenomena supported in the literature, they are insufficient to reproduce the explanatory power of BTQ, 
and that our $q$ variable captures quantity variation beyond performance-chasing alone. 

The second exercise examines whether macroeconomic variables or aggregate financial metrics, used individually or combined linearly, can replace the $q$ variables and provide predictive power comparable to the BTQ model. We explore the FRED-MD dataset \citep{mccracken2016fred}, which contains a comprehensive set of 126 monthly macroeconomic variables, including dividend yield, default spread, and personal income growth. These are frequently used as conditioning variables in conditional factor models \citep{jagannathan1996conditional, lettau2001resurrecting, petkova2005value, daniel2011testing}.\footnote{We pre-process the Fred-MD series following standard procedures, detailed in Appendix \ref{appsubsec:macro}.}

\begin{table}[t!]
	\caption{BTQ vs. ``beta times macro variables''}
	\label{tab:macro}
	\vspace*{-18pt}
	\begin{center}
	\begin{tabular*}{\textwidth}{l l}
		\toprule
        \multicolumn{2}{l}{A. BTQ (benchmark: market beta times market $q$)} \\[4pt] 
        \hspace{1em}
        0.75& ~\\[10pt] 
    \end{tabular*}

    \begin{tabular*}{\textwidth}{l l l l l l}
        \multicolumn{6}{l}{B. Market beta times each macro variable (126 FRED-MD variables one at a time)} 
		\\[4pt] 
		\hspace{1em}
		best of the 126 & 95th percentile & 75th percentile & 50th percentile & 25th percentile \\
        \hspace{1em}
        0.28&	0.18&	0.00&	-0.12&	-0.63 
		\\[10pt] 
    \end{tabular*}
    \begin{tabular*}{\textwidth}{l@{\hspace{0.136\textwidth}}l@{\hspace{0.136\textwidth}}l@{\hspace{0.136\textwidth}}l@{\hspace{0.136\textwidth}}l}
        \multicolumn{5}{l}{C. Market beta times the principal component of the macro variables} \\[4pt]
        \hspace{1em}
        PC1 & PC2 & PC3 & PC4 & PC5 \\
        \hspace{1em}
        -0.17& 0.01 & -1.05 & -0.09 & -0.46 
		\\[10pt]
    \end{tabular*}
    \begin{tabular*}{\textwidth}{l l}
        \multicolumn{2}{l}{D. Market beta times the best linear combination of the macro variables} \\[4pt]
        \hspace{1em}
		OLS (best in-sample fit, no regularization) & LASSO (best among reg.\ param.) \\
        \hspace{1em}
        $<-100$ %
		& 0%
		\\[4pt] \hline
    \end{tabular*}
		\vspace*{-15pt}
	\end{center}
	\footnotesize
	Note: Predicting stock returns with conditional CAPM predictors in the form of ``market beta times ...''. All numbers are OOS $R^2$ in percentage. Multi-factor results in Appendix Table \ref{tab:macro_OOSR2_SummaryStats} are similar. Details of Panel D results across regularization parameters are in Appendix Figure \ref{fig:macroLASSO}.
\end{table}%

The results show that neither these variables nor their linear combinations come close to $q$'s predictive power. When tested individually in the form of ``market beta times a macro variable,'' even the best-performing macroeconomic variable among the 126 series achieves an OOS predictive $R^2$ that is only a fraction of the ``market beta times market $q$'' specification. Most macroeconomic variables yield negative OOS $R^2$ (Table \ref{tab:macro} Panels A vs.\ B).\footnote{The best macro variable, with $R^2 = 0.28\%$, is unemployment insurance initial claims. The second and third are housing starts in the northeast and personal consumption expenditures on services, respectively. These variables offer no coherent economic explanation, as they were chosen ex post.} For linear combinations of the macroeconomic variables, neither unsupervised principal components (Panel C) nor supervised combinations fitted on stock returns (Panel D, fitting details in Appendix Figure \ref{fig:macroLASSO}) yield any predictive power.\footnote{Principal components of the macroeconomic series are used for forecasting in, for example, \cite{stock2002forecasting} and \cite{ludvigson2009macro} for the bond market.} Multi-factor specifications report similar results in Appendix Table \ref{tab:macro_OOSR2_SummaryStats}.

\section{Conclusion}\label{sec:conclusion}

This paper considers a new but important aspect of risk's economic role in determining asset prices---the \emph{quantity} variation in investors' risk holdings induced by trading flows. The economic rationale is simple: when sophisticated investors hold more of a systematic risk factor, they require greater compensation for bearing that risk, which in turn drives the expected return of every stock exposed to the factor. Yet the empirical model yields a compelling risk-based explanation for expected stock returns.

We show that incorporating quantity into canonical factor pricing has important implications for asset pricing studies with three new findings. First, quantity variation elicits risk-return tradeoff relationships, which have been hard to capture with $\beta$ only and thereby cast doubt on whether risk explains expected returns. We find the cross-sectional relationship between factor exposures and expected returns ($\beta$-$\E r$ relationship) strongly depends on factor quantity variation, and the previous null result is a mixed average unconditional on quantity. Second, quantity enables a risk-based predictive model (termed beta times quantity, BTQ) for monthly stock returns. The model delivers high prediction accuracy in this hard empirical task dominated by unstructured machine learning models and firm characteristics. Third, incorporating quantity provides a new way for factor selection and, thereby, new answers to the factor zoo problem. Instrumenting factor premiums with quantity variation has not only greater identification power but also more economic relevance than traditional factor premium tests. We find that a few factors out of the factor zoo are selected for the model's high predictive power, and in a latent factor setting, the first two principal components overwhelmingly dominate the remaining components.

Besides showing the improvements against the $\beta$-only baseline, we also implement various versions of the ``quantity-only'' model, which directly relates stock-level quantity to expected stock returns. We find this alternative baseline also falls short by far in explaining expected stock returns.	This result implies that the stock returns' factor structure and the no-arbitrage pricing condition are important for modeling expected returns, even in the presence of significant price impacts from noise flows.

In summary, we show both quantity and risk should work together for modeling expected stock returns. At a high level, this is a natural result given the interaction between sophisticated investors and noise traders. It bridges factor pricing (which emphasizes rational agents' aversion to risk) and the price impact of noise flows (which emphasizes the price dislocation effects of non-fundamental flows and inelastic demand). The contribution of this empirical paper is providing a simple and actionable way to integrate quantity information into canonical factor models and showing its significant improvement to the factor model's empirical relevance.

We are confident that future research can similarly incorporate quantity information into other existing asset pricing methods to yield new insights for various research questions. Another interesting direction for further research is to explore a richer set of asset holdings information to construct other quantity variables; a concurrent paper \citep{gabaix2023asset} is highly relevant for this potential direction.

\clearpage
\singlespacing

\clearpage

\doublespacing

\begin{appendices}

    \renewcommand{\theequation}{A.\arabic{equation}}
    \renewcommand{\thefigure}{A.\arabic{figure}}
    \renewcommand{\thetable}{A.\arabic{table}}
    \renewcommand{\thepage}{A.\arabic{page}}
    
    \setcounter{equation}{0}
    \setcounter{figure}{0}
    \setcounter{table}{0}
    \setcounter{page}{1}
    \setcounter{footnote}{0}
    
    \begin{center}
        \LARGE{Online Appendix of ``Quantity, Risk, and Return''}
    \end{center}

	\section{Equilibrium theory and microfoundation of quantity-factor premium relationship}\label{sec:microfound}

	In this section, we develop a theoretical model to provide a formal and explicit interpretation of the empirical framework and findings. This model formalizes key concepts such as sophisticated investors' inelastic demand, noise traders' flows, and how their interaction in equilibrium determines factor premiums. The microfoundation delivers the main empirical specification that observed factor premiums (which are the equilibrium outcomes) are positively related to factor quantities (Eq.\ \ref{eq:fp_specification}). The model highlights two theoretical underpinnings that support the strong explanatory power of BTQ for expected stock returns (i.e., high $R^2$):
	1) sophisticated investors exhibit sufficiently inelastic demand, driven by two primitive conditions: high factor-level risk aversion and limited risk-bearing capacity relative to the aggregate stock market; and 2) the noise traders' flows are indeed noisy and exhibit sufficient variation.

	\subsection{Factor pricing identities when factor premium is an equilibrium outcome}

	We first provide basic factor pricing identities when the factor premium is allowed to be an endogenous equilibrium outcome. We show that a high factor premium and a low price of the factor risk are two sides of the same coin. The model has two periods $t$ and $t+1$, and the risk-free rate is $\rf$.

	Random state variable $\ftld_{t+1}$, with $\E_t[\ftld_{t+1}] = 0$, represents the physical systematic risk of a factor, independent of time-$t$ equilibrium trading outcomes. For simplicity and without loss of generality, we omit subscript $k$ for variables $\ftld$ as well as $q$, $\mu$ etc.\ below, as the theory applies generically to any factor.

	Let $f_{t+1}(q_t)$ be the payoff of a zero-cost factor portfolio that has a unit exposure to $\ftld_{t+1}$.
    This variable represents the observed long-short portfolio \textit{return} (e.g., the SMB series from French's website). With ``$(q_t)$,'' we allow for a general form where the payoff random variable depends on the equilibrium quantity variable $q_t$. By construction,
    \begin{align}
        f_{t+1}(q_t) = \ftld_{t+1} + \mu_t(q_t), &  & \forall q_t,
        \label{eq:factordecomp}
    \end{align}
    where $\mu_t(q_t) = \E_t [f_{t+1}(q_t)]$ is the factor premium, the focus of this section. In the cross section, by the APT, a stock with exposure $\beta_t$ to systematic risk $\ftld_{t+1}$ has expected excess return $\beta_t \mu_t(q_t)$ (assuming no exposure to other systematic risks).

	Turning to time-$t$ prices, we define $P^{\ftld}_t (q_t) \defeq \Price_t[\ftld_{t+1}](q_t)$ as the price of the state contingent payoff $\ftld_{t+1}$, where $\Price_t[~\cdot~](q_t) \defeq \E_t[M_{t+1}(q_t)~\cdot~]$ is the payoff pricing operator with the stochastic discount factor (SDF) $M_{t+1}(q_t)$. Once again, ``$(q_t)$'' indicates that the prices are endogenous to $q_t$.

	By construction, the price of the factor portfolio payoff is 0: $\Price_t[f_{t+1}(q_t)](q_t) = 0, \forall q_t$. This leads to the intuitive identity between equilibrium factor premium and factor risk price. Applying the pricing operator to both sides of the factor return decomposition \eqref{eq:factordecomp}, by the law of one price, we obtain $0 = \Price_t \Brack{f_{t+1}(q_t)} = P^{\ftld}_t (q_t) + \mu_t(q_t)/(1 + \rf)$. Therefore,
	\begin{align}
		\mu_t(q_t) = -(1 + \rf) P^{\ftld}_t (q_t), &  & \forall q_t.
		\label{eq:two_demand}
	\end{align}
	This expression captures the canonical inverse relationship between the zero-cost factor premium and the price of zero-mean factor risk. Our model extends the canonical relationship by allowing both sides of the equation to be endogenous equilibrium outcomes.

	\subsection{Demand functions and the equilibrium}

\begin{figure}[t!]
        \caption{Demand functions and the equilibrium}
        \hspace*{-1em}%
        \begin{subfigure}[t]{0.5\textwidth}
            \captionsetup{justification=centering}
            \caption{sophisticated hold more $q$, price drops...}
            \centering
            \vspace*{.2em}%
            \includegraphics[scale=0.62]{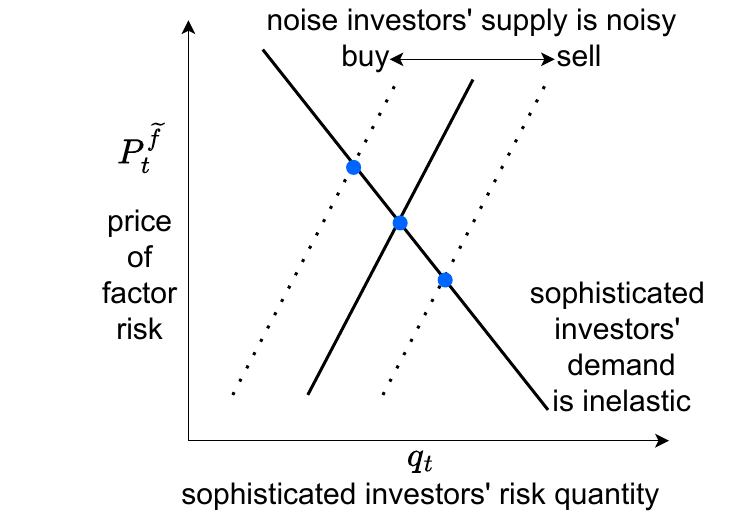}
        \end{subfigure}%
        \begin{subfigure}[t]{0.5\textwidth}
            \captionsetup{justification=centering}
            \caption{... and factor premium rises}
            \centering
            \includegraphics[scale=0.62]{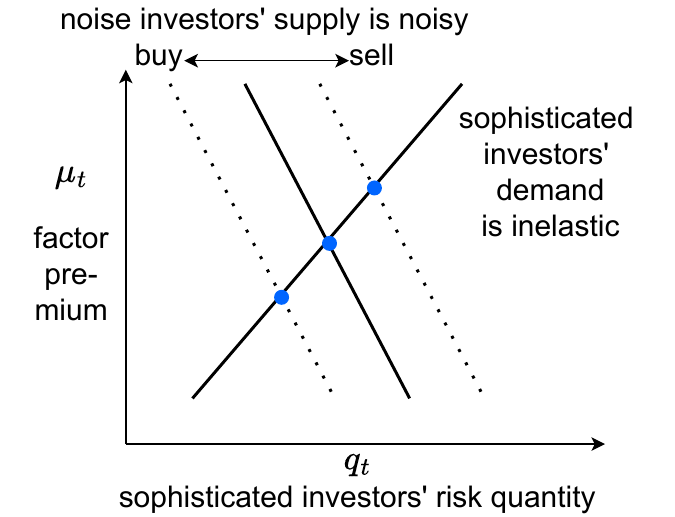}
        \end{subfigure}%
        \vspace{1em}
        \begin{minipage}{\textwidth}
            { \footnotesize Note: Two forces for strong quantity-factor premium empirical relationship are 1) inelastic sophisticated investors' demand (sloped, not flat), and 2) significant fluctuations in noise traders' supply (shifts left and right). The two panels are equivalent illustrations with price or factor premium---just flip everything upside down according to Eq.\ \ref{eq:two_demand}.\par}
        \end{minipage}
        \label{fig:equilibrium}
    \end{figure}

	Next, we introduce the demand and supply functions of the sophisticated and noise investors, along with their equilibrium interactions, as illustrated in Figure \ref{fig:equilibrium}.

	Sophisticated investors obey factor pricing and enforce the law of one price. The aforementioned SDF, $M_{t+1}(q_t)$, and the pricing operator are interpreted as theirs, which are influenced by their quantity holdings: $q^\textrm{sophi}_t$. Hence, their demand function is: $P^{\ftld}_t (q_t^\textrm{sophi}) = \Price_t [\ftld_{t+1}](q^\textrm{sophi}_t) = \E_t[M_{t+1}(q^\textrm{sophi}_t) \ftld_{t+1}]$.

	Both $P^{\ftld}_t (~\cdot~)$ and $\mu_t(~\cdot~)$ are equivalent representations of demand, connected by Eq.\ \ref{eq:two_demand}.
	The theory can be equivalently stated with either one. This equivalence indicates that the realized zero-cost returns $f_{t+1}(q_t)$, such as those from Fama-French, should be modeled as the equilibrium outcome of time-$t$ trading activities, rather than exogenous variables.

	Empirically, typical demand-based asset pricing studies target {\it concurrent} price impacts. They model investors' demand curves as $\Price_t [\widetilde{X}_{t+1}](q_t)$, where a payoff $\widetilde{X}_{t+1}$ (in our case $\ftld_{t+1}$) is independent of the time-$t$ equilibrium quantity. In this paper, we shift the focus from concurrent price impact to risk premiums, aligning more closely with factor pricing literature's emphasis on expected future returns. %
	With the identity in  Eq.\ \ref{eq:two_demand}, we explicitly show how our and the literature's focuses are related.
	Shifting from concurrent price impact to future returns also dispenses with the endogeneity issue between $q_t$ and $P^{\ftld}_t$ (such as flow chasing concurrent return), which necessitates instrumental variable methods in \citet{gabaix2024granular}. Nonetheless, our model still requires the exogeneity of time-$t$ trading to future risk $\ftld_{t+1}$ realization, which is easier to justify.

	Inelastic demand by sophisticated investors is key to the reported empirical relationships. That is, $P^{\ftld}(~\cdot~)$ is downward sloping (or $\mu(~\cdot~)$ is upward sloping). Perfectly elastic demand (a flat line where price remains constant regardless of quantity) would imply that the sophisticated investors have unlimited risk-bearing capacity or ``deep pockets,'' inconsistent with real-world observations. %
    The microfoundation in the next subsection connects inelasticity to sophisticated investors' risk aversion and their capital share in the market.

	Since sophisticated investors trade with noise traders, the ``supply'' function of noise traders (how much $q$ they sell) is $q_t^{\textrm{noise}}(P^{\ftld}_t) = \textrm{DeterministicSupply}(P^{\ftld}_t) + \eta_t$, where $\eta_t$ is the noise supply component and unspecified function $\textrm{DeterministicSupply}(P^{\ftld}_t)$ is the deterministic component of supply that can respond elastically to price.

	We are agnostic about most aspects of the noise traders' supply, but require that the noise component $\eta_t$ exhibits significant variation and is indeed noisy in the sense of not predicting the future factor risk $\ftld_{t+1}$. It does not affect the result whether $\textrm{DeterministicSupply}(\cdot)$ is perfectly inelastic (meaning noise traders are completely insensitive to price, vertical in Figure \ref{fig:equilibrium}) or somewhat elastic (meaning they partially adjust supply to price). This is because the observed equilibrium quantity is $q_t$, and that $\eta_t$ is unobserved and unmodelled anyway. It is implausible that $\textrm{DeterministicSupply}(\cdot)$ is or near perfectly elastic (horizontal in Figure \ref{fig:equilibrium}), as it would imply that the noise traders are not only ``sophisticated'' about price but also ``deep-pocketed.'' We are also agnostic about the sources of the ``noise,'' which could be driven by various factors such as investor sentiment, beliefs, or media narratives (e.g., \citealp{zhou2024active}).

	The equilibrium is obtained by market clearing: $q_t = q_t^{\textrm{sophi}} = q_t^{\textrm{noise}}$, given the two demand (supply) functions. Equilibrium $q_t$ is determined by noise trading shock $\eta_t$ (with an unspecified function).

	The empirical model additionally assumes that sophisticated investors' demand curve is time-invariant: $P^{\ftld}_t (q_t) = P^{\ftld}(q_t)$,  or equivalently \ $\mu_t(q_t) = \mu(q_t)$ for any $t$ and $q_t$. This is largely an empirical restriction stating that, besides $q^\textrm{sophi}_t$, the demand function is not affected by any other time-varying variables. Section \ref{sec:channels} explores alternative variables, such as factor momentum and macro variables, but finds none with reliable empirical power for pricing the cross section of stock returns. The next subsection provides a theoretical demand function with this property, under appropriate assumptions.

	With all these setups, we conclude that the observed equilibrium outcomes, $\{q_t, P^{\ftld}_t\}$, lie on the sophisticated investors' demand curve $P^{\ftld} (~\cdot ~)$, and equivalently, the observed $\{q_t, \mu_t\}$ on $\mu(~\cdot ~)$. This establishes that, across periods, the factor premium $\mu_t$ is positively related to the factor quantity $q_t$.

	\subsection{A microfoundation of the inelastic demand function}\label{sec:inelastic}

	In this subsection, we provide a specific microfoundation for sophisticated investors' pricing kernel, which results in an inelastic, downward-sloping demand function with a closed-form expression,
	$\mu(q_t) = \mu + \lambda q_t$, consistent with the empirical model in Eq. \ref{eq:fp_specification}. The inelasticity arises from two key factors: 1) the risk aversion of sophisticated investors and 2) their limited total capital for absorbing the factor risk. Here, we derive an analytical expression for these two forces, providing one simple mechanism for inelastic demand, though other mechanisms, such as investment mandates \citep{gabaix2022search}, are also possible.

	Suppose sophisticated investors' total wealth (AUM) is $\$ W_t$ and their existing portfolio has a random payoff $\$ W_{t+1}$, with a return $r^\mathrm{W}_{t+1} \defeq W_{t+1}/W_t - 1$. A representative sophisticated investor with \$1 AUM has CARA utility with risk aversion $\gamma$, $\E_t[-\exp(-\gamma   r^\mathrm{W}_{t+1})]$. The utility and demand have identical functional forms across sophisticated investors and scale proportionally with their individual AUMs, under the standard assumption that CARA risk aversion scales inversely with the AUM.

	Taking zero-cost factor payoff $f_{t+1}(q_t)$ as given, the representative sophisticated investor optimally allocates additional exposure to systematic risk factors $b$ by solving:
	\begin{align}\label{eq:eqm}
		b_t = \arg\max_{b} \E_t[-\exp(-\gamma   (r^\mathrm{W}_{t+1}+b f_{t+1}(q_t)).  ]
	\end{align}
	Assuming that the wealth return $r^\mathrm{W}_{t+1}$ and factor return $ f_{t+1}(q_t)$ are jointly normally distributed, the first-order condition of Eq.\ \ref{eq:eqm} implies:
	\begin{align}%
		\mu_t(q_t) =\mathbb{E}_t[ f_{t+1}(q_t)] & = \gamma\mbox{cov}_t(r^\mathrm{W}_{t+1}, f_{t+1}(q_t)) + \gamma b_t \mbox{var}_t(f_{t+1}(q_t)) \nonumber \\
		& =\gamma\mbox{cov}_t(r^\mathrm{W}_{t+1}, \ftld_{t+1}) + \gamma b_t \mbox{var}_t(\ftld_{t+1}),\label{eq:demand_b}
	\end{align}
	where the last equality uses Eq.\ \ref{eq:factordecomp}. Given per AUM demand $b_t$, the aggregated demand of additional factor exposure is $b_t W_t$.

	Up to this point, we have established the demand function connecting the factor premium to the sophisticated investors' risk holdings measured in terms of their factor exposure: $b_t W_t$. The empirical counterpart of this measure is the flow-induced factor beta: $\sum_i \flowstock_{i, t} \betahat_{i, k, t}$, the intermediate term in constructing the quantity variable in Eq.\ \ref{eq:porttheory}.
	To connect the demand function in this measure to the quantity variable $q_t$ in the empirical model, we assume that the aggregate stock market capitalization at time $t$ is $\mbox{AGG}_t$, and sophisticated investors' total wealth $W_t$ is a constant fraction $\pi$ of $\mbox{AGG}_t$, such that $W_t = \pi \mbox{AGG}_t$. Therefore, the factor quantity defined in Section \ref{subsec:qconstruction} can be simplified as
	\begin{equation}
		\qtld_t = b_t W_t \mbox{var}_t[f_{t+1}(q_t)]/\mbox{AGG}_t = b_t \mbox{var}_t[f_{t+1}(q_t)] \pi.
	\end{equation}
	Substituting into Eq. \ref{eq:demand_b}, we have the demand function in terms of $\qtld_t$:
	\begin{align}
		\mu_t(q_t) = \gamma\mbox{cov}_t(r^\mathrm{W}_{t+1}, \ftld_{t+1}) + \frac \gamma \pi \qtld_t.
	\end{align}
	In the empirical implementation, we also standardize $\qtld_t$ as $q_t = \qtld_t / \sigma(\qtld_t)$, so that the estimated linear coefficient $\lambda$ can be more intuitively interpreted as the price effect per standard-deviation shock in $\qtld_t$. (The magnitude of $\qtld_t$ tends to be very small, because this raw measure is normalized by the aggregate stock market capitalization.)

	Finally, the demand function matches the empirical specification in Eq.\ \ref{eq:fp_specification}:
	\begin{align}\label{eq:match}
		\mu(q_t) & = \gamma\mbox{cov}_t(r^\mathrm{W}_{t+1}, \ftld_{t+1}) + \frac \gamma \pi \sigma(\qtld_t) q_t\\
		& =\mu + \lambda q_t,
	\end{align}
	with $\mu \defeq \gamma\mbox{cov}_t(r^\mathrm{W}_{t+1}, \ftld_{t+1})$ and $\lambda \defeq (\gamma /\pi) \sigma(\qtld_t)$.

	Both terms in Eq.\ \ref{eq:match} have clear interpretations. In the first term, we assume $\gamma\mbox{cov}_t(r^\mathrm{W}_{t+1}, \ftld_{t+1}) = \mu$. That is, the background factor risk premium (the risk premium when no additional factor exposure is taken, $b_t = q_t = 0$) is constant. Empirically, it implies that no conditioning variables other than $q_t$ affect the factor premium. Section \ref{sec:channels} supports this assumption and finds no reliable alternative conditioning variables for pricing the cross section of stock returns. In fact, the main empirical results show, via the ``beta-only'' benchmark, that $\mu$ is indistinguishable from zero, suggesting that other conditioning variables have effects too weak to be empirically detected.

	The second term, $\lambda \defeq  (\gamma /\pi) \sigma(\qtld_t)$, shows the conditions underlying the strong empirical relationship between quantity and factor premium. The first condition requires that sophisticated investors' demand is sufficiently inelastic. This inelasticity is governed by two primitive parameters: 1) high risk-aversion (high $\gamma$) and 2) a small share of sophisticated investors relative to the stock market (low $\pi$). Specifically, the factor premium multiplier is $\gamma /\pi$; and the price multiplier, according to Eq.\ \ref{eq:two_demand}, is $\gamma /(\pi(1+\rf))$, inversely related to the elasticity of the demand function. This aligns with \citet{gabaix2022search}, who also highlight the limited share of hedge funds capital (around 5\% of total investors) as a key driver of the inelastic demand. The second condition requires noise traders' supply to exhibit significant variation (large $\sigma(\qtld_t)$). Greater variation in the noise trading---reflected in large left-right shifts in the supply curve---translates to greater variation in the factor premium attributed to the quantity channel, which this paper captures empirically.

    \section{Technical details}

    \subsection{Construction and cleaning of mutual fund flows}\label{sec:flowcleaning}

	In this appendix, we present details related to constructing and cleaning mutual fund flows. Our primary data source is the CRSP Survivorship-Bias-Free Mutual Fund database. We start with all funds' return and total net assets (TNA) data at the share-class level. A mutual fund may include multiple share classes. We first drop observations with no valid CRSP identifier, \texttt{crsp\_fundno}. A few fund-share classes report multiple TNAs in a given month. These are likely data duplicates, so we keep only the first observation of the month.
	In what follows, we discuss the cleaning steps for returns and TNA at the share-class level. After cleaning, we aggregate the share-class level data to the fund level.

	\subsubsection{Return cleaning}
    
	We first correct data errors in monthly mutual fund net returns, \texttt{mret}.

	First, we address extremely positive returns. We study the case in which a particular fund share has returns greater than $100\%$ and has existed for more than one year.\footnote{We use the one-year threshold because mutual fund return and TNA during the first year are sometimes inaccurate in the CRSP database. For example, return and TNA can be stale or reported using a placeholder number such as 0.1. We address these issues by cross-checking with the alternative database.} We manually check the entire time series of each share class in this subsample. Most of these extreme returns reflect misplaced decimal points, which confound returns in decimal and percentage formats. For these cases, we divide the faulty returns by 100.

	Second, we address extreme negative returns. Similarly, we study the case in which a particular fund share has existed for more than one year and has returns lower than $-50\%$. With extremely negative returns, we need to distinguish data errors from significantly negative returns before a fund's closure. Thus, we manually check only the subsample of negative returns that occur at least one year prior to the last observation of a closed fund. We manually check whether these extreme returns reflect data-input errors for this subsample. For the cases with misplaced decimal points, we divide the faulty returns by 100.

	\subsubsection{TNA cleaning}

	Unlike many prior studies that construct percentage mutual fund flows, we study dollar-value flows to preserve the cross-sectional relative magnitudes. The mutual fund size distribution features a very long right tail. Winsorizing the extreme dollar-value TNA likely removes both valid large values and input errors, generating significant bias. We devise an algorithm to identify and correct erroneous observations of TNA:
    \begin{enumerate}
		\item Using the sample with corrected returns, we calculate dollar flows as in Eq.\ \ref{eq:mutualfundflow} at the share-class level.
		\item We study the top and bottom 0.5\% of all dollar flows.\footnote{The choice of the top and bottom 0.5\% is motivated by the distribution of dollar flows, where most extreme values tend to occur at these tails.} We manually check this subsample's TNA time series of all share classes. We identify several common errors:
          \begin{itemize}
              \item Misplaced decimal points (usually by hundredths or thousandths).
              \item Stale TNA observations from CRSP, typically when a fund reorganizes its share class offering (e.g., adding a new share class and moving assets into a single share class).
              \item CRSP sometimes sets $\mbox{TNA}=0.1$ for the first few months of a new fund or a new share class.
          \end{itemize}
          We correct the misplaced decimal issue. For funds suffering from the latter two problems, we obtain their TNA from Morningstar.\footnote{We merge the CRSP and Morningstar databases using a fund's ticker.} Morningstar's TNA data (\texttt{Net\_Assets\_ShareClass\_Monthly}) suffer to a lesser extent from these issues than CRSP's TNA data. We conclude by further cross-checking other third-party vendors (e.g., Yahoo Finance and Bloomberg Terminal). Hence, whenever a fund's CRSP TNA deviates more than 50\% from its Morningstar TNA, we use the Morningstar TNA.
		\item We repeat the previous steps one more time to ensure that we have accounted for most, if not all, extreme errors.
		\item We compare our cleaned TNA with total assets (\texttt{assets}) from Thomson/Refinitiv Holdings data.\footnote{We merge the two databases via the linking table MFLINKS, which WRDS provides.} Following \citet{coval2007asset} and \citet{lou2012flow}, we drop observations whenever  our cleaned TNA deviate more than 50\% from \texttt{assets} from Thomson/Refinitiv.
	\end{enumerate}
	Using cleaned return and TNA data, we calculate dollar flows at the share-class level using equation \eqref{eq:mutualfundflow}. We obtain a fund's flows by adding up the flows of all share classes in the same fund.
	The final sample contains 1,707,742 fund$\times$month observations. %

	\subsubsection{Cross-validating the data-cleaning procedure}\label{sec:crossvalid}

	We cross-validate our data-cleaning procedure by comparing our aggregated mutual fund flows with alternative sources. We compute the quarterly aggregate flows in dollar amounts from our main sample and compare them with data from the Investment Company Institute (ICI) and the Flow of Funds (FoF).

	The ICI publishes aggregate monthly mutual fund flows, from which we extract quarterly data spanning from 2007 to 2022. Specifically, we use the ICI’s Total Equity mutual fund flows, which align closely with the coverage of mutual funds in our sample. Additionally, we draw on data from the Federal Reserve Board’s Financial Accounts of the United States -- Z.1 (formerly known as the Flow of Funds or FoF) from the same time period, providing quarterly observations. For our analysis, we focus on mutual fund flows (Line 28) within Corporate Equities (Table 223) and use unadjusted flows (FU).

	\begin{figure}[!t]
		\centering
		\caption{Time series of aggregate mutual fund flows from various sources} %
		\hspace*{-1em}%
		\centering
		\includegraphics[scale=0.8]{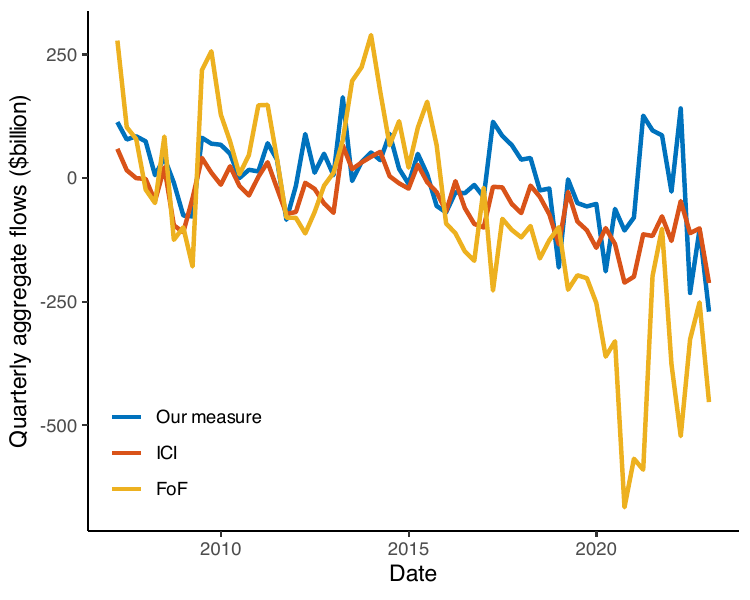}
		\begin{minipage}{\textwidth} %
			{ \footnotesize Note: The figure plots the quarterly time series of our measure, ICI flows, and Flow of Funds (FoF) flows.\par}
		\end{minipage}
		\label{fig:flow_comp}
	\end{figure}

	Figure \ref{fig:flow_comp} plots the quarterly time series of aggregate mutual fund flows from all three sources. Our measure of aggregate mutual fund flows is broadly consistent with the other two sources. The correlation between our aggregate flow measure and ICI flow is 0.63, while the correlation between our measure and FoF flow is 0.47.

	The differences observed in Figure \ref{fig:flow_comp} among the three measures likely reflect variations in mutual fund coverage. Specifically, the ICI flow tracks virtually all U.S. equity mutual funds that invest in both domestic and world equity markets.\footnote{The ICI is a trade association for the mutual fund industry, and virtually all U.S. mutual funds are ICI members (\citealp{warther1995aggregate}).} The FoF flow, sourced from unpublished ICI data, aggregates unadjusted flows into and out of all U.S. mutual funds (including variable annuity long-term mutual funds). It is calculated based on mutual fund assets in common stock, preferred stock, and rights and warrants.\footnote{See  \url{https://www.federalreserve.gov/apps/fof/SeriesAnalyzer.aspx?s=FA653064100&t=F.223&suf=Q}.} In comparison, our mutual fund sample contains U.S. mutual funds covered by CRSP, which collects historical data from various sources.\footnote{The sources include the Fund Scope Monthly Investment Company Magazine, the Investment Dealers Digest Mutual Fund Guide, Investor's Mutual Fund Guide, the United and Babson Mutual Fund Selector, and the Wiesenberger Investment Companies Annual Volumes.} Due to the nature of the data collection process, CRSP's coverage is smaller than ICI's coverage.

\subsubsection{Alternative method to construct factor-level flow directly from mutual fund flow}\label{subsec:altermutual}

In an alternative method, flows into mutual funds, $\flowfund_{m,t}$, can be directly aggregated into flows into factors, $\flowfactor_{k,t}$, by substituting Eq.\ \ref{eq:flowhat} into Eq.\ \ref{eq:porttheory}. This approach approximately aggregates $\flowfund_{m,t}$ based on each fund's beta to each factor.
Specifically, we have 
\begin{align}
    \flowfactor_{k,t}=-\sum_{i} \sum_{\mbox{\scriptsize fund } m}\flowfund_{m,t}\weightfund_{i,m,\text{quarter}(t)-2} \betahat_{i,k,t}\varhat_t(f_{k,t}).
\end{align}
The alternative is 
\begin{equation}
    -\sum_{\mbox{\scriptsize fund } m}\flowfund_{m,t}\betahat_{m,k,t}\varhat_t(f_{k,t}),
\end{equation}
where $\betahat_{m,k,t}$ is the beta of fund $m$ to factor $k$ in month $t$. These two are approximately the same because $\sum_{i} \weightfund_{i,m,\text{quarter}(t)-2} \betahat_{i,k,t}\approx \sum_{i} \weightfund_{i,m,t} \betahat_{i,k,t} = \betahat_{m,k,t}$.

We do not use the alternative method for two reasons. First, our method follows the literature and can help exclude potential informed trading by mutual fund managers.
The direct approach is only an approximation because we use lagged, not current, mutual fund holdings to construct $\flowstock_{i,t}$, an important detail to exclude potential informed trading by mutual fund managers (\citealp{lou2012flow}).  Second, starting from stock-level flows and building upward (Eq.\ \ref{eq:porttheory}) is more general and extends beyond mutual fund flow-induced trading. For example, retail investor order flow imbalance, a widely used measure of noise trading in the literature (\citealp{lee1991inferring}; \citealp{boehmer2021tracking}; \citealp{li2022price}), is constructed directly at the stock level.

	\subsection{Technical details of Lasso implementation}\label{app:technical_details_lasso}

	In optimization \eqref{eq:lasso}, adding the term ``$1/\sigma(\qtld_{k,t})$'' is technically necessary because we have already standardized $\qtld_{k,t}$ to $q_{k,t} = \qtld_{k,t} / \sigma(\qtld_{k,t})$ (see Section \ref{subsec:qconstruction}). Optimization \eqref{eq:lasso} therefore is equivalent to running the standard Lasso on the pre-standardized BTQ ($\betahat \times \qtld$)
	\begin{equation}\label{eq:lassotrue}
		\min_{\tilde{\lambda}_1 \dots \tilde{\lambda}_K}\frac{1}{2|\text{IS}|}\sum_{i, t \in \text{IS}}\Paren{r_{i,t+1}-\sum_{k=1}^K \lamtld_k \betahat_{i,k,t} \qtld_{k,t}}^2 + \omega \sum_{k=1}^K \Abs{\lamtld_k},
	\end{equation}
	and then standardizing the coefficients for economic interpretation: $\lambda_k = \lamtld_k \sigma(\qtld_{k,t})$. Although we standardize $\qtld_{k,t}$ for interpretability, we do not want to lose the information contained in the original quantity $\qtld_{k,t}$ during the Lasso selection process. A factor with greater variation in $\qtld_{k,t}$ will have an inflated $\lambda_k$ after being standardized to $q_{k,t}$, but it should not be additionally penalized for this reason. Standard Lasso implementation where the economic interpretation is not a priority would typically standardize the predictor (BTQ together) across the $\{i, t\}$ panel. Here, we customize the standardization based on the required economic interpretation.

	Similarly, for the $\beta$-only model, the Lasso implementation is
	\begin{equation}\label{eq:lassobetaonly}
		\min_{\mu_1 \dots \mu_K}\frac{1}{2|\text{IS}|}\sum_{i, t \in \text{IS}}\Paren{r_{i,t+1}-\sum_{k=1}^K \mu_k \betahat_{i,k,t}}^2+ \omega \sum_{k=1}^K \Abs{\mu_k}.
	\end{equation}

	We perform ten-fold cross-validation to tune hyperparameter $\omega$ based on only in-sample information from 2000 to 2009. In each fold, we exclude one year of observations and solve the Lasso problem \eqref{eq:lassotrue} using the remaining nine years of in-sample data. The model is evaluated in the left-out year to form predicted returns $\widehat{r}_{i,t+1}^\text{[cv]}$. After enumerating all folds and forming predicted returns for all in-sample observations, we calculate the cross-validated (CV) in-sample mean squared errors (MSE) as $\sum_{i, t \in \text{IS}}\Paren{r_{i,t+1}-\widehat{r}_{i,t+1}^\text{[cv]}}^2$. Hyperparameter $\omega$ is tuned by choosing the one with the minimum CV MSE.
	
    \section{Additional empirical results}

    \subsection{Additional properties of the quantity variable $q_{k,t}$}\label{appsec:qproperties}

\begin{figure}[!t]
		\begin{center}
			\caption{Degree of comovement among the quantity variables ($q$) of different factors}
			\label{fig:qcomove}
			\begin{minipage}[t]{0.38\textwidth}
				\centering
				A. Correlation matrix
				\\[34pt]%
				\resizebox{\linewidth}{!}{\begin{tabular*}{\textwidth}{l@{\extracolsep{\fill}}cccc}
    \toprule
    & MKT & SMB & HML &MOM\\
    MKT	&1	&&&\\
    SMB&	0.55&	1	&&\\
    HML	&0.47	&0.57	&1&\\
    MOM	&-0.47&	-0.23&	-0.75&	1\\
    \bottomrule
\end{tabular*}}
			\end{minipage}
			~~
			\begin{minipage}[t]{0.58\textwidth}
				\centering
				B. PC variances for $q$'s of 153 JKP factors
				\\[4pt]%
				\includegraphics[scale=0.7]{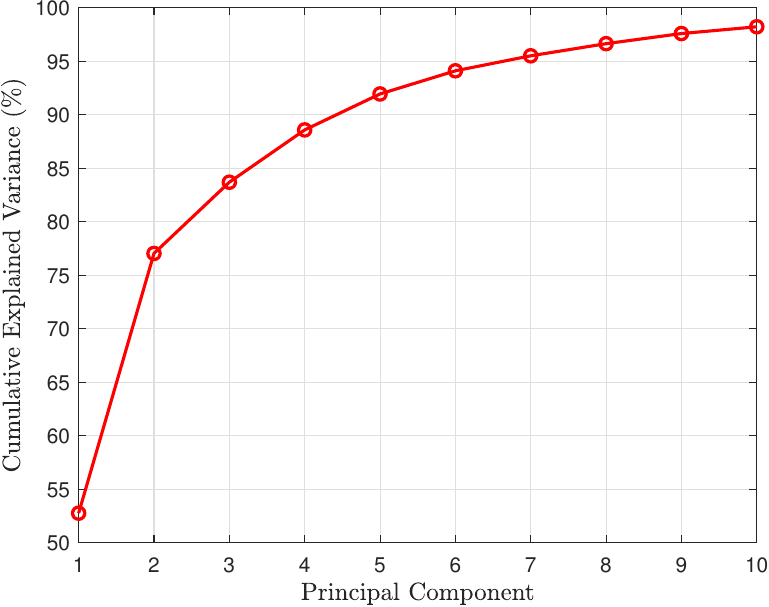}
			\end{minipage}
		\end{center}
		\footnotesize Note: Panel A: pairwise correlation for $q_{k,t}$ of the four Fama-French-Carhart factors. Panel B: cumulative explained variances by principal components of $q_{k,t}$ series of the 153 JKP factors.
	\end{figure}

	Figure \ref{fig:qcomove} reports various statistics of the constructed quantity variables $q_{k,t}$ to show the extent to which these time-series variables comove. Panel A shows the pairwise correlation matrix of the four Fama-French-Carhart factors. The results reveal some comovement (both positive and negative) among the four variables, which is also evidenced in the time series plot in Figure \ref{fig:ff3c_flow} in the main text. HML-MOM has the greatest (in absolute value) correlation of $-0.75$. Nonetheless, all pairwise correlations are far from $\pm1$, indicating that the $q_{k,t}$ variables are far from collinear, and each captures unique information about the underlying quantity variations.

	Panel B shows a similar pattern of limited comovement among the 153 JKP factors. Instead of reporting pairwise correlations, we conduct a principal component analysis (PCA) on the $q_{k,t}$ variables of the 153 JKP factors and report the cumulative explained variances by principal components. The plot reveals a clear factor structure among the 153 factor-level $q_{k,t}$ variables, yet also indicates significant unique information across different dimensions of $q_{k,t}$. The first principal component explains around half of the total variance, and the first two principal components in total explain around 77\% of the total variance. It requires five principal components to explain 90\% of the total variance, and seven to explain 95\% of the total variance. We also note that these in-sample PC statistics are likely exaggerated due to overfitting.

	In summary, the $q_{k,t}$ variables are not collinear and capture unique information about the underlying quantity variations. This feature indicates the paper's main result on BTQ's predictive power is not driven by a few special $q_{k,t}$ variables. The fact that BTQ's predictive power is consistent across various factor specifications speaks to the robustness of the underlying economic mechanism.

	\subsection{Predicting factor returns with factor quantity}\label{sec:factor_predict}

	This appendix subsection presents the results of using factor quantities to predict factor returns. While time-series predictability is not the main focus of this paper, it is naturally implied by the paper's theoretical framework, particularly the factor premium modeling in Eq.\ \ref{eq:fp_specification}. Empirically, we successfully detect the predictability to a certain extent, consistent with the theoretical motivation. However, we also note the apparent methodological limitations of predicting factor returns with simple time-series regressions.

    \begin{table}[!t]
		\caption{{Predicting factor return $f_{k,t+1}$ using quantity $q_{k,t}$} }
		\vspace*{-1em}
		\renewcommand{\arraystretch}{1.3}
		\begin{center}
			\scalebox{1}{ \begin{tabular*}{\textwidth}{l@{\extracolsep{\fill}}ccccccc}
    \toprule
        & \multicolumn{4}{c}{Fama-French-Carhart factors} & \multicolumn{3}{c}{Across 153 JKP factors} \\
    \cmidrule(lr){2-5} \cmidrule(lr){6-8}
        & MKT & SMB & HML & MOM & Q25 & Median & Q75 \\
\midrule
$\lambda_k$ (\%) & 1.04&	0.49&	0.82&	1.10	&0.25	&0.66&	1.00\\
$t$-stat & (3.25)	&(2.45)	&(2.89)&	(1.76)&	(1.41)	&(2.00)&	(2.64) \\
$\mu_k$ (\%) & 0.38&	0.19&	0.08&	0.36&	-0.20&	-0.01&	0.27\\
$t$-stat & (1.39)&	(1.16)&	(0.38)&	(1.41)&	(-1.41)&	(-0.10)&	(1.63) \\
$R^2$ (\%)& 5.05&	2.48&	5.59&	4.35&	1.45&	4.74&	7.36\\
\midrule
OOS $R^2$ (\%)& 6.74&	-1.05&	-14.70	&-1.29	&-7.08	&-0.95	&2.07\\
    \bottomrule
\end{tabular*}
}
		\end{center}
		\label{tab:factor_predict_results}
		\footnotesize
		{Note: Factor return predictive regressions ($ f_{k,t+1} = \mu_k + \lambda_k q_{k,t} + error_{k,t+1}$) for $k=$ each of the Fama-French-Carhart factors and JKP factors. The point estimates are in percentage terms. That is, the first cell indicates a one standard deviation increase in $q_{k,t}$ predicts a 1.04\% increase in market return in the following month. The $t$-statistics are based on Newey-West standard errors. The first five rows are full-sample regressions (2000-2022) with $R^2$ evaluated in the same full sample. The ordinary IS $R^2$ with a constant term is reported: $R^2 = 1 - \sum_{t} (f_{k,t+1} - \widehat{f}_{k,t+1})^2 / \sum_{t} (f_{k,t+1} - \widehat \mu_k)^2$. The last row ``OOS $R^2$'' is with the regressions estimated in 2000-2009 and evaluated in 2010-2022, and we benchmark the $R^2$ against predicting zero: OOS $R^2 = 1 - \sum_{t} (f_{k,t+1} - \widehat{f}_{k,t+1})^2 / \sum_{t} f_{k,t+1}^2$.}
	\end{table}%

	Table \ref{tab:factor_predict_results} presents the results of the time-series regression $f_{k,t+1} = \mu_k + \lambda_k q_{k,t} + error_{k,t+1}$ for various factors. The estimated $\lambda_k$ coefficients are predominantly positive and statistically significant for all Fama-French-Carhart factors and the majority of JKP factors. This indicates that each factor's expected return is positively related to its quantity, consistent with our theoretical motivation. The full-sample $R^2$ values are around 5\%, which is relatively high for factor return prediction at the monthly frequency (see \citealp{welch2008comprehensive}).

    However, the OOS $R^2$ values are mostly negative, with the exception of the market factor, which achieves a higher $R^2$ of $6.8\%$. 
    Nonetheless, we should interpret this high $R^2$ with caution as the time-series $R^2$ metric may contain significant noise. These results underscore the apparent limitations of using simple univariate time-series regression for predicting aggregate factor returns. The construction of the factor quantity series is not designed for time-series return prediction, which is understood to be a challenging task that requires more sophisticated methods and richer predictor data \citep*{kelly2013market, kelly2024virtue}.\footnote{Dynamic regime shift models can further enhance prediction accuracy by accounting for structural breaks (\citealp{smith2020equity,smith2021break,gao2023optimizing}).}

	\subsection{Additional results on stock return forecasting}

    \begin{table}[!t]
		\caption{{Table \ref{tab:xsreg_ff_factors} continued, $\beta$-only model's coefficient estimates}}
		\vspace*{-1em}
		\renewcommand{\arraystretch}{1}
		\begin{center}
			\scalebox{1}{ \newcolumntype{C}{>{\centering\arraybackslash}p{0.15\textwidth}}

\begin{tabular*}{\textwidth}{@{\extracolsep{\fill}}cCCCCC}
    \toprule
    &CAPM& FF3 & FF3C & FF5 & FF5C\\
    \midrule
\multicolumn{5}{l}{$\beta$-only model: $\mu_k$ (\%, monthly), t-statistics in parentheses}\\[6pt]
MKT& 0.38& 	0.45& 	0.35& 	0.55& 	0.50\\
& (1.07)& 	(0.90)& 	(0.75)	& (1.21)	& (1.15)
\\
SMB&&-0.05	& 0.06	& -0.04& 	0.09\\
& & (-0.15)& 	(0.19)& 	(-0.11)& 	(0.27)\\
HML&& 0.58& 	0.51& 	0.56& 	0.44\\
& & (1.74)& 	(1.59)	& (1.49)& 	(1.23)\\
MOM&&&	-0.41&&		-0.48\\
&&	&(-1.09)	&&	(-1.26)\\
CMA&&		&&0.04&	0.10\\
&&	&&	(0.17)&	(0.51)\\
RMW&&	&&	0.09	&0.13\\
&&	&&	(0.34)&	(0.52)\\
\bottomrule
\end{tabular*}
}
		\end{center}
		\label{tab:xsreg_ff_factors_coef}
		\footnotesize
		{Note: Table \ref{tab:xsreg_ff_factors} in the main text reports the $R^2$ values for the BTQ and the $\beta$-only models, as well as the coefficients for the BTQ model. This table reports the $\beta$-only model's coefficient estimates. Same as the main text table, the $\mu_k$ coefficients are in percentage terms, and the $t$-statistics are based on standard errors clustered by month.}
	\end{table}%

	This appendix subsection contains additional empirical results on stock return forecasting that are omitted in the main text Subsection \ref{subsec:retpred}. Table \ref{tab:xsreg_ff_factors_coef} completes Table \ref{tab:xsreg_ff_factors} by providing the full-sample coefficient estimates for the $\beta$-only model.

	Table \ref{tab:xsreg_ff_factors} in the main text already shows that the $\beta$-only model has weak predictive power, with low and even negative $R^2$ values in some OOS cases. Table \ref{tab:xsreg_ff_factors_coef} further shows that the $\mu$ coefficients in the $\beta$-only model are either statistically insignificant or negative in various factor specifications. This, once again, shows the empirical difficulty in establishing a positive risk-return association using $\beta$ only without quantity information.

	\subsection{Additional results on factor selection}\label{appsubsec:factorzoo}

\begin{figure}[!t]
		\centering
		\caption{Factor importance in Lasso factor selection}
		\centering
		\includegraphics[scale=1]{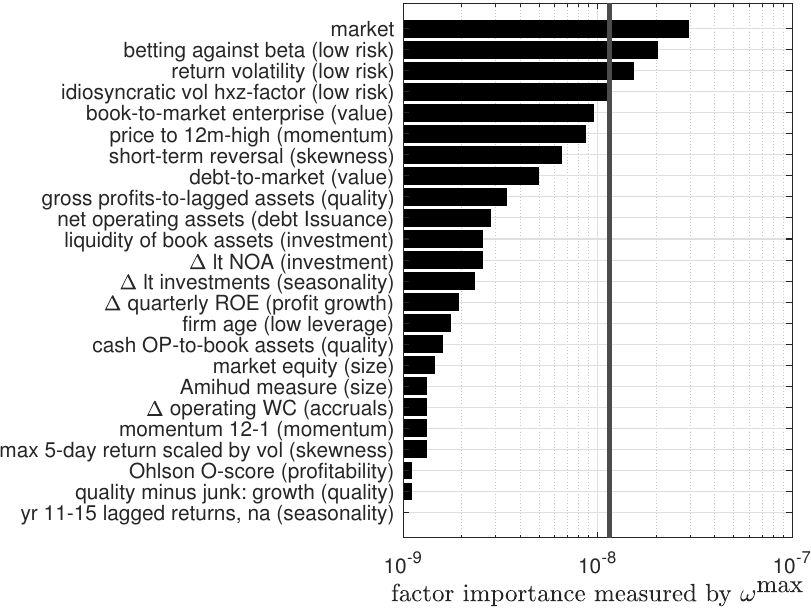}
		\begin{minipage}{\textwidth}
			{\medskip \footnotesize Note: We measure the importance of factor $k$ by $\omegamax_k$, the largest $\omega$ value at which factor $k$ is still selected. Specifically, $\omegamax_k \defeq \sup\{\omega: \widehat{\lambda}_k(\omega)\ne 0\}$, where $\widehat{\lambda}_k(\omega)$ is the Lasso estimate of $\lambda_k$ at hyperparameter $\omega$. This figure reports $\omegamax_k$ for the top 24 factors in the JKP factor zoo, omitting the rest with $\omegamax_k < 10^{-9}$. The vertical black line indicates the tuned $\omega$ based on ten-fold cross-validation.\par}
		\end{minipage}
		\label{fig:lasso_full_lagged_6_importance}
	\end{figure}

	This appendix subsection presents a more formal factor importance analysis and reports other important factors besides the top five reported in Section \ref{subsec:factorzoo} for the Lasso estimation (Eq.\ \ref{eq:lasso}). Factor importance is measured using a feature selection metric from the Lasso regressions. In particular, we measure the importance of factor $k$ by $\omegamax_k$, the largest $\omega$ value at which factor $k$ is still selected. Specifically, $\omegamax_k \defeq \sup\{\omega: \widehat{\lambda}_k(\omega)\ne 0\}$, where $\widehat{\lambda}_k(\omega)$ is the Lasso estimate of $\lambda_k$ at hyperparameter $\omega$. Figure \ref{fig:lasso_full_lagged_6_importance} reports $\omegamax_k$ for the top 24 factors in the JKP factor zoo, omitting the rest with $\omegamax_k < 10^{-9}$.

	The most significant factor is unambiguously the market factor, followed by the low-risk factors constructed with technical (past return) information, the value factor constructed with fundamental information, and a version of the momentum factors. The remaining less important factors are related to investment style clusters such as the value, quality, investment, seasonality, etc. Specifically, the top 24 factors' full names, the factor clusters they belong to, and code names (as in JKP data) are detailed below:
	\vspace{-20pt}
	\begin{singlespace}
		\setlength{\baselineskip}{1pt} %
		\begin{tabbing}
		\hspace{8cm}\=\hspace{4cm}\=\hspace{4cm}\kill %
			market \> -- \> \texttt{mkt}, \\
			betting against beta \> low risk \> \texttt{betabab\_1260d}, \\
			return volatility \> low risk \> \texttt{rvol\_21d}, \\
			idiosyncratic volatility q-factor \> low risk \> \texttt{ivol\_hxz4\_21d}, \\
			book-to-market enterprise \> value \> \texttt{bev\_mev}, \\
			current price to high price over last year \> momentum \> \texttt{prc\_highprc\_252d}, \\
			short-term reversal \> skewness \> \texttt{ret\_1\_0}, \\
			debt-to-market \> value \> \texttt{debt\_me}, \\
			gross profits-to-lagged assets \> quality \> \texttt{gp\_atl1}, \\
			net operating assets \> debt issuance \> \texttt{noa\_at}, \\
			liquidity of book assets \> investment \> \texttt{aliq\_at}, \\
			change in long-term net operating assets \> investment \> \texttt{lnoa\_gr1a}, \\
			change in long-term investments \> seasonality \> \texttt{lti\_gr1a}, \\
			change in quarterly return on equity \> profit growth \> \texttt{niq\_be\_chg1}, \\
			firm age \> low leverage \> \texttt{age}, \\
			cash-based operating profits-to-book assets \> quality \> \texttt{cop\_at}, \\
			market equity \> size \> \texttt{market\_equity}, \\
			Amihud measure \> size \> \texttt{ami\_126d}, \\
			change in current operating working capital \> accruals \> \texttt{cowc\_gr1a}, \\
			price momentum t-12 to t-1 \> momentum \> \texttt{ret\_12\_1}, \\
			highest 5 days of return scaled by volatility \> skewness \> \texttt{rmax5\_rvol\_21d}, \\
			Ohlson O-score \> profitability \> \texttt{o\_score}, \\
			quality minus junk: growth \> quality \> \texttt{qmj\_growth}, \\
			years 11-15 lagged returns, nonannual \> seasonality \> \texttt{seas\_11\_15na}. 
		\end{tabbing}
    \end{singlespace}

    \subsection{Interpreting the magnitude of $\lambda$ estimates and connection to the literature}\label{sec:comparelambda}
    
    This appendix shows that the magnitude of the $\lambda$ estimates---reported at the factor level in Tables \ref{tab:xsreg_individual_factor} and \ref{tab:xsreg_ff_factors} and at the stock level in Table \ref{tab:totalflow}---is consistent with the market reality and recent estimates from the literature. This alignment further supports our interpretation of the economic channel, although identifying the coefficient per se is not our goal; rather, the focus is on demonstrating the predictive power.

	The estimated market-level $\lambda_{\text{MKT}}$, presented in Tables \ref{tab:xsreg_individual_factor} and \ref{tab:xsreg_ff_factors}, ranges from $1.2$ to $1.8$ across various univariate and multivariate settings. This implies that for a one standard deviation increase in the quantity ($q$) of the market factor, the expected return of a stock with a market beta of 1 increases by $1.2\%$ to $1.8\%$ per month. 

	Our $\lambda$ coefficient can be converted to the ``price multiplier'' in the demand-based asset pricing literature. The $\lambda$ coefficient is the sensitivity between risk premium ($\mu$) and the quantity ($q$, constructed in Section \ref{sec:construct}). In the literature, the price multiplier is typically defined as $(\Delta P / P)/(\Delta Q / Q)$, where $\Delta Q / Q$ is the percentage change in the security's quantity (\citealp{gabaix2022search}). 

    To relate the magnitude of our $\lambda$ coefficient to that of the price multiplier, consider the following conversions. The monthly standard deviation of $q$ is $1.88\times 10^{-6}$ in terms of the pre-standardized $\qtld$ measure (see Table \ref{tab:factorflow_stats}). According to Eq.\ \ref{eq:porttheory}, this is roughly $0.22\%$ of ($\beta$-aggregated) monthly dollar flow shock as a fraction of the total U.S. stock market capitalization, after adjusting for the market factor's monthly variance: $0.22\% = \sqrt{6}\times 1.88\times 10^{-6} / (16\%^2 \div 12)$.\footnote{The annualized standard deviation of the market factor return during our sample period is $16\%$. We multiply by $\sqrt{6}$ because our pre-standardized $\qtld$ measure in Eq.\ \ref{eq:qtildeconstruction} uses a 6-month lagged average of flow shocks, and we assume these monthly shocks are independent.} In terms of the price multiplier measure, let the numerator be the flow shock, $d Q / Q = 0.22\%$, then the price multiplier itself is between $1.2\% / 0.22\% = 5.5$ and $1.8\% / 0.22\% = 8.2$, which matches in magnitude with the estimate of $5$ reported in \cite{gabaix2022search}.

    The magnitude of the $q$ variation is also consistent with market realities.\footnote{Significant noise in $q$ fluctuation is also one of the two requirements for our channel to have a significant empirical impact, as shown in Appendix \ref{sec:microfound}.} As shown above, a one-standard-deviation increase in market $q$ corresponds to a 0.22\% increase in dollar flow as a fraction of the total stock market capitalization. Considering that the mutual fund sector holds about $20 \sim 30\%$ of the total U.S.\ stock market capitalization \citep*{ben2022ratings}, this one standard deviation flow shock corresponds to about $0.7\% = 0.22 / 30 \sim 1.1\% = 0.22 / 20$ of the mutual fund sector's total AUM, which is a reasonable level.    
    
    From the opposite direction, we can also justify the magnitude of our predictability using the literature's price multiplier estimates. Assuming that a one-standard-deviation change in $q$ captures a flow shock of about 1\% of the total mutual fund sector's AUM, which multiplies to about $0.2\% = 1\% \times 20\%$ of the total stock market capitalization. With a market-level price multiplier of 5 \citep{gabaix2022search}, this implies a monthly market factor premium variation of about 1\% = 0.2\% $\times$ 5. Under a one-factor model, this factor premium variation matches the observed total stock return variation, with a monthly standard deviation of 10\%, at an $R^2 \approx 1\% = (1\%)^2 / (10\%)^2$.

	At the stock level, the coefficients in Table \ref{tab:totalflow} column $\lamstock$, are also comparable to the price multipliers of stocks to flows reported in the literature.  This is because we have normalized $\flowstock_{i,t}$ by the stock's own market cap in constructing the $\qstock_{i,t}$ predictors (see Eq.\ \ref{eq:qstockh}). The magnitude of our $\lamstock$ stabilizes between $0.5$ and $0.8$ for larger $h$ values, which is close to the multiplier estimates in the literature, generally ranging from $0.5$ to $1$ (e.g., \citealp{lou2012flow}; \citealp{chang2015regression}; \citealp{da2018destabilizing}; \citealp{hartzmark2022predictable}; \citealp{pavlova2023benchmarking}; \citealp{li2024impact}).

    \subsection{Additional robustness checks on return predictability} \label{app:add_rob}

	Table \ref{tab:OOS_R2_FF} evaluates the robustness of the BTQ model's predictive power across different size and time sub-samples for the Fama-French-Carhart factors. Panel A breaks down the stock-month observations in the OOS evaluation panel into five size groups using concurrent NYSE market capitalization breakpoints. The same OOS predicted returns ($\rhat_{i, t+1}$) are respectively evaluated in each size group. Panel B breaks down the OOS panel by time into three sub-periods: 2010-2014, 2015-2018, and 2019-2022, and reports sub-sample $R^2$ similarly. Panel C repeats the original joint OOS (2010-2022) evaluation reported in the main text Table \ref{tab:xsreg_ff_factors} Panel B Line ``BTQ'' for ease of reference.

    \begin{table}[!t]
		\caption{{BTQ OOS prediction accuracy ($R^2$ in \%) in size and time sub-samples} }
		\vspace*{-1em}
		\renewcommand{\arraystretch}{1.3}
		\begin{center}
			\scalebox{1}{ \begin{tabular*}{\textwidth}{l@{\extracolsep{\fill}}cccccc}
    \toprule
    evaluation sample & \# of obs.  &CAPM& FF3 & FF3C & FF5 & FF5C\\
    &~ & \hspace{-2.5em}$K=1$ & 3 & 4 & 5 & 6\\
    \midrule
    \multicolumn{6}{l}{
    Panel A: size group evaluation}\\
1 (small)	&323,617&	0.69&	0.72&	0.72&	0.58&	0.64\\
2	&165,059	&0.99&	1.37	&1.44	&0.51&	0.79\\
3	&141,153	&1.16	&1.74&	1.83&	0.42&	0.82\\
4	&115,763&	0.76&	1.97&	2.20&	-0.33	&0.46\\
5 (big)	&103,927	&-0.56	&1.66	&2.00&	-1.18	&-0.17\\
\midrule
    \multicolumn{6}{l}{
    Panel B: sub-period evaluation}\\
2010-2014	&321,425	&1.10&	1.33	&1.34	&1.03	&0.98\\
2015-2018	&255,959	&0.17&	0.11&	0.11&	0.07	&0.07\\
2019-2022	&272,135&	0.90&	1.38	&1.47&	0.37	&0.81
\\
    \midrule
    \multicolumn{6}{l}{
    Panel C: original OOS evaluation (in Table \ref{tab:xsreg_ff_factors} Panel B)}\\
    OOS (2010-2022) & 849,519  & 0.75&1.03&1.07&0.44&0.65\\
    \bottomrule
\end{tabular*}}
		\end{center}
		\label{tab:OOS_R2_FF}
		\footnotesize
		{Note:} OOS $R^2$ evaluated in different size and time sub-samples for the Fama-French-Carhart factors. Panel A breaks down the OOS panel into five size groups according to NYSE market capitalization quintiles and reports the OOS $R^2$ in each size group. Panel B breaks down the OOS evaluation into three sub-periods: 2010-2014, 2015-2018, and 2019-2022. Panel C repeats the original joint OOS (2010-2022) evaluation reported in the main text Table \ref{tab:xsreg_ff_factors} Panel B Line ``BTQ'' for ease of reference.
	\end{table}%

	Table \ref{tab:OOS_R2_FF} shows that the BTQ model's predictive power reported in Table \ref{tab:xsreg_ff_factors} is robust in most size and time sub-samples. In particular, the FF3 and FF3C specifications perform even better in large stocks, which are usually the most challenging group for stock return prediction. Across sub-periods, the BTQ model's predictive power is relatively stable over time. The first and the last sub-periods (2010-2014 and 2019-2022) have higher $R^2$ values than the middle sub-period (2015-2018) across various model specifications. These size and time sub-sample results for the Fama-French-Carhart factors are very similar to those reported for the factors selected from the factor zoo and the selected PC factors in main text Table \ref{tab:OOS_R2_LASSOPC}.

	Table \ref{tab:totalflowfull} extends the main text Table \ref{tab:totalflow} to all $h$ from $1$ to $12$ and confirms that the ``quantity-only'' alternative model does not forecast stock returns.

	\begin{table}[t!]
		\caption{``Quantity-only'' alternative model does not forecast stock returns: expanding Table \ref{tab:totalflow} to all $h$ from $1$ to $12$}
		\label{tab:totalflowfull}
		\vspace*{-1em}
		\renewcommand{\arraystretch}{1.3}
		\begin{center}
			\scalebox{1}{ \begin{tabular*}{\textwidth}{c@{\extracolsep{\fill}}cccccccc}
    \toprule
    & \multicolumn{4}{l}{A. constant $\lamstock$} & \multicolumn{2}{l}{B. $\lamstock$ by size quintile}& \multicolumn{2}{l}{C. $\lamstock_i$ by stock}\\
    \cmidrule(lr){2-5} \cmidrule(lr){6-7} \cmidrule(lr){8-9}
    $h$~ &IS $R^2$(\%) \hspace*{-12pt} & OOS $R^2$(\%) \hspace*{-12pt} & $\lamstock$ & \hspace*{6pt}$t$-stat  \hspace*{10pt}
    &IS $R^2$(\%) \hspace*{-16pt}  & OOS $R^2$(\%)
    &\hspace*{-6pt}IS $R^2$(\%) \hspace*{-12pt}  & OOS $R^2$(\%) \hspace*{-18pt}  \\
    \midrule
    1	&		0.000	&0.000&	-0.10	&-0.27&0.003&	-0.001&0.47	&-232\\
    2	&		0.000&	-0.001&	0.05	&0.12&0.002&	-0.003&0.44&	-215\\
    3		&	0.000	&0.000	&0.17&	0.29&0.003&	0.000&0.39	&-155\\
    4		&	0.000	&-0.001	&0.20&	0.33&0.004&	-0.002&0.39&	-156\\
    5	&		0.003&	0.001	&0.51&	0.75&0.006&	0.002&0.41	&-150\\
    6		&	0.005&	0.006	&0.75&	1.01&0.007&	0.006&0.41	&-107\\
    7	&		0.006&	0.008	&0.82	&1.12&0.008&	0.008&0.41	&-107\\
   8	&		0.003	&0.003&	0.66	&0.87&0.006	&0.005&0.37	&-142\\
    9	&		0.004	&0.004	&0.75&	0.99&0.006&	0.006&0.38	&-96\\
    10	&		0.003&	0.000	&0.69&	0.96&0.006&	0.003&0.38	&-101\\
    11	&		0.003&	-0.003&	0.73&	0.95&0.006&	0.000&0.38	&-98\\
    12		&	0.003	&-0.007	&0.77	&1.01&0.006	&-0.004&0.38&	-81\\
    \bottomrule
\end{tabular*}}
		\end{center}
		\footnotesize
		{Note: Panel A: univariate predictive regression, $r_{i,t+1} = \lamstock \qstockh_{i,t} + error_{i,t+1}$. B: size-dependent predictive regression, $r_{i,t+1} = \lamstock_{\text{size\_quintile}(i, t)} \qstockh_{i,t} + error_{i,t+1}$, where $\lamstock_{\text{size\_quintile}(i, t)}$ is indexed by the NYSE size quintile of the stock. C: stock-specific predictive regression, $r_{i,t+1} = \lamstock_i \qstockh_{i,t} + error_{i,t+1}$. The $R^2$ values are expressed as percentages, e.g., $0.005$ in row $h=6$ means $0.005\%$, a very small value.}
	\end{table}%

    \subsection{Details for macroeconomic variables as alternatives to factor quantity}\label{appsubsec:macro}

    The second exercise in Section \ref{sec:channels} examines the use of macroeconomic variables as alternatives to factor quantity ($q$) in the BTQ model. This subsection reports the accompanying technical details and additional results.

	We preprocess the 126 FRED-MD macro series following standard procedures. We first transform each raw series to a stationary process according to the transformation code provided by FRED-MD. We then standardize these macro variables by demeaning them and dividing by their standard deviations, using the mean and standard deviation estimated during the in-sample period (2000-2009).

	Table \ref{tab:macro} in the main text reports the comparisons in the single-factor setting by predicting stock returns with predictors in the form of market beta times a macro variable (or linear combinations of the macro variables). In this appendix, Table \ref{tab:macro_OOSR2_SummaryStats} expands the analysis to the multi-factor settings. The predictors are $\Brack{\betahat_{i, \text{MKT}, t} x_{j, t},\betahat_{i, \text{SMB}, t} x_{j, t},\betahat_{i, \text{HML}, t} x_{j, t}, \dots}$, where $x_{j, t}$ is the $j$-th macroeconomic variable, or a full-sample principal component (PC) of the 126 macroeconomic variables. We observe that beta times macro variables in multi-factor settings also do not predict stock returns.

    \begin{table}[t!]
		\caption{``Beta times macro variables'', expanding Table \ref{tab:macro} to multi-factor settings}
		\label{tab:macro_OOSR2_SummaryStats}
		\vspace*{-1em}
		\renewcommand{\arraystretch}{1.3}
		\begin{center}
			\scalebox{1}{ \newcolumntype{C}{>{\centering\arraybackslash}p{0.1\textwidth}}
\begin{tabular*}{\textwidth}{@{\extracolsep{\fill}}CCCCCC}
    \toprule
    &CAPM& FF3 & FF3C & FF5 & FF5C\\
    \midrule
    \multicolumn{6}{l}{A. Benchmark}\\
    BTQ & 0.75&1.03&1.07&0.44&0.65\\
    \midrule
    \multicolumn{6}{l}{B. Multi-factor beta times each macro-variable}\\
Mean	&-0.63&	-1.85	&-2.11	&-8.69	&-8.60\\
Q10&	-1.88&	-6.08	&-6.81	&-19.43	&-18.27\\
Q25	&-0.63	&-1.67&	-2.65	&-4.04	&-5.92\\
Q50	&-0.12	&-0.47&	-0.57	&-1.25	&-1.28\\
Q75&	0.00&	-0.08	&-0.12	&-0.31&	-0.38\\
Q90	&0.13&	0.02&	0.01	&-0.07&	-0.12\\
Q95&	0.18&	0.11&	0.12&	0.01&	0.00\\
Max	&0.28&	0.38&	0.37	&0.41&	0.39\\
\midrule
\multicolumn{6}{l}{C. Multi-factor beta times PCs of macro-variables}\\
PC1	&-0.17	&-0.67	&-2.82&	-20.07&	-15.37\\
PC2&	0.01&	0.02&	-0.06&	-0.26	&-0.26\\
PC3	&-1.05	&-5.93&	-5.90&	-5.70	&-4.43\\
PC4	&-0.09&	-1.88	&-1.87	&-6.41	&-9.73\\
PC5&	-0.46&	-0.82	&-0.81	&0.04	&-0.24\\
    \bottomrule
\end{tabular*}
}
		\end{center}
		\footnotesize
		{Note: Predicting stock returns using conditional multi-factor predictors in the form of ``market beta times ...'', ``SMB beta times ...'', ``HML beta times ...'', and so on. All numbers are OOS $R^2$ in percentage.}
	\end{table}%

	Table \ref{tab:macro} Panel D searches for the optimal linear combinations of the 126 FRED-MD macro variables that can substitute for the $q_{\text{mkt}, t}$ variables in the single-factor CAPM BTQ model. In particular, we minimize the following predictive least squares loss:
    \begin{align}
		\min_{\Curly{b_j}}\sum_{i, t\in \text{IS}} \Paren{r_{i,t+1} - \betahat_{i, \text{mkt}, t} \Paren{\sum_{j \in \text{FRED-MD}} b_j x_{j, t}}}^2,
    \end{align}
    where $x_{j, t}$ is the $j$-th macro variable at time $t$, and $b_j$ is the coefficient to be estimated. This search is equivalent to an OLS regression of $r_{i,t+1}$ on 126 predictors $\betahat_{i, \text{mkt}, t} x_{j, t}$.

	This OLS setup drastically overfits with an extremely negative OOS predictive $R^2$, as shown in Table \ref{tab:macro} Panel D, which is expected given the large number of predictors. To address the overfitting issue, we apply regularization to the linear combination search by regressing $r_{i,t+1}$ on $\betahat_{i, \text{mkt}, t} x_{j, t}$ with LASSO penalty on the coefficients. Note that $x_{j, t}$ has already been demeaned and standardized using the in-sample period estimates. In detail, the Lasso optimization problem is:
    \begin{align}
    \min_{\Curly{b_j}}\frac 1 {2|\text{IS}|}\sum_{i, t\in \text{IS}} \Paren{r_{i,t+1} - \betahat_{i, \text{mkt}, t} \Paren{\sum_{j \in \text{FRED-MD}} b_j x_{j, t}}}^2+ \omega \sum_{j \in \text{FRED-MD}} |b_j|,
    \end{align}
    where $\omega$ is the regularization parameter.

    \begin{figure}[t!]
		\centering
		\caption{Market beta times the best linear combination of the macro variables, prediction accuracy along regularization path}
		\label{fig:macroLASSO}
		\includegraphics[scale=0.7]{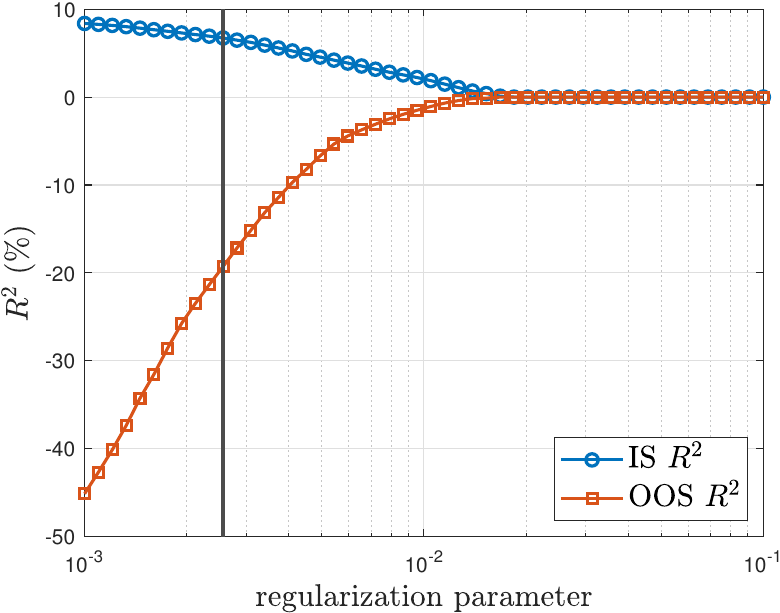}
		\begin{minipage}{\textwidth}
			\vspace{1em}
			{ \footnotesize Note: Model fit and parameter estimates as the regularization parameter ($\omega$, horizontal axis) varies. The IS $R^2$ is evaluated in the training window (2000-2009), and the OOS $R^2$ is the same model evaluated in the testing window (2010-2022). The vertical black line indicates the tuned $\omega$ based on ten-fold cross-validation.\par}
		\end{minipage}
    \end{figure}

	Figure \ref{fig:macroLASSO} plots both the IS and OOS $R^2$ values across different regularization parameters. The OOS $R^2$ values remain below $0$ across all $\omega$ values and only approach $0$ at large $\omega$ values, where all coefficients are shrunk to zero, resulting in a model that predicts zero returns for all stock-months. This result suggests the fitted dimension reduction combinations of macro variables do not have predictive power either.

    \clearpage
    \renewcommand{\thepage}{A.\arabic{page}}

\end{appendices}

\end{document}